\documentclass[structabstract]{aa}  
\usepackage{graphicx}
\usepackage{txfonts}
\usepackage{natbib}
\usepackage{float}
\usepackage{lscape}
\usepackage{nicefrac}

\begin{document}
   \title{Far-infrared molecular lines from Low- to High-Mass \\ Star Forming Regions
    observed with \textit{Herschel}}

   \author{A.~Karska\inst{1,2}, F.~Herpin\inst{3,4}, S.~Bruderer\inst{1}, J.R.~Goicoechea\inst{5}, G.J.~Herczeg\inst{6},
   E.F.~van Dishoeck\inst{1,2},  I.~San~Jos\'{e}-Garc\'{i}a\inst{2},
   A.~Contursi\inst{1}, H.~Feuchtgruber\inst{1}, D.~Fedele\inst{1}, A.~Baudry\inst{3,4},
   J.~Braine\inst{3,4}, L.~Chavarr\'{i}a\inst{5}, J.~Cernicharo\inst{5}, F.F.S.~van der Tak\inst{7,8}, 
   and F.~Wyrowski\inst{9}}

	\institute{
   			  $^{1}$ Max-Planck Institut f\"{u}r Extraterrestrische Physik (MPE),
   			  Giessenbachstr. 1, D-85748 Garching, Germany\\
   			  $^{2}$ Leiden Observatory, Leiden University, P.O. Box 9513,
          	  2300 RA Leiden, The Netherlands\\
          	  $^{3}$ Universit\'{e} de Bordeaux, Observatoire Aquitain des Sciences de l'Univers, 2 rue de
l'Observatoire, BP 89, F-33271 Floirac Cedex, France \\
			  $^{4}$ CNRS, LAB, UMR 5804, F-33271 Floirac Cedex, France \\
        	  $^{5}$ Centro de Astrobiolog\'{\i}a. Departamento de Astrof\'{\i}sica. CSIC-INTA. Carretera de Ajalvir,
				Km 4, Torrej\'{o}n de Ardoz. 28850, Madrid, Spain\\
              $^{6}$ Kavli Institut for Astronomy and Astrophysics, Yi He Yuan Lu 5, HaiDian Qu, Peking University, Beijing,
          	  100871, PR China\\
          	  $^{7}$ SRON Netherlands Institute for Space Research, PO Box 800, 9700 AV, Groningen, The Netherlands\\
              $^{8}$ Kapteyn Astronomical Institute, University of Groningen, PO Box 800, 9700 AV, Groningen, The
				Netherlands\\
              $^{9}$ Max-Planck-Institut f\"{u}r Radioastronomie, Auf dem H\"{u}gel 69, 53121 Bonn, Germany\\
             \email{karska@mpe.mpg.de}
             }

   \date{Received May 24, 2013; accepted November 26, 2013}
	\titlerunning{Far-infrared molecular lines from Low- to High-Mass Star Forming Regions
    observed with Herschel}
	\authorrunning{A.~Karska et al. 2013}

  \abstract
  {}
  {Our aim is to study the response of the gas to energetic processes associated with high-mass 
  star formation and compare it with previously published studies on low- and 
  intermediate-mass young stellar objects (YSOs) using the same methods. The quantified far-infrared 
  line emission and absorption of CO, H$_2$O, OH, and [\ion{O}{i}] reveals the excitation and 
  the relative contribution of different atomic 
   and molecular species to the gas cooling budget.}
  {Herschel-PACS spectra covering 55--190 $\mu$m are analyzed for ten high-mass star 
  forming regions of luminosities $L_\mathrm{bol}\sim10^{4}-10^{6}$ $L_{\odot}$ and 
  various evolutionary stages at spatial scales of $\sim10^4$ AU. Radiative transfer 
  models are used to determine the contribution of the quiescent envelope to the far-IR 
  CO emission.}
  {The close environments of high-mass protostars show strong far-infrared emission from molecules,
  atoms, and ions. Water is detected in all 10 objects even up to high excitation lines, often in 
   absorption at the shorter wavelengths and in emission at the longer wavelengths.
    CO transitions from $J=14-13$ up to typically $29-28$ ($E_\mathrm{u}/k_\mathrm{B}\sim580-2400$ K)
    show a single temperature component
    with a rotational temperature of $T_\mathrm{rot}\sim$300 K. 
    Typical H$_2$O excitation temperatures are $T_\mathrm{rot}\sim$250 K, while OH has $T_\mathrm{rot}\sim$80 K. 
    Far-IR line cooling is dominated by CO ($\sim75$\%) and to a smaller extent by 
     [\ion{O}{i}] ($\sim$20 \%), which becomes more important for the most evolved sources. 
     H$_2$O is less important as a coolant for high-mass sources due to the fact that many lines are 
     in absorption.}
    {Emission from the quiescent envelope is responsible for $\sim45-85$ \% of the total CO luminosity 
    in high-mass sources compared with only $\sim$10\% for low-mass YSOs. The highest$-J$ lines ($J_\mathrm{up}\geq$20) 
    originate most likely from shocks, based on the strong correlation 
    of CO and H$_2$O with physical parameters ($L_\mathrm{bol}$, $M_\mathrm{env}$) of the sources from low- to high-mass YSOs. 
    Excitation of warm CO described by $T_\mathrm{rot}\sim$300 K is very similar for all mass regimes, whereas H$_2$O temperatures are 
    $\sim$100 K higher for high-mass sources compared with low-mass YSOs. The total 
    far-IR cooling in lines correlates strongly with bolometric luminosity, consistent with 
    previous studies restricted to low-mass YSOs. Molecular cooling (CO, H$_2$O, and OH) 
    is $\sim$4 times more 
    important than cooling by oxygen atoms for all mass regimes. The total far-IR line 
    luminosity is about 10$^{-3}$ and 10$^{-5}$ times lower than the dust luminosity for the 
    low- and high-mass star forming regions, respectively.}
   

   \maketitle
%
\section{Introduction}
High-mass stars ($M>8$ M$_{\odot}$) play a central role in the energy budget, the shaping,
and the evolution of galaxies \citep[see review by][]{ZY07}. They are the main source of 
UV radiation in galaxy disks. Massive outflows and \ion{H}{ii} regions are powered by massive stars and 
are responsible for generating turbulence and heating of the interstellar medium (ISM). 
At the end of their lives, they inject heavy elements into the ISM that
  form the next generation of molecules and dust
  grains. These atoms and molecules are the main cooling channels of the ISM.
   The models of high-mass star formation are still 
  strongly debated; the two competing scenarios are 
  turbulent core accretion and `competitive accretion' \citep[e.g.][]{Ce05}. 
Molecular line observations are crucial to determine the impact of 
UV radiation, outflows, infall and turbulence on the formation and evolution of the high-mass 
protostars and ultimately distinguish between those models. 
 
Based on observations, the `embedded phase' of high-mass star formation may 
empirically be divided into several stages \citep[e.g.][]{HvD97,vT00,Be07}:
 (i) massive pre-stellar cores (PSC); (ii) high-mass protostellar objects (HMPOs); 
 (iii) hot molecular cores (HMC); and (iv) ultra-compact \ion{H}{ii} regions (UC\ion{H}{ii}). 
The pre-stellar core stage represents initial conditions of high-mass star formation,
 with no signatures of outflow / infall or maser activity. During the high-mass protostellar objects
  stage, infall of a massive envelope onto the central star and strong outflows 
  indicate the presence of an active protostar. In the hot molecular core stage, 
  large amounts of warm and dense gas and dust are seen. The temperature of $T>100$ K
  in subregions $<0.1$ pc in size is high enough to evaporate molecules off the grains. 
 In the final, ultra-compact \ion{H}{ii} regions stage, a considerable amount of ionized 
 gas is detected surrounding the central protostar.  

The above scenario is still debated \citep{Be07}, in particular whether stages (ii) and 
(iii) are indeed intrinsically different. The equivalent sequence for the
 low-mass Young Stellar Objects (hereafter YSO) is better established \citep{Shu87,An93,An00}. 
 The `embedded phase' of low-mass protostars consists of: (i) pre-stellar core phase; 
(ii) Class 0; and (iii) Class I phase. The Class 0 YSOs are surrounded 
by a massive envelope and drive collimated jets/ outflows. In the more evolved Class I 
objects the envelope is mostly dispersed and more transparent for UV radiation; the outflows 
are less powerful and have larger opening angles. 

Low mass sources can be probed at high spatial resolution due to a factor 
of 10 smaller distances, which allow us to study well-isolated 
sources and avoid much of the confusion due to clouds along the line of sight. The line emission 
is less affected by foreground extinction and therefore provides a good tool to 
study the gas physical conditions and chemistry in the region. 
The slower evolutionary timescale results in a larger number of low-mass YSOs compared
 to their high-mass counterparts, which is also consistent with observed stellar/ core mass functions.

 While low-mass YSOs are extensively studied in the far-infrared,
 first with the Infrared Space Observatory \citep[ISO,][]{ISO} and now
 with \textit{Herschel} \citep{Herschel}\footnote{Herschel is an ESA
   space observatory with science instruments provided by European-led
   Principal Investigator consortia and with important participation
   from NASA.}, the same is not the case for high-mass sources
 \citep[see e.g.][]{HvD97,Va01,Bo03}.  For those, the
 best-studied case is the relatively nearby Orion BN-KL region
 observed with ISO's Long- and Short-Wavelength Spectrometers
 \citep{LWS,SWS}. Spectroscopy at long-wavelengths (45-197 $\mu$m)
 shows numerous and often highly-excited H$_2$O lines in emission
 \citep{Ha98}, high-$J$ CO lines \citep[e.g., up to $J$=43-42 in OMC-1
 core,][]{Se00} and several OH doublets \citep{Go06}, while the shorter
 wavelength surveys reveal the CO and H$_2$O vibration-rotation bands
 and H$_2$ pure rotational lines \citep{vD98,Ro00}.
 Fabry-Perot (FP) spectroscopy ($\lambda$/$\Delta\lambda\sim$10,000,
 30 km s$^{-1}$) data show resolved P-Cygni profiles for selected
 H$_2$O transitions at $\lambda<100$ $\mu$m, with velocities extending
 up to 100 km s$^{-1}$ \citep{Wr00,Ce06}.  
At the shortest wavelengths ($<45$ $\mu$m) all pure rotational H$_2$O lines show
absorption \citep{Wr00}. ISO spectra towards
 other high-mass star forming regions are dominated by atomic and ionic lines \citep[see review by][]{EvDISO},
  similar to far-IR spectra of extragalactic sources \citep{Fi99,St02}.

The increased sensitivity, spectral and spatial resolution of the Photodetector 
Array Camera and Spectrometer (PACS) \citep{Po10} onboard \textit{Herschel} 
now allows the detailed study of the molecular content of a larger sample of high-mass star forming regions. 
 In particular, the more than an order of magnitude improvement in the spectral resolution 
 in comparison with the ISO-LWS grating observing mode allows the routine detections of 
 weak lines against the very strong continuum of high-mass sources with \textit{Herschel}, with line-to-continuum ratios below 1\%.
 
 The diagnostic capabilities of far-infrared lines have been
 demonstrated by the recent results on low- and intermediate-mass
 YSOs and their outflows
 \citep{Fi10,He12,Go12,Ma12,Wa13,Ka13,Gr13}.  The CO ladder from
 $J$=14-13 up to 49-48 and a few tens of H$_2$O lines with a range of
 excitation energies are detected towards the Class 0 sources, NGC1333
 IRAS4B and Serpens SMM1 \citep{He12,Go12}.  The highly-excited H$_2$O
 8$_{18}$--7$_{07}$ line at 63.3 $\mu$m
 ($E_\mathrm{u}/k_\mathrm{B}=1071$ K) is seen towards almost half of
 the Class 0 and I sources in the \citet{Ka13} sample, even for
 bolometric luminosities as low as \mbox{$\sim1$ L$_{\odot}$}.
 Non-dissociative shocks and to a smaller extent UV-heating are
 suggested to be the dominant physical processes responsible for the
 observed line emission \citep{vK10,Vi11,Ka13}. The contribution from
 the bulk of the quiescent warm protostellar envelope to the PACS
 lines is negligible for low-mass sources. Even for the
 intermediate-mass source NGC7129 FIRS2, where the envelope
 contribution is higher, the other processes dominate \citep{Fi10}.

In this paper, we present Herschel-PACS spectroscopy of 10 sources
that cover numerous lines of CO, H$_2$O, OH, and [\ion{O}{i}] 
lines obtained as part of the `Water in star forming regions with Herschel' (WISH) key program
 \citep{WISH}. WISH observed in total about 80 protostars at
different evolutionary stages (from prestellar cores to circumstellar
disks) and masses (low-, intermediate- and high-mass) with both the
Heterodyne Instrument for the Far-Infrared \citep[HIFI;][]{dG10} and
PACS \citep{Po10}. This paper focusses only on PACS observations of high-mass YSOs. 
It complements the work by \citet{vT13}, which describes our source sample and uses HIFI
to study spectrally resolved ground-state H$_2$O lines towards all our objects.
That paper also provides updated physical models of their envelopes.

The results on high-mass YSOs will be compared with those for low- and intermediate-mass
 young stellar objects, analyzed in a similar manner \citep{Ka13,Wa13,Fi10} 
in order to answer the following questions:
How does far-IR line emission/absorption differ for high-mass protostars at 
different evolutionary stages? What are the
dominant gas cooling channels for those sources? What physical components 
do we trace and what gas conditions cause the excitation of the observed lines?
Are there any similarities with the low- and intermediate-mass protostars?
 
The paper is organized as follows: \S 2 introduces the source sample
and explains the observations and reduction methods; \S 3 presents
results that are derived directly from the observations; \S 4 focuses
on the analysis of the data; \S 5 discusses our results in the context of the
available models and \S 6 summarizes the conclusions.\\
\begin{figure*}[tb]
\begin{center}
\includegraphics[angle=90,height=13cm]{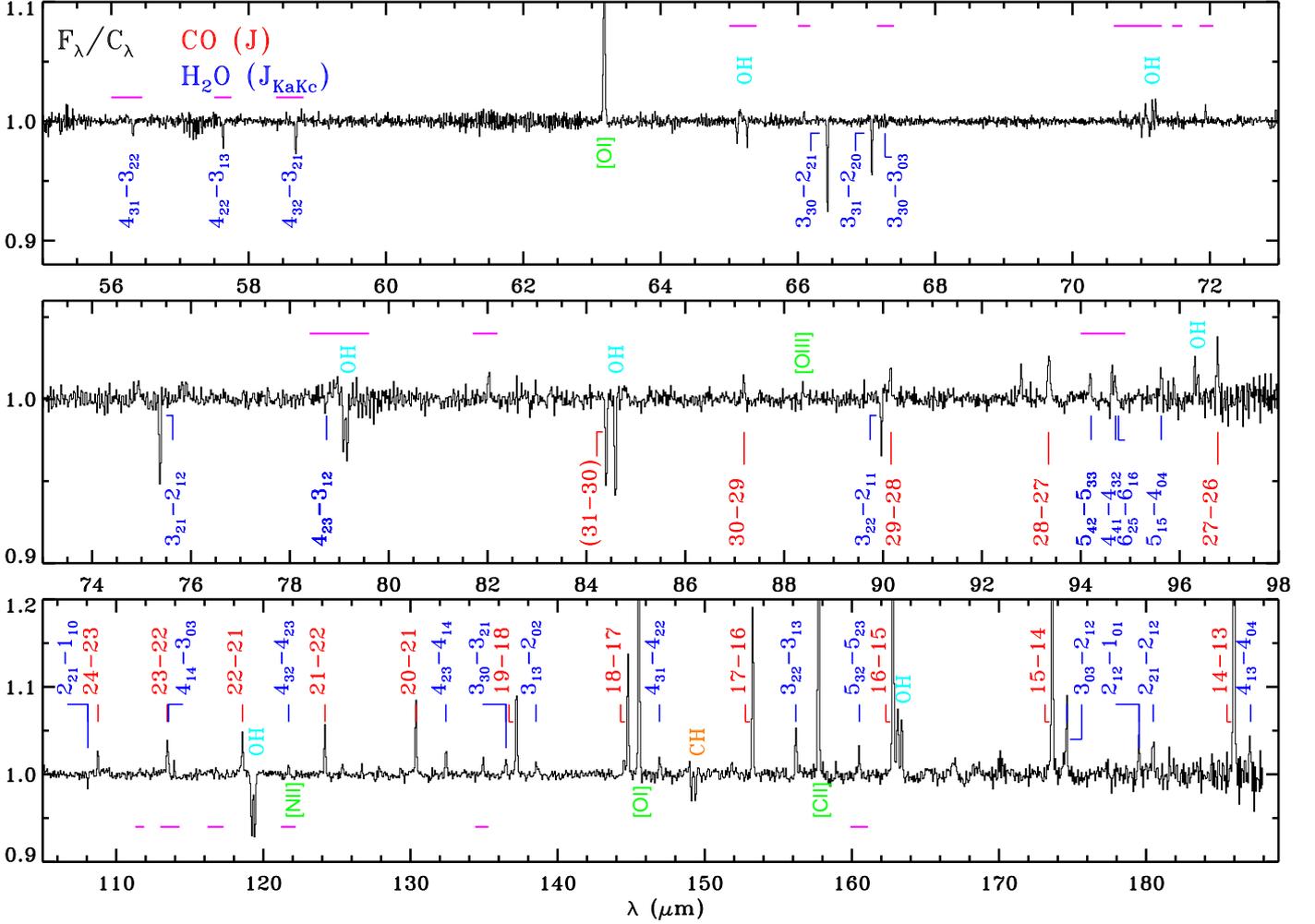}
\vspace{0.3cm}

\caption{\label{spec} Herschel-PACS continuum-normalized spectrum of W3 IRS5 at the central position. Lines of 
CO are shown in red, H$_2$O in blue, OH in light blue, CH in orange, and atoms and ions in 
green. Horizontal magenta lines show spectral regions zoomed in Figure \ref{wblow}.}
\end{center}
\end{figure*}

\section{Observations and data reduction}
\begin{figure}[!tb]
\begin{center}
 \includegraphics[angle=0,height=10cm]{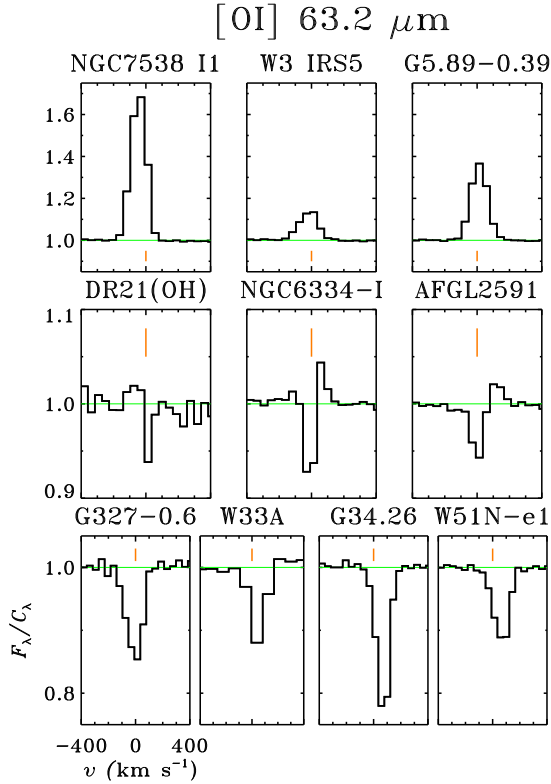} 
\vspace{0.3cm}

\caption{\label{oxyprofiles} Herschel-PACS profiles of the [\ion{O}{i}] 63.2 $\mu$m line at 
central position.}
\end{center}
\end{figure}
We present spectroscopy observations of ten high-mass star forming regions collected 
with the PACS instrument on board \textit{Herschel}
in the framework of the `WISH' program. The sources have an average distance of $\langle$D$\rangle$=2.7 kpc, 
and represent various stages of evolution, from the classical high-mass 
protostellar objects (HMPOs) to hot molecular cores (HMCs) and ultra-compact \ion{H}{ii}
regions (UC \ion{H}{ii}). The list of sources and their basic properties are given in 
Table \ref{catalog}. Objects are shown in the sequence of increasing value of
 an evolutionary tracer, $L^{0.6}M_\mathrm{env}^{-1}$, introduced in \citet{Bo96}. 
The sequence does not always correspond well with the evolutionary stages most commonly assigned
 to the sources in the literature (last column of Table 2), perhaps because
multiple objects in different evolutionary stages are probed 
 within our spatial resolution element \citep[see e.g.][for the case of G327-0.6]{Wy06,Le13}.

\begin{table}
\begin{minipage}[t]{\columnwidth}
\caption{Catalog information and source properties.}
\label{catalog}
\centering
\renewcommand{\footnoterule}{}  
\begin{tabular}{lllrrrrrrrrr}
\hline \hline
Object & $D$ & $L_\mathrm{bol}$ & 
$M_\mathrm{env}$ & $L^{0.6}M_\mathrm{env}^{-1}$ & Class \\
~ &  (kpc) &    ($L_\mathrm{\odot}$) & ($M_\mathrm{\odot}$) 
&  ($L_\mathrm{\odot}^{0.6}$$M_\mathrm{\odot}^{-1}$)  & \\
\hline
G327-0.6 	& 3.3  & 7.3 10$^4$ & 2044 & 0.41 & HMC \\
W51N-e1     & 5.1  & 5.2 10$^5$ & 4530 & 0.59 & UC\ion{H}{ii}\\
DR21(OH)    & 1.5  & 1.3 10$^4$ & 472 & 0.62 & HMPO \\ 
W33A	    & 2.4  & 3.0 10$^4$ & 698 & 0.70 & HMPO \\
G34.26+0.15 & 3.3  & 1.9 10$^5$ & 1792 & 0.82 & UC\ion{H}{ii}\\
NGC6334-I   & 1.7  & 1.1 10$^5$ & 750 & 1.41  & HMC\\
NGC7538-I1  & 2.7  & 1.1 10$^5$ & 433 & 2.45 & UC\ion{H}{ii}\\
AFGL2591 	& 3.3  & 1.2 10$^5$ & 373 & 2.99 & HMPO \\
W3-IRS5	    & 2.0  & 2.1 10$^5$ & 424 & 3.68 & HMPO \\
G5.89-0.39  & 1.3 & 4.1 10$^4$ & 140 & 4.18 & UC\ion{H}{ii}\\
\hline
\end{tabular}
\end{minipage}
\tablefoot{Source coordinates 
with references and their physical parameters are taken from \citet{vT13}.}
\end{table}
PACS is an integral field unit with a $5\times5$ array of spatial pixels (hereafter
\textit{spaxels}). Each spaxel covers $9\farcs4\times9\farcs4$,
providing a total field of view of $\sim47''\times47''$. The focus of this work is on 
the central spaxel only. The central spaxel probes similar physical scales as the full 
$5\times5$ array in the 10 times closer low-mass sources. Full $5\times5$ maps
for the high-mass sources, both in lines and continuum, will be discussed in future papers.

The range spectroscopy mode on PACS uses large grating steps to quickly scan the full 
50-210 $\mu$m wavelength range with Nyquist sampling of the spectral elements.
The wavelength coverage consists of three grating orders (1st: 102-210
$\mu$m, 2nd: 71-105 $\mu$m or 3rd: 51-73 $\mu$m), two of which are
always observed simultanously (one in the blue, \mbox{$\lambda<105$ $\mu$m},
and one in the red, \mbox{$\lambda>102$ $\mu$m}, parts of the spectrum). 
The spectral resolving power is $R=\lambda/\Delta \lambda \approx$1000-1500
 for the 1st order, 1500-2500 for the 2nd order, and 2500-5500 for the 3rd order 
 (corresponding to velocity resolution from \mbox{$\sim$75 to 300 km s$^{-1}$}).

Two nod positions were used 
for chopping 3$^\prime$ on each side of the source. The comparison of the two positions 
was made to assess the influence of the off-source flux of observed species from the 
off-source positions, in particular for atoms and ions. The [\ion{C}{ii}] fluxes are 
strongly affected by the off-position flux and saturated for most sources -- we therefore 
limit our analysis of this species to the two sources with comparable results for both nods,
AFGL2591 and NGC7538-IRS1 (see Table 2).

Typical pointing accuracy is better than 2$^{\prime\prime}$. However, two sources 
(G327-0.6 and W33A) were mispointed by a larger amount as indicated by the location
of the peak continuum emission on maps at different 
wavelengths (for the observing log see Table \ref{log} in the Appendix). In order to account 
for the non-centric flux distribution on the integral field unit due to mis-pointing and to improve 
the continuum smoothness, for these sources two spaxels with maximum continuum levels are used 
(spaxel 11 and 21 for G327 and 23 and 33 for W33A). Summing a larger number of spaxels was 
not possible due to a shift of line profiles from absorption to emission. The spatial extent 
of line emission / absorption will be analyzed in future papers.

We performed the basic data reduction with the Herschel Interactive Processing Environment 
v.10 \citep[HIPE,][]{Ot10}. The flux was normalized to the telescopic
background and calibrated using Neptune observations. Spectral flatfielding within HIPE was
used to increase the signal-to-noise \citep[for details, see][]{He12,Gr13}. 
The overall flux calibration is accurate to $\sim
20\%$, based on the flux repeatability for multiple observations of
the same target in different programs, cross-calibrations with HIFI
and ISO, and continuum photometry. 

Custom IDL routines were used to further process the datacubes. The
line fluxes were extracted from the central spaxel (except G327-0.6
and W33A, see above) using Gaussian fits with fixed line width
\citep[for details, see][]{He12}. Next, they were corrected for the
wavelength-dependent loss of radiation for a point source (see PACS
Observer's Manual\footnote{http://herschel.esac.esa.int/Docs/PACS/html/pacs\_om.html}).
 That approach is not optimal for the cases where
emission is extended beyond the central spaxel, but that is mostly the
case for atomic lines, which will be presented in a companion paper 
by Kwon et al. (in preparation, hereafter Paper II). The uncertainty
introduced by using the point-source correction factors for extended
sources depends on the amount of emission in the surrounding ring of
spaxels. The continuum fluxes are calculated using all 25
  spaxels, except G327-0.6 where 1 spaxel was excluded due to
  saturation. In most cases, the tabulated values are at wavelengths
  near bright lines. They are calculated using spectral regions
  on both sides of the lines (but masking any features) and interpolated
  linearly to the wavelength of the lines. The fluxes are presented in
Table \ref{tab:cont} in the Appendix. Our continuum fluxes are
  included in the spectral energy distribution fits presented in
  \citet{vT13}, who used them to derive physical models for all our sources.
  Those models and associated envelope masses are used in this work in
  Sections 5.1 and 5.3.
\begin{figure*}[tb]
\begin{center}
\includegraphics[angle=90,height=13cm]{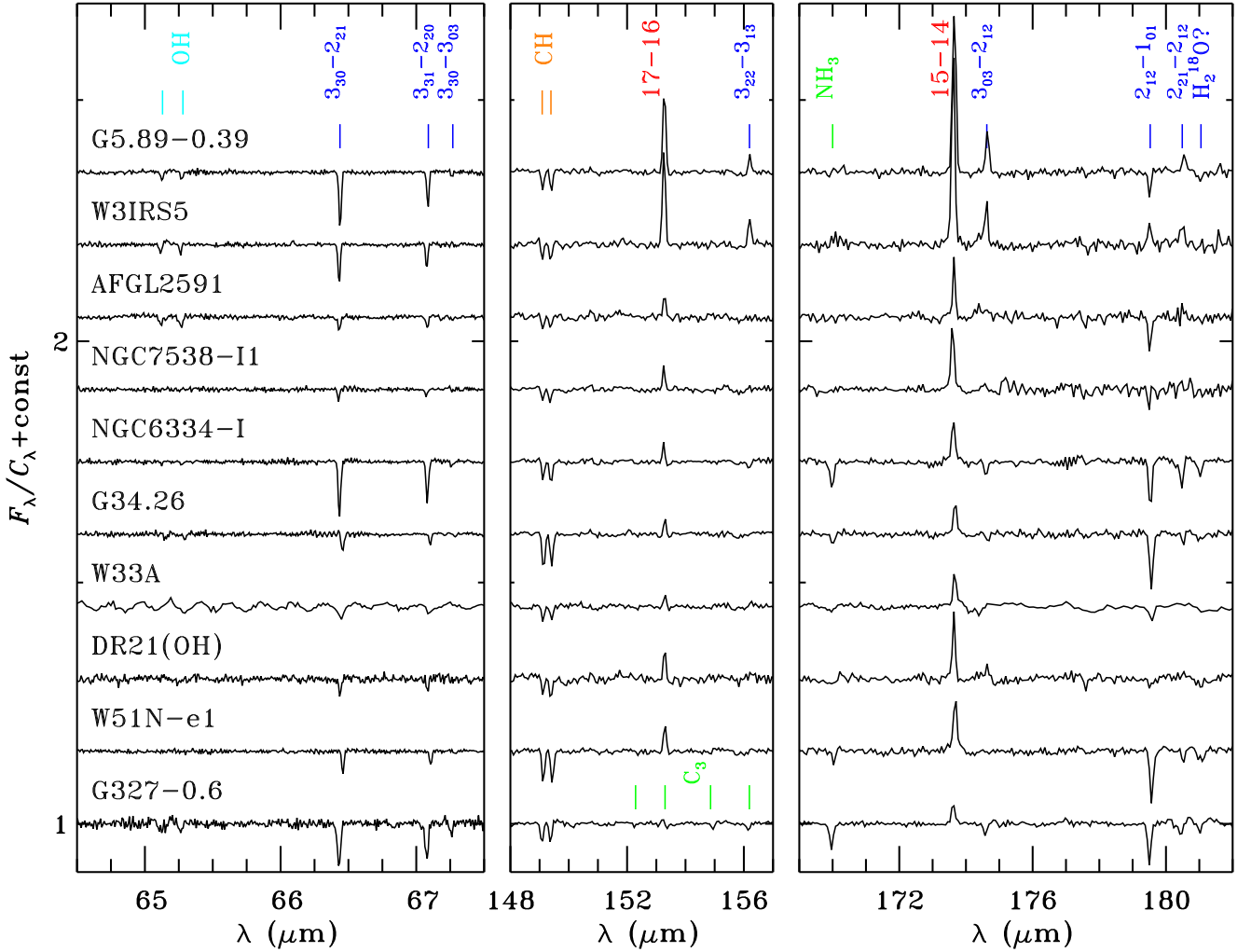}
\vspace{0.3cm}

\caption{\label{widma} Normalized spectral regions of all our sources at the central position
at 64-68, 148-157, and 169-182 $\mu$m. Objects are shown in the evolutionary sequence, 
with the most evolved ones on top. Lines of 
CO are shown in red, H$_2$O in blue, OH in light blue, CH in orange, and NH$_\mathrm{3}$ and 
C$_\mathrm{3}$ in green. Spectra are shifted vertically to improve the clarity of the figure.}
\end{center}
\end{figure*}
\section{Results}
Figure \ref{spec} shows the full normalized PACS spectrum with line identifications for W3 IRS5, 
a high-mass protostellar object with the richest molecular emission among our sources. 
Carbon monoxide (CO) transitions from $J$=14-13 to $J$=30-29  
are detected, all in emission (see blow-ups of high-$J$ CO lines in Figures \ref{wblow} and \ref{cozoom}).
Water vapor (H$_2$O) transitions up to $E_\mathrm{up}\sim1000$ K 
are detected (e.g. $7_{16}-6_{25}$ at 66.1 $\mu$m, see blow-ups in Figure \ref{wblow}). 
At wavelengths shortwards of $\sim$90 $\mu$m 
many H$_2$O lines are seen in absorption, but those at longer wavelengths are primarily in emission. 

Seven hydroxyl (OH) doublets up to $E_\mathrm{up}\approx618$ K are
seen\footnote{The highest-excited OH doublet at 71 $\mu$m is not
  considered further in the analysis, because the 70-73 $\mu$m region
  observed with PACS is affected by spectral leakage and thus is
  badly flux-calibrated.}. All lines within the
$^{2}\Pi_{\nicefrac{3}{2}}$ ladder (119, 84, and 65 $\mu$m doublets;
see Figure 1 in Wampfler et al. 2013) are strong absorption lines. The
$^{2}\Pi_{\nicefrac{1}{2}}$ ladder lines (163 and 71 $\mu$m) are seen
in emission.  The cross-ladder transitions at 79 $\mu$m
(OH $\nicefrac{1}{2}$,$\nicefrac{1}{2}$-$\nicefrac{3}{2}$,$\nicefrac{3}{2}$, $E_\mathrm{up}\approx180$ K) 
and 96 $\mu$m (OH $\nicefrac{3}{2}$,$\nicefrac{1}{2}$-$\nicefrac{5}{2}$,$\nicefrac{3}{2}$, $E_\mathrm{up}\approx270$
K) are absorption and emission lines, respectively.  Only the
ground-rotational lines of methylidyne (CH) are detected at 149 $\mu$m
in absorption.  The [\ion{O}{i}] transitions at 63 $\mu$m and 145
$\mu$m are both strong emission lines in W3 IRS5.  That is not always
the case for other sources in our sample. The [\ion{O}{i}] line at 63
$\mu$m, where the velocity resolution of PACS is at its highest
($\sim$90 km s$^{-1}$), shows a variety of profiles (see Figure
\ref{oxyprofiles}): pure absorption (G327-0.6, W51Ne1, G34.26, W33A),
regular P-Cygni profiles (AFGL2591, NGC6334-I), hints of inverse
P-Cygni profiles in DR21(OH) and pure emission (W3 IRS5, NGC7538-I1,
G5.89). The [\ion{O}{i}] line at 145 $\mu$m, however, is always
detected in emission. The P-Cygni profiles resolved at velocity scales
of $\sim100$ km s$^{-1}$ resemble the high-velocity line wings observed in
ro-vibrational transitions of CO in some of the same sources
\citep{Mi90,He11}, suggested to originate in the wind impacting the
outflow cavities.  Note however that not all absorption needs to be associated
with the source: it can also be due to foreground
absorption (e.g. outflow lobe or the ISM). 
For example, ISO Fabry Perot and new Herschel/ HIFI
observations of H$_2$O and OH in Orion also show line wings in
absorption / emission extending to velocities up to $\sim$100 km
s$^{-1}$ \citep[][Choi et al. in prep.]{Ce06,Go06}, but the ISO-LWS
Fabry-Perot observations of Orion did not reveal P-Cygni profiles in
the [\ion{O}{i}] 63 $\mu$m line.

A comparison of selected line-rich parts of the spectra for all our sources is presented 
in Figure \ref{widma}. The spectra are normalized by the continuum emission to better visualize
 the line absorption depths. However, the unresolved profiles of PACS underestimate the true 
 absorption depths and cannot be used to estimate the optical depths.

The 64-68 $\mu$m segment covers the highly-excited H$_2$O lines at 66.4, 67.1, and 67.3 $\mu$m 
($E_\mathrm{up}\approx410$ K); high-$J$ CO lines at 65.7 ($J=$40-39) and 67.3 ($J=$39-38); and the
OH $^{2}\Pi_{\nicefrac{3}{2}}$ $J=\nicefrac{9}{2}-\nicefrac{7}{2}$ ($E_\mathrm{up}\approx510$ K) 
doublet at 65 $\mu$m. The low-lying H$_2$O lines are detected for all sources. 
The high-$J$ CO lines are not detected in this spectral region. The OH doublet 
is detected for 6 out of 10 sources (see also Figure \ref{ohprof} in the Appendix).  

The main lines seen in the 148-157 $\mu$m region are: CH \mbox{$^2\Pi_{\nicefrac{3}{2}}$ $J=\nicefrac{3}{2}$ --
$^2\Pi_{\nicefrac{1}{2}}$ $J=\nicefrac{1}{2}$} transition at 149 $\mu$m
(in absorption), CO 17-16, and H$_2$O $3_{22}-3_{13}$ line at 156.2 $\mu$m ($E_\mathrm{up}\approx300$ K). 
Weak absorption lines at $\sim155-156$ $\mu$m seen towards the hot core G327-0.6 are most likely 
C$_3$ ro-vibrational transitions \citep[][Paper II]{Ce00}.

The most commonly detected lines in the 170-182 $\mu$m spectral region include the CO 15-14 line 
and the H$_2$O lines at 174.6, 179.5 and 180.5 $\mu$m ($E_\mathrm{up}\approx$100-200 K). The 
profiles of H$_2$O lines change from object to object: the H$_2$O $2_{12}-1_{01}$ line at 
179.5 $\mu$m is in absorption for all sources except W3 IRS5; the H$_2$O $2_{21}-1_{12}$ 
at 180.5 $\mu$m is in emission for the three most evolved sources (top 3 spectra on Figure 
\ref{widma}). The ammonia line at $\sim170$ $\mu$m, NH$_3$ (3,2)a-(2,2)s, is detected towards 4 sources 
(G327-0.6, W51N-e1, G34.26, and NGC6334-I). An absorption line at $\sim181$ $\mu$m 
corresponds to H$_{2}^{18}$O $2_{12}-1_{01}$ and/ or H$_3$O$^+$ $1_{1}^{-}-1_{1}^{+}$ lines
\citep{Go01}. Extended discussion of ions and molecules other than CO, H$_2$O, and OH will 
appear in Paper II.

Line profiles of H$_2$O observed with HIFI show a variety of emission and absorption 
components that are not resolved by PACS \citep{Ch10,Kr12,vT13}. The only H$_2$O lines observed in 
common by the two instruments within the WISH program are the ground-state transitions: 
$2_{12}-1_{01}$ at 179.5 $\mu$m (1670 GHz) and $2_{21}-1_{12}$ at 180.5 $\mu$m (1661 GHz), both dominated by 
absorptions and therefore not optimal to estimate to what extent a complex water profile is diluted 
at the PACS spectral resolution. However, HIFI observations of the lines between excited rotational states, 
which dominate our detected PACS lines, are generally in emission at the longer wavelengths 
probed by HIFI. In the case of $^{12}$CO 10--9, the line profiles of YSOs observed with HIFI consist 
of a broad outflow and a narrow quiescent component with the relative fraction of the integrated 
intensity of the narrow to broad components being typically 30-70\% for low-mass sources \citep{Yi13}.
For the single case of a high-mass YSO, W3 IRS5, this fraction is about 50\% \citep{IreneCO}.
    
To summarize, PACS spectra of high-mass sources from our sample show detections of many 
molecular lines up to high excitation energies. CO, H$_2$O, OH, and CH lines are seen towards 
all objects, whereas weaker lines of other molecules 
are detected towards less than half of the sources. CO lines are always seen in emission, CH 
in absorption, whereas other species show different profiles depending on the transition and
 the object. Table \ref{obs1} shows the CO line fluxes for all lines in the PACS range. 
\section{Analysis}
\subsection{Far-IR line cooling}
\begin{figure}[!tb]
\begin{center}
\includegraphics[angle=90,height=6cm]{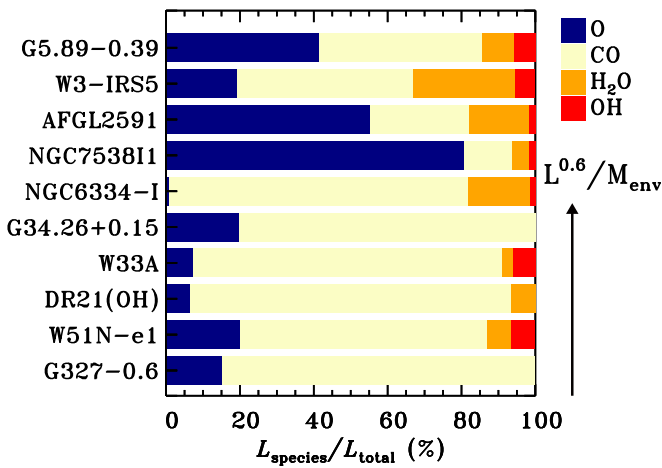}
\caption{\label{hmcool} Relative contributions of [\ion{O}{i}] (dark blue), CO (yellow), 
 H$_2$O (orange), and OH (red) cooling to the total far-IR gas cooling at central position 
 are shown from left to right horizontally for each source. The objects follow the evolutionary 
 sequence with the most evolved sources on top.}
\vspace{0.2cm}
\end{center}
\end{figure}
Emission lines observed in the PACS wavelength range are used to calculate the contribution
of different species to the total line cooling from high-mass protostars. Our goal is to 
compare the cooling of warm gas by molecules and atoms with the cooling by dust and 
connect them with the evolutionary stages of the objects. Relative contributions to the cooling
 between different
 molecules are also determined, which can be an indicator of the physical processes in the 
 environments of young protostars \citep[e.g.][]{Ni02,Ka13}.

We define the total far-IR line cooling ($L_\mathrm{FIRL}$) as the sum of all emission line luminosities from 
the fine-structure [\ion{O}{i}] lines (at 63 and 145 $\mu$m) and the detected molecules, 
following \citet{Ni02} and \citet{Ka13}. [\ion{C}{ii}], the most important line coolant of diffuse interstellar gas,
 is also expected to be a significant cooling agent in high-mass star forming regions and extragalactic sources. It is
not explicitly included in our analysis, however,
because the calculated fluxes are strongly affected by off-source emission and often saturated 
(see also Section 2). The emitted [\ion{C}{ii}] luminosity is shown below for only two sources, where 
reliable fluxes were obtained. Cooling in other ionic lines such as [\ion{O}{iii}], [\ion{N}{ii}], and [\ion{N}{iii}]
is also excluded, due to the off-position 
contamination and the fact that those lines trace a different physical component than the molecules 
and [\ion{O}{i}]. Since the only molecules with emission lines are CO, H$_2$O, and OH, the equation 
 for the total far-IR line cooling can be written as: 
 $L_\mathrm{FIRL}=L_\mathrm{OI}+L_\mathrm{CO}+L_\mathrm{H_{2}O}+L_\mathrm{OH}$. 
 
Table \ref{cool} summarizes our measurements. The amount of cooling by dust is described 
by the bolometric luminosity and equals $\sim10^4$-10$^5$ L$_{\odot}$ for our sources \citep{vT13}.
 The total far-IR line cooling ranges from $\sim1$ to $\sim40$ L$_{\odot}$, several
orders of magnitude less than the dust cooling. Relative contributions of  
oxygen atoms and molecules to the gas cooling are illustrated in Figure \ref{hmcool}. 
Atomic cooling is the largest for the more evolved sources in our sample (see Table 1), in particular 
for NGC7538 IRS1 and AFGL2591, where it is the dominant line cooling channel. For those two 
sources, additional cooling by [\ion{C}{ii}] is determined and amounts to $\sim$10-25\% of 
$L_\mathrm{FIRL}$ (Table \ref{cool}). Typically, atomic cooling accounts for $\sim$20 \% of the total line far-IR line cooling.
Molecular line cooling is dominated by CO, which is responsible for $\sim15$ up to $85$\% of 
$L_\mathrm{FIRL}$, with a median contribution of 74\%. H$_2$O and OH median contributions to 
the far-IR cooling are less than 1\%, because many of their transitions are detected in absorption. Assuming that the absorptions arise in the same gas, as found for the case of Orion \citep{Ce06}, they
therefore do not contribute to the cooling, but heating of the gas. Still, the 
contribution of H$_2$O to the total FIR cooling increases 
slightly for more evolved sources, from $\sim$5\% (DR21(OH)) to 30\% (W3IRS5), 
whereas no such trend is seen for OH.
\begin{table*}
\caption{\label{cool} Far-IR line cooling by molecules and atoms in units of L$_{\odot}$.}
\centering
\renewcommand{\footnoterule}{}  
\begin{tabular}{lc|ccc|ccc|c}
\hline \hline
Source & $L_\mathrm{bol}$ & $L_\mathrm{FIRL}$ & $L_\mathrm{mol}$ & $L_\mathrm{OI}$ & $L_\mathrm{CO}$ &  $L_\mathrm{H2O}$  & $L_\mathrm{OH}$ & $L_\mathrm{CII}$ \\
~ & (L$_{\odot}$) &  \multicolumn{3}{c|}{(L$_{\odot}$)} &  \multicolumn{3}{c|}{(L$_{\odot}$)} &  (L$_{\odot}$) \\
\hline
G327-0.6  & 7.3 10$^4$ & 2.1(0.7)  &   1.8(0.6)  &   0.3(0.1)  &   1.8(0.6)  &    --  &     --    &     --     \\
W51N-e1 & 5.2 10$^5$    & 37.8(10.8)  &  30.2(8.9)  &  7.6(1.8)  &  25.3(6.8)  &   2.5(1.1)  &   2.4(1.0) &    --    \\
DR21(OH) & 1.3 10$^4$ & 1.4(0.4)  &   1.3(0.4)  &   0.09(0.03)  &   1.2(0.3)  &   0.09(0.04)  &    --   &     --   \\
W33A     & 3.0 10$^4$ & 0.7(0.2)  &   0.6(0.2)  &   0.05(0.02)  &   0.6(0.2)  &   0.02(0.01)  &   0.04(0.01) &    --   \\
G34.26+0.15 & 1.9 10$^5$ & 9.6(2.9)  &   7.7(2.4)  &   1.9(0.5)  &   7.7(2.4)  &    --   &     --  &     --     \\
NGC6334-I &  1.1 10$^5$ & 4.2(1.3)  &   4.2(1.2)  &   0.04(0.03)  &  3.4(1.0)  &   0.7(0.2)  &   0.06(0.03) &    --     \\
NGC7538-IRS1 & 1.1 10$^5$  & 13.6(3.3)  &  2.6(0.8)  &  11.0(2.5)  &   1.8(0.5)  &   0.6(0.2)  &   0.2(0.1) &   2.0(0.4)   \\
AFGL2591 & 1.2 10$^5$  & 6.1(1.8)  &   2.7(0.8)  &   3.4(0.9)  &   1.6(0.5)  &   1.0(0.4)  &   0.11(0.04) & 1.9(0.4) \\
W3-IRS5  & 2.1 10$^5$  &  22.0(6.0)  &  17.8(5.0)  &   4.2(1.0)  &  10.5(2.5)  &   6.1(2.2)  &   1.2(0.4) &     -- \\
G5.89-0.39 & 4.1 10$^4$ &  8.8(2.2)  &   5.1(1.3)  &   3.7(0.9)  &   3.9(0.9)  &   0.8(0.3)  &   0.5(0.2) &    -- \\
\hline 
\end{tabular}
\tablefoot{Columns show: (1) bolometric luminosity, $L_\mathrm{bol}$, (2) total FIR line cooling, 
$L_\mathrm{FIRL}$ ($L_\mathrm{mol}$+$L_\mathrm{OI}$), (3) molecular cooling, $L_\mathrm{mol}$, and 
(4) cooling by oxygen atoms, $L_\mathrm{OI}$. Cooling by individual molecules is shown in column (5) CO, (6) H$_2$O, and (7) OH. 
 Absence of emission lines that would contribute to the cooling are shown with "--" (absorption lines 
 are detected for H$_2$O and OH). Errors are written in brackets and include 20\% calibration error
  on individual line fluxes.}
\end{table*}

\subsection{Molecular excitation}
\begin{figure*}[!tb]
\begin{center}
\includegraphics[angle=90,height=8cm]{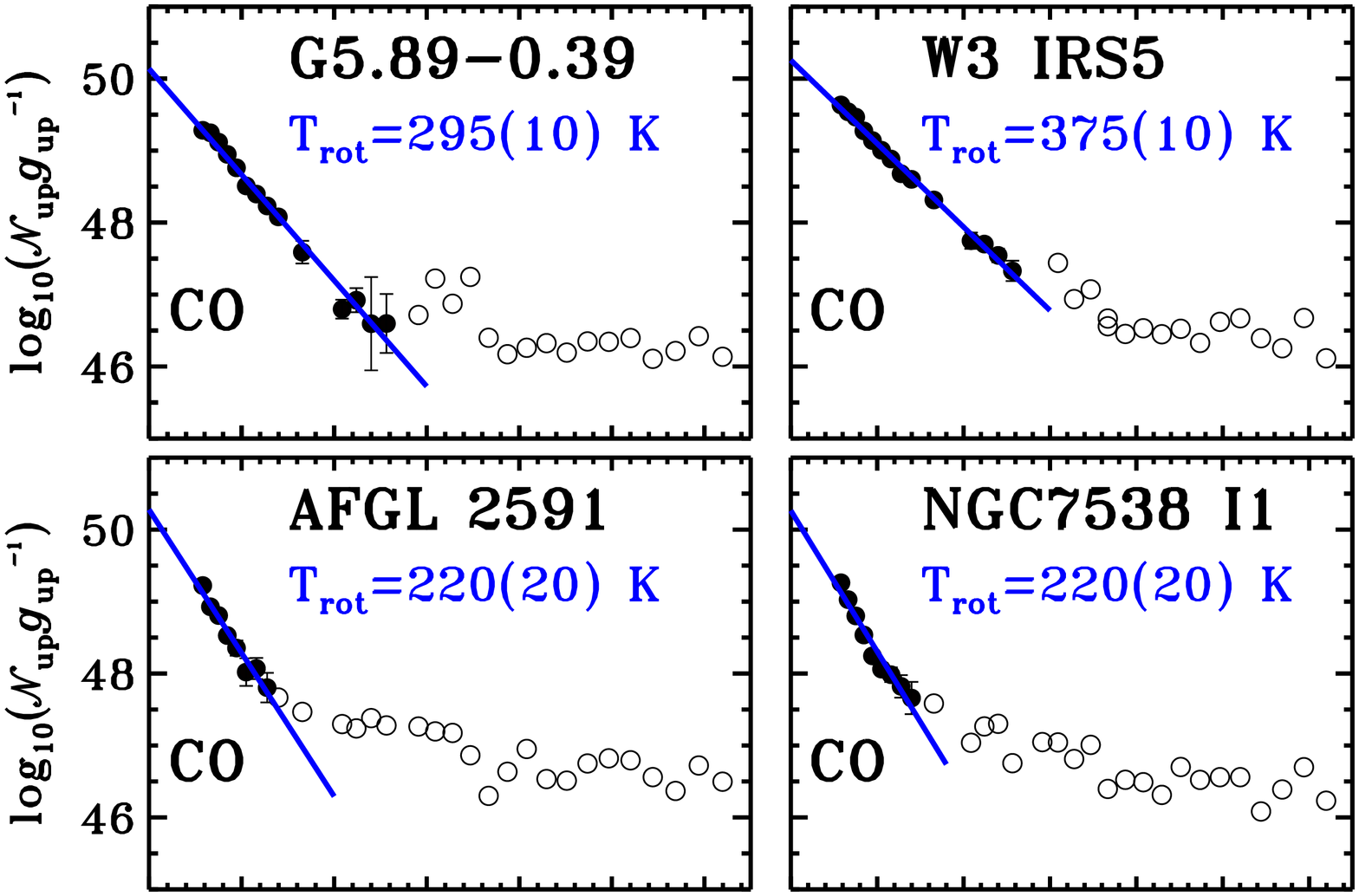}
\vspace{-0.4cm}

\includegraphics[angle=90,height=8cm]{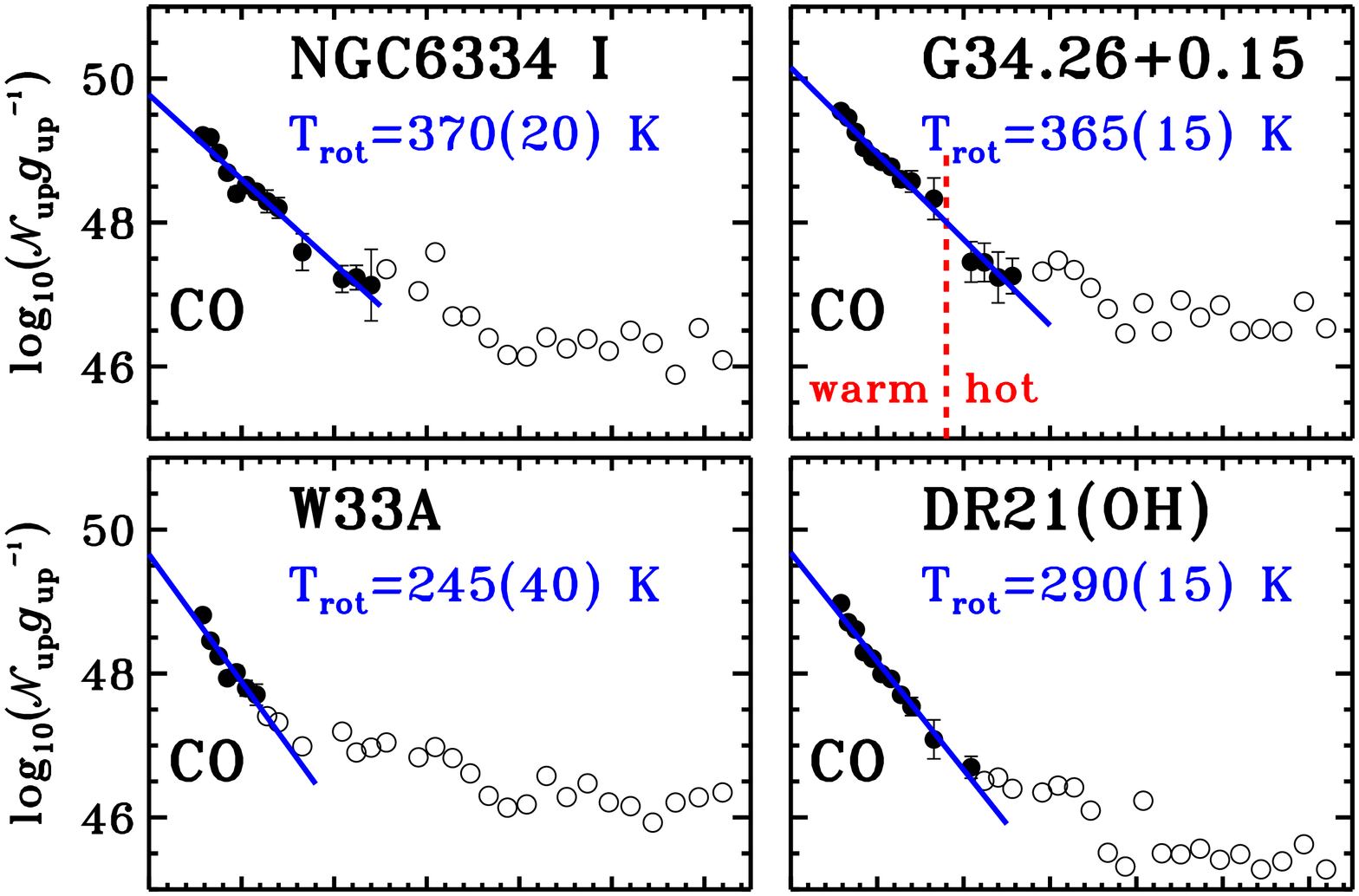}
\vspace{-0.4cm}

\includegraphics[angle=90,height=8cm]{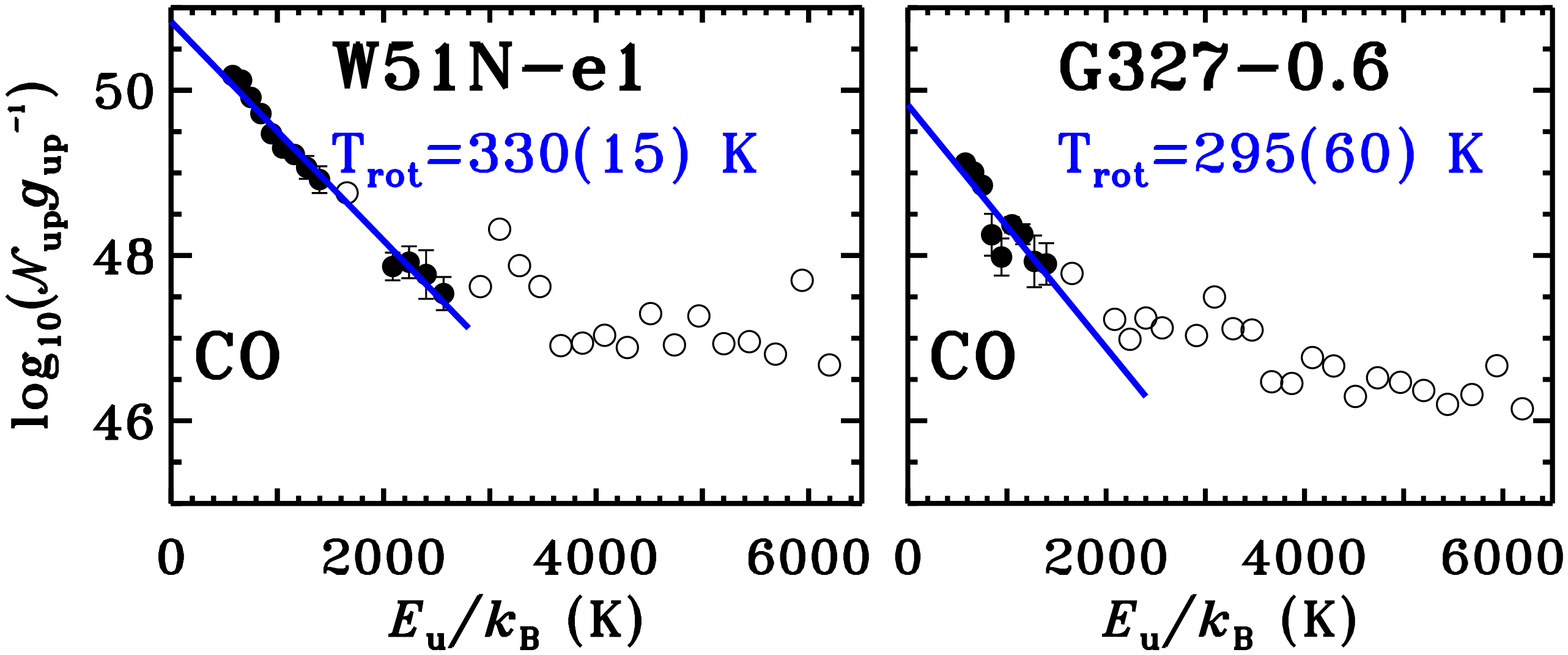}
\vspace{-2.5cm}

\caption{\label{coex} Rotational diagrams of CO for all objects in our
  sample.  The base 10 logarithm of the number of
  emitting molecules from a level $u$,
  $\mathcal{N}_\mathrm{u}$, divided by the degeneracy of the level,
  $g_\mathrm{u}$, is shown as a function of energy of the upper level
  in kelvins, $E_\mathrm{up}$. Detections are shown as filled
  circles, whereas three sigma upper limits are shown as empty
  circles. Blue lines show linear fits to the data and the
    corresponding rotational temperatures. The vertical red line in
    the G34.26+0.15 panel shows the dividing line between the warm and
    hot components as seen in rotational diagrams of low-mass YSOs.
  Errors associated with the fit are shown in brackets.}
\end{center}
\end{figure*}
\begin{figure*}[!tb]
\begin{center}
\includegraphics[angle=0,height=20cm]{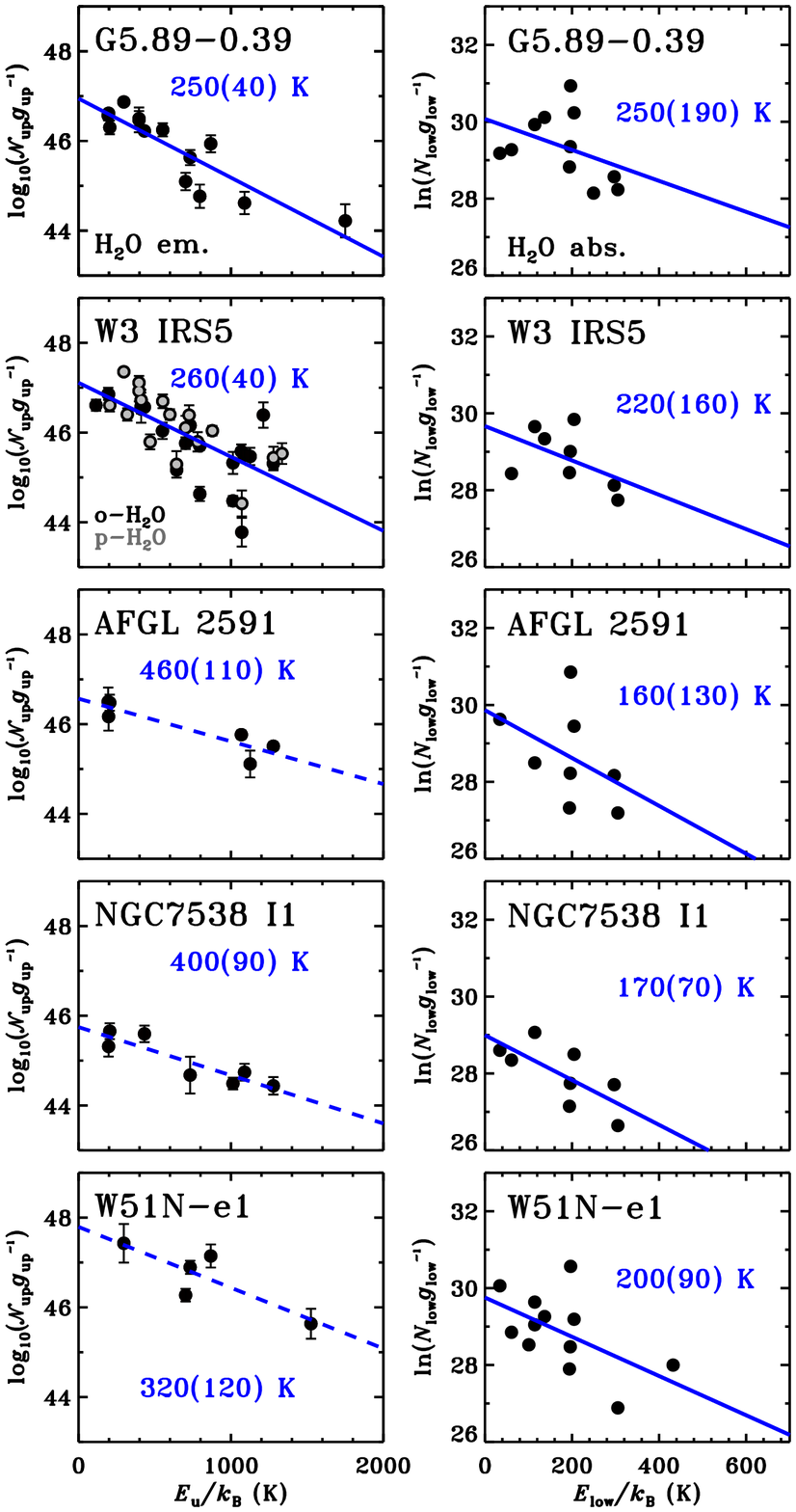}
\vspace{0.6cm}

\caption{\label{wex} Rotational diagrams of H$_2$O calculated using emission lines (left 
column) and absorption lines (right column), respectively. For emission line diagrams, 
the logarithm with base 10 of total number of molecules in a level $u$, $\mathcal{N}$, 
divided by the degeneracy of the level, $g_\mathrm{u}$, is shown as a function of energy
 of the upper level in Kelvins, $E_\mathrm{low}$. For 
absorptions lines, the natural logarithm of the column density in a level $l$, $N_\mathrm{l}$, divided by 
the degeneracy of the level, $g_\mathrm{l}$, is shown in y-axis. A one component linear fit is 
shown with the corresponding value of rotational temperature and error of the fit in brackets. 
A solid line is used for the cases where at least 10 lines are detected. In the W3 IRS5 emission panel,
 para-H$_2$O lines are shown in grey and ortho-H$_2$O lines in black, respectively.}
\end{center}
\end{figure*}
Detections of multiple rotational transitions of CO, H$_2$O and OH allow us to determine
 the rotational temperatures of the emitting or absorbing gas using Boltzmann diagrams 
\citep[e.g.][]{GL99}. For H$_2$O and OH, the densities are likely not high 
enough to approach a Boltzmann distribution and therefore the diagrams presented below 
are less meaningful.

Emission line fluxes are used to calculate the number of emitting
molecules, $\mathcal{N}_\mathrm{u}$, for each molecular transition
using Equation (\ref{rot1}), assuming that the lines are optically thin.
Here, $F_{\lambda}$ denotes the flux of the line at wavelength
$\lambda$, $d$ is the distance to the source, $A$ is the Einstein
coefficient, $c$ is the speed of light and $h$ is Planck's constant:
\begin{equation}
\label{rot1}
\mathcal{N}_\mathrm{u}=\frac{4\pi d^{2}F_{\lambda}\lambda}{hcA}
\end{equation}
The base 10 logarithm of $\mathcal{N}_\mathrm{u}$ over degeneracy of the upper level $g_\mathrm{u}$
 is shown as a function of the upper level energy, $E_\mathrm{u}$, in Boltzmann diagrams 
 (Figures \ref{coex} and \ref{wex}). The rotational temperature is calculated in a standard way, from the 
 slope $b$ of the linear fit to the data in the natural logarithm units, 
 $T_\mathrm{rot}=-1/b$. 
 
Because the size of the emitting region is not resolved by our instrument, the calculation of column densities
 requires additional assumptions and therefore only the total numbers of emitting molecules is determined.
 The formula for the total number of emitting molecules, $\mathcal{N}_\mathrm{tot}$, is derived 
 from the expression for total column density, $N_\mathrm{tot}=Q\cdot \mathrm{exp}(a)$, where 
 $Q$ is the partition function for the temperature and $a$ is the y-intercept. Correcting 
 for a viewing angle, $\Omega=d^2/\pi R^2$, and multiplying by the gas emitting area of radius $R$, yields:
\begin{equation}
\label{rot3}
\mathcal{N}_\mathrm{tot}=Q\cdot \mathrm{exp}(a)\cdot d^2
\end{equation}
 For details, see e.g. \citet{Ka13} and \citet{Gr13}.

For absorption lines, column densities, $N_\mathrm{l}$, are calculated 
 from line equivalent widths, $W_{\lambda}$, using Equation \ref{rot2} \citep[e.g.][]{Wr00}.
\begin{equation}
\label{rot2}
N_\mathrm{l}=\frac{8\pi c W_{\lambda} g_\mathrm{l}}{\lambda^{4} A g_\mathrm{u}}
\end{equation}
This relation assumes that the lines are optically thin, the covering factor is unity
and the excitation temperature $T_\mathrm{ex}\ll hc/k\lambda$ for all lines.

Some of the emission / absorption lines are 
 likely P-Cygni profiles \citep{Ce06} that are not resolved with PACS and which we assume to be
 pure emission / absorbing lines.

\begin{figure}[tb]
\begin{center}
\includegraphics[angle=90,width=10cm]{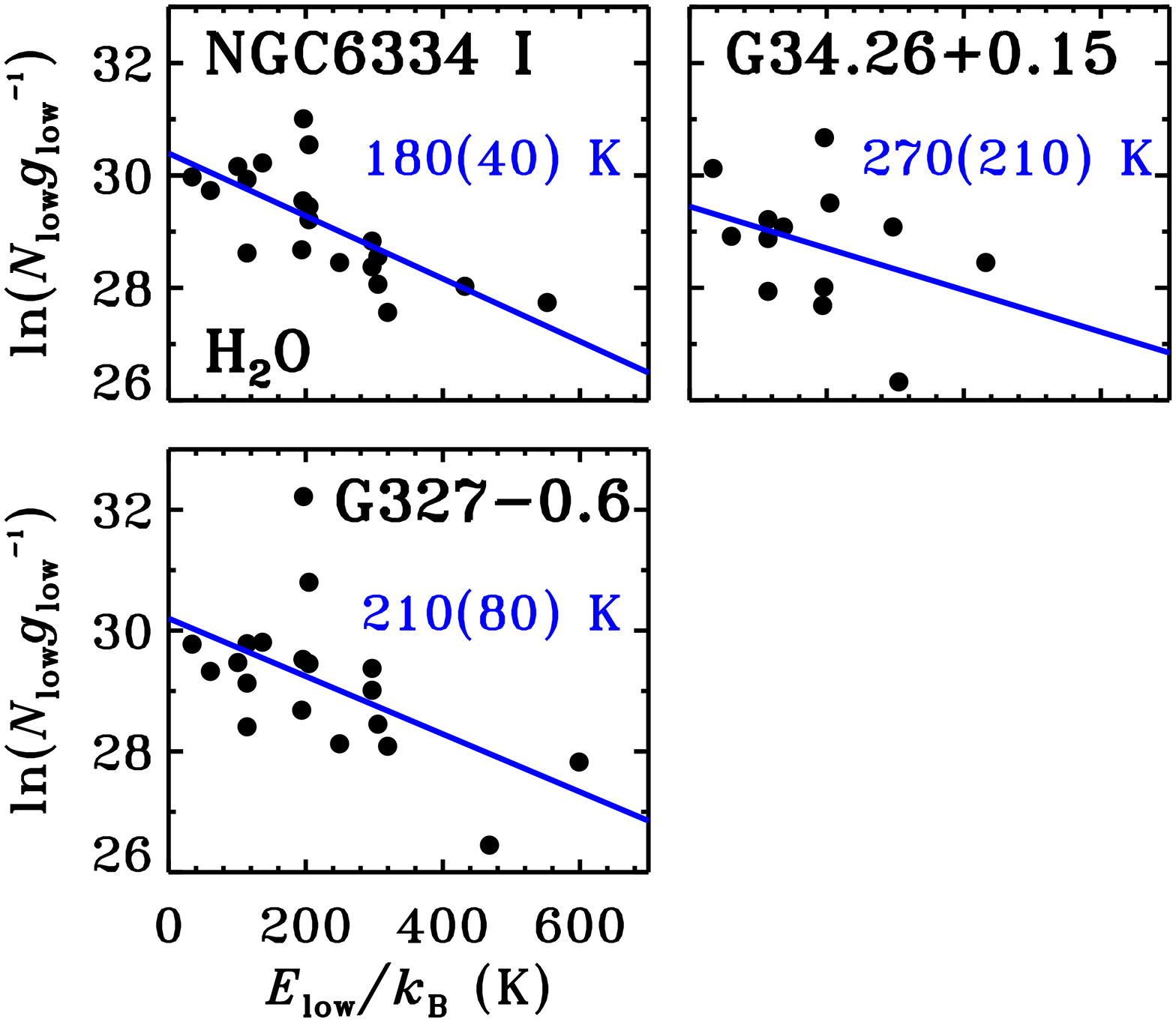}
\vspace{0.3cm}

\caption{\label{wabs}Rotational diagrams of H$_2$O calculated using absorption lines. 
A single component fit is used to calculate the temperature shown in the panels.}
\end{center}
\end{figure}

 The natural logarithm of $N_\mathrm{l}$ over the degeneracy of the
 lower level, $g_\mathrm{l}$, is shown on the Boltzmann diagrams as a
 function of the lower level energy, $E_\mathrm{l}$ (Figures
 \ref{wex}, \ref{wabs} and \ref{ohabs}). Rotational temperatures and
 total column densities are calculated in the same way as for the
 emission lines (see above).
			
In the following sections, the excitation of CO, H$_2$O, and OH are discussed separately. 
Table \ref{tab:exc} summarizes the values of rotational temperatures, $T_\mathrm{rot}$, 
and total numbers of emitting molecules or column densities, for those species.
\begin{table*}
\caption{\label{tab:exc}CO, H$_2$O, and OH rotational excitation}
\centering
\renewcommand{\footnoterule}{}  
\begin{tabular}{lcccccccccccccc}
\hline \hline
Source & \multicolumn{2}{c}{Warm CO} & \multicolumn{2}{c}{H$_2$O (em.)} & \multicolumn{2}{c}{H$_2$O (abs.)} & $R$\tablefootmark{a} & 
\multicolumn{1}{c}{OH $^{2}\Pi_{\nicefrac{3}{2}}$\tablefootmark{b}} & \multicolumn{1}{c}{OH\tablefootmark{c}}\\

~ & $T_\mathrm{rot}$(K) & $\mathrm{log}_\mathrm{10}\mathcal{N}$ & $T_\mathrm{rot}$(K) & $\mathrm{log}_\mathrm{10}\mathcal{N}$ &
$T_\mathrm{rot}$(K) & $\mathrm{ln}N_\mathrm{low}$ &  (arc sec) & $T_\mathrm{rot}$(K) & $T_\mathrm{rot}$(K) &  \\
 \hline
G327-0.6 & 295(60) & 51.9(0.3) &  \ldots & \ldots & \textbf{210(80)} & 34.5(0.5) & \ldots & \textbf{105(12)} & 77(29) \\
W51N-e1 & 330(15) & 52.9(0.1)  & 320(120) & 50.1(0.5) & \textbf{200(90)}  & 34.0(0.5) & 3.4 & 66(-)  & 54(29)   \\
DR21(OH) & 290(15) & 51.7(0.1)   & \ldots & \ldots & \ldots & \ldots  & \ldots &  \textbf{98(27)}   & 79(30)    \\
W33A     & 245(40) & 51.6(0.2)    & \ldots  & \ldots & \ldots & \ldots  & \ldots &  \ldots  & \ldots\\
G34.26+0.15 & 365(15) & 52.3(0.1)  &  \ldots  & \ldots &  \textbf{270(210)}  & 34.0(0.6) & \ldots & \textbf{89(5)}  &  71(15) \\
NGC6334-I & 370(20) & 51.9(0.1)   & \ldots   & \ldots & \textbf{180(40)}  & 34.6(0.3)  & \ldots & 93(-)  & 71(50)  \\
NGC7538-IRS1 & 220(20) & 51.7(0.2) & 400(90) & 48.2(0.2) &  170(70)  & 33.1(0.5) & 1.1 & 52(-)   & 54(37)  \\
AFGL2591 & 220(20) & 51.7(0.2)   & 460(110) & 49.1(0.2) &  160(130)  & 33.9(1.1) & 1.6 & \textbf{105(29)}    & 89(21)  \\
W3-IRS5  & 375(10) &  52.4(0.1)   & \textbf{260(40)} & 49.3(0.2) &  220(160)  & 34.1(0.6) & 3.2 & \textbf{109(3)}   & 83(22)   \\
G5.89-0.39 & 295(10) & 52.2(0.1)  & \textbf{250(40)}  & 49.1(0.2) & \textbf{250(190)}  & 34.6(0.6) & 3.0 & \textbf{96(1)}   & 77(16) \\
\hline
\end{tabular}
\tablefoot{Rotational temperatures of H$_2$O and OH within the $^{2}\Pi_{\nicefrac{3}{2}}$ ladder are calculated 
using at least 10 and 3 lines / doublets, respectively, and are shown in boldface. 
Non-detections are marked with dots. For OH temperatures determined using only 2 transitions, 
the associated error is not given and marked with "-".\\
\tablefoottext{a}{Size of the H$_2$O emitting region assuming that all H$_2$O lines trace the same physical component 
(see Section 4.2.2.).}
\tablefoottext{b}{Rotational temperature of OH calculated using the OH $^{2}\Pi_{\nicefrac{3}{2}}$ ladder transitions only, see Figure 9 and Section 4.2.3.}
\tablefoottext{c}{Rotational temperature of OH calculated using all lines detected in absorption.}
}
\end{table*}
\subsubsection{CO}

Figure \ref{coex} shows CO rotational diagrams for all our
sources. CO detections, up to $J=$30-29 ($E_\mathrm{u}=2565$ K), are well described by 
single rotational temperatures in the range from 220 K (AFGL2591 and 
NGC7538 IRS1) to $\sim$370 K (W3IRS5 and G34.26+0.15) and the average of 
$T_\mathrm{rot,CO}\sim$300(23)$\pm$60 K.\footnote{The value in the brackets (23) shows 
the average error of rotational temperatures for different sources, whereas $\pm60$ 
is the standard deviation of rotational temperatures.} The highest temperatures are
seen for objects where high-$J$ CO transitions with $E_\mathrm{u}>2000$ K are detected.

Temperatures of $\sim$300 K are attributed to the `warm' component 
in low-mass YSOs \citep[e.g.][]{Go12,Ka13,Gr13}, where they are calculated using 
transitions from $J_\mathrm{up}=14$ ($E_\mathrm{u}=580$ K) to 24 ($E_\mathrm{u}=1660$ K).
 In those sources, a break around $E_\mathrm{u}\sim1800$ K in the rotational diagram is 
noticable (see vertical line on Figure \ref{coex}) and $J_\mathrm{up}\ge$25 transitions are 
attributed to the `hot' component. Such a turning point is not seen on the diagrams of
 our high-mass sources with detections extending beyond the $J=24-23$ transition.

In NGC7538 IRS1, a possible break is seen around
  $E_\mathrm{u}\sim1000$ K. A two component fit to the data results in
  rotational temperatures $T_\mathrm{rot1}\sim$160$\pm$10 K and
  $T_\mathrm{rot2}\sim$370$\pm$35 K. The latter temperature is
  consistent within errors with the `warm' component seen towards
  low-mass sources. The colder temperature resembles the $\sim70-100$
  K `cool' component seen in $J_\mathrm{up}\le$14 transitions
  \citep[e.g.][]{Go12,Ka13,Wi13}, detected at wavelengths longer than
  the PACS range.
  
The absence of the hot component towards all our sources is not
significant according to the calculated upper limits. In addition to
limited $S/N$ and line-to-continuum ratio, there are other effects
that may prevent the hot component from being detected. These include
the fact that the continuum becomes more optically thick at the
shorter wavelengths (see also below) and/or a smaller filling factor
of the hot component in the PACS beam compared with low-mass sources.
More generally, both the `warm' and `hot' components could still be part of a single
physical structure such as proposed in \citet{Ne12}.

The average logarithm of the number of emitting (warm) CO molecules,
$\mathrm{log}_\mathrm{10}\mathcal{N}$, is similar for all objects, and
equals 52.4(0.1)$\pm$0.5. Values of
$\mathrm{log}_\mathrm{10}\mathcal{N}$ in the range from 51.6 to 53.1
are derived. DR21(OH), one of the lowest bolometric luminosity sources
in our sample, exhibits one of the lowest $\mathcal{N}$(CO) contents,
whereas the highest CO content is found for W51N-e1, the most luminous
source.

\begin{figure}[!tb]
\begin{center}
\includegraphics[angle=90,width=8.5cm]{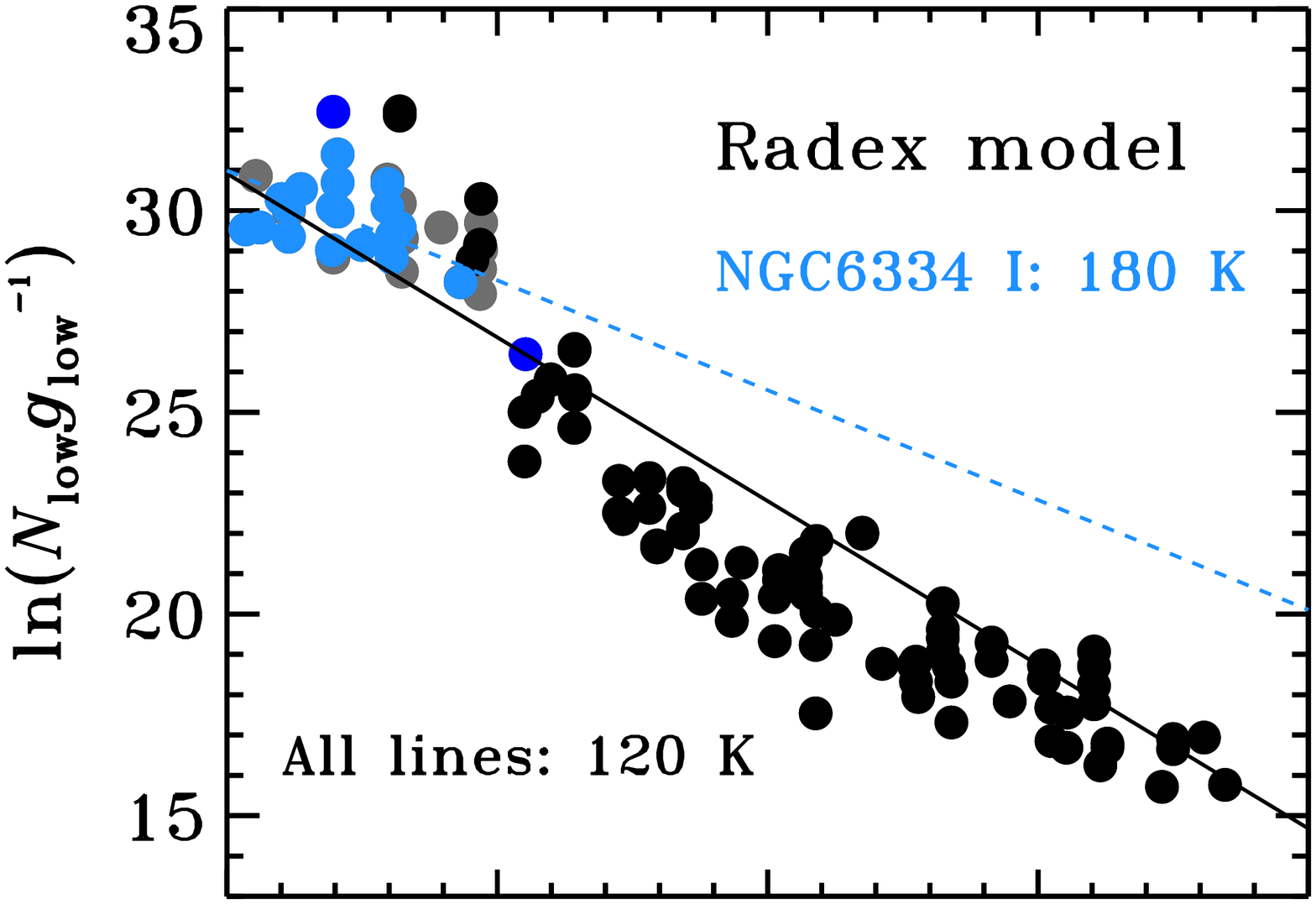}
\vspace{-1cm}

\includegraphics[angle=90,width=8.5cm]{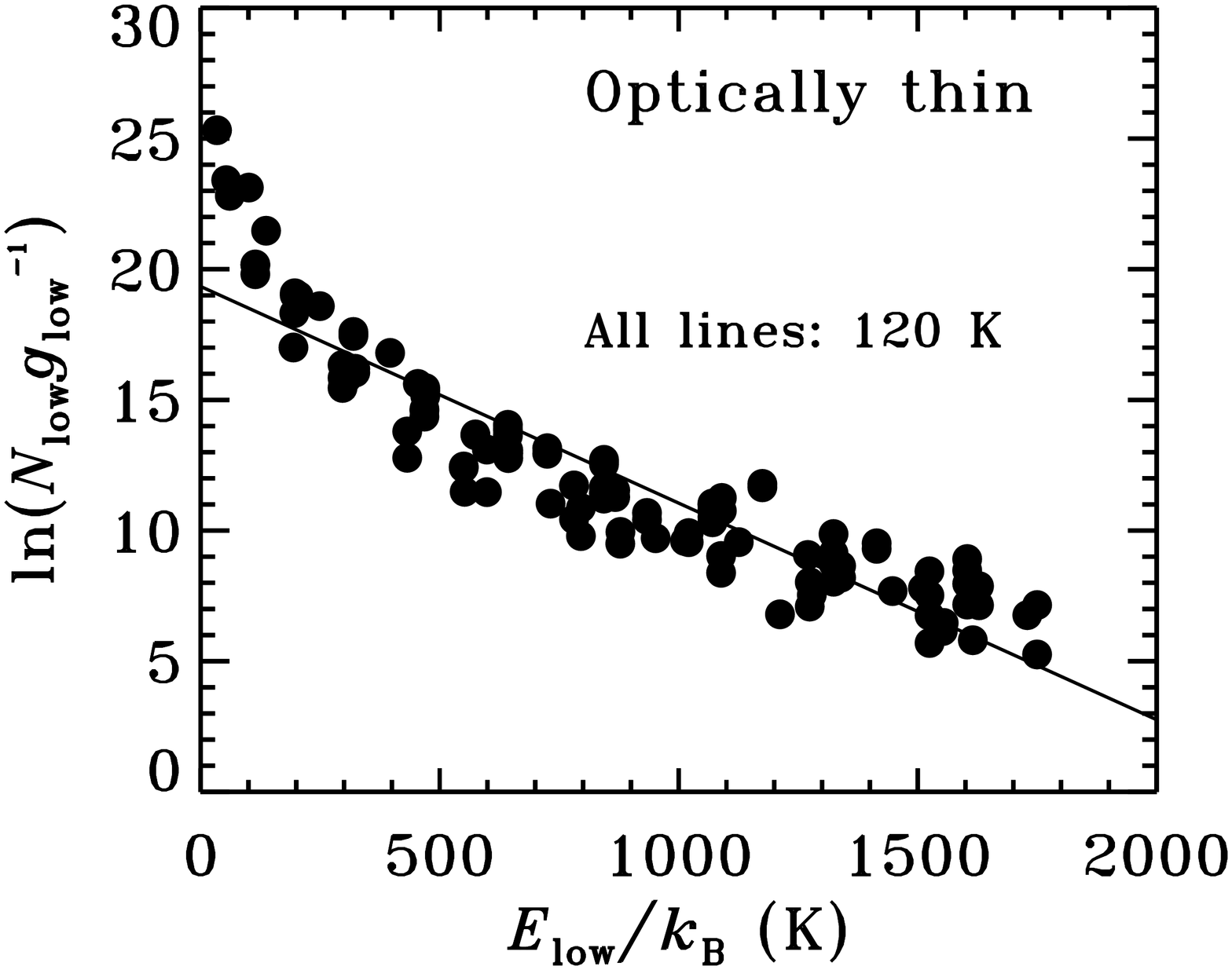}
\caption{\label{wmod} \textit{Top}: Rotational diagram of H$_2$O calculated
    assuming a kinetic temperature $T=1000$ K, H$_2$ density $n=10^6$
    cm$^{-3}$, H$_2$O column density $2\cdot10^{16}$ cm$^{-2}$, and line
    width $\Delta V=5$ km s$^{-1}$. Lines observed in absorption in NGC6334-I 
    are shown in blue. Fits are done to all lines in the PACS range 
    included in the LAMDA database \citep{LAMDA} (in black) and to the 
    NGC6334-I lines (in blue). Darker shades denote optically thin lines. 
    \textit{Bottom}: The same as above, but 
    assuming a H$_2$O column density $10^{12}$ cm$^{-2}$, such that all lines are 
    optically thin.}
\end{center}
\end{figure}
\subsubsection{H$_2$O}
Figure \ref{wex} shows rotational diagrams of H$_2$O calculated for five sources with 
H$_2$O lines detected both in emission and in absorption. Because the emitting region 
is not resolved by our observations, the diagrams calculated using the emission and 
absorption lines are shown separately. Figure \ref{wabs} shows H$_2$O diagrams for the 
three sources where all H$_2$O lines are seen in absorption.

The rotational diagrams show a substantial scatter, larger
  than the errors of individual data points, caused by large
  opacities, subthermal excitation (see \citealt{He12} and
  below) and possible radiation excitation by far-IR dust emission pumping. 
  The determination of rotational temperatures is therefore 
  subject to significant errors when only a limited number of lines is
  detected. In Table \ref{tab:exc} H$_2$O temperatures calculated
  using at least 10 lines are indicated in boldface.
Ro-vibrational spectra of H$_2$O from ISO-SWS
 towards massive protostars (four of them in common with our sample)
 show rotational temperatures of H$_2$O as high as 500 K, in agreement
 within the errors with our measurements \citep{Bo03}.
\begin{figure*}[!tb]
\begin{center}
\includegraphics[angle=90,height=8cm]{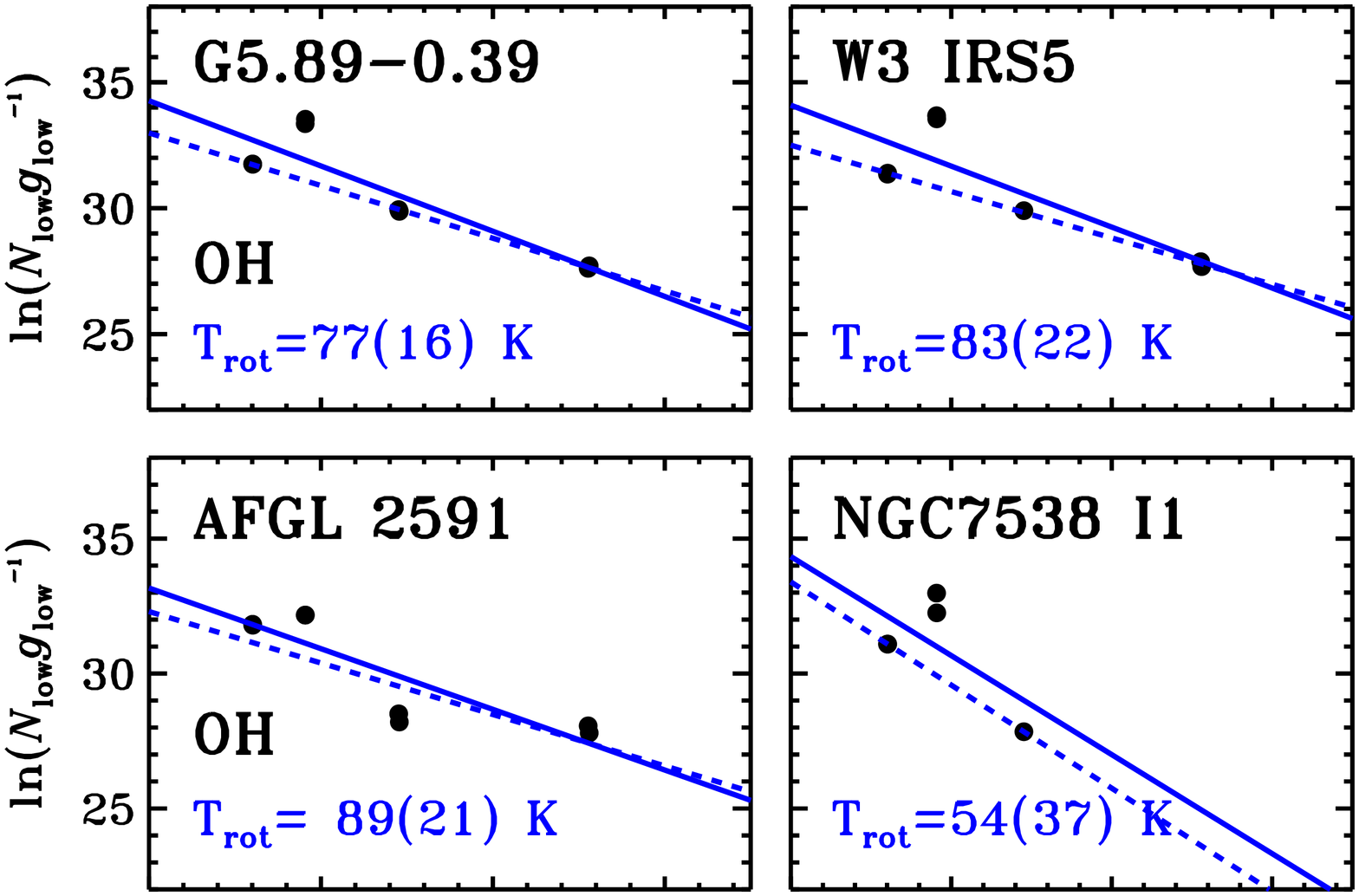}
\vspace{-0.4cm}

\includegraphics[angle=90,height=8cm]{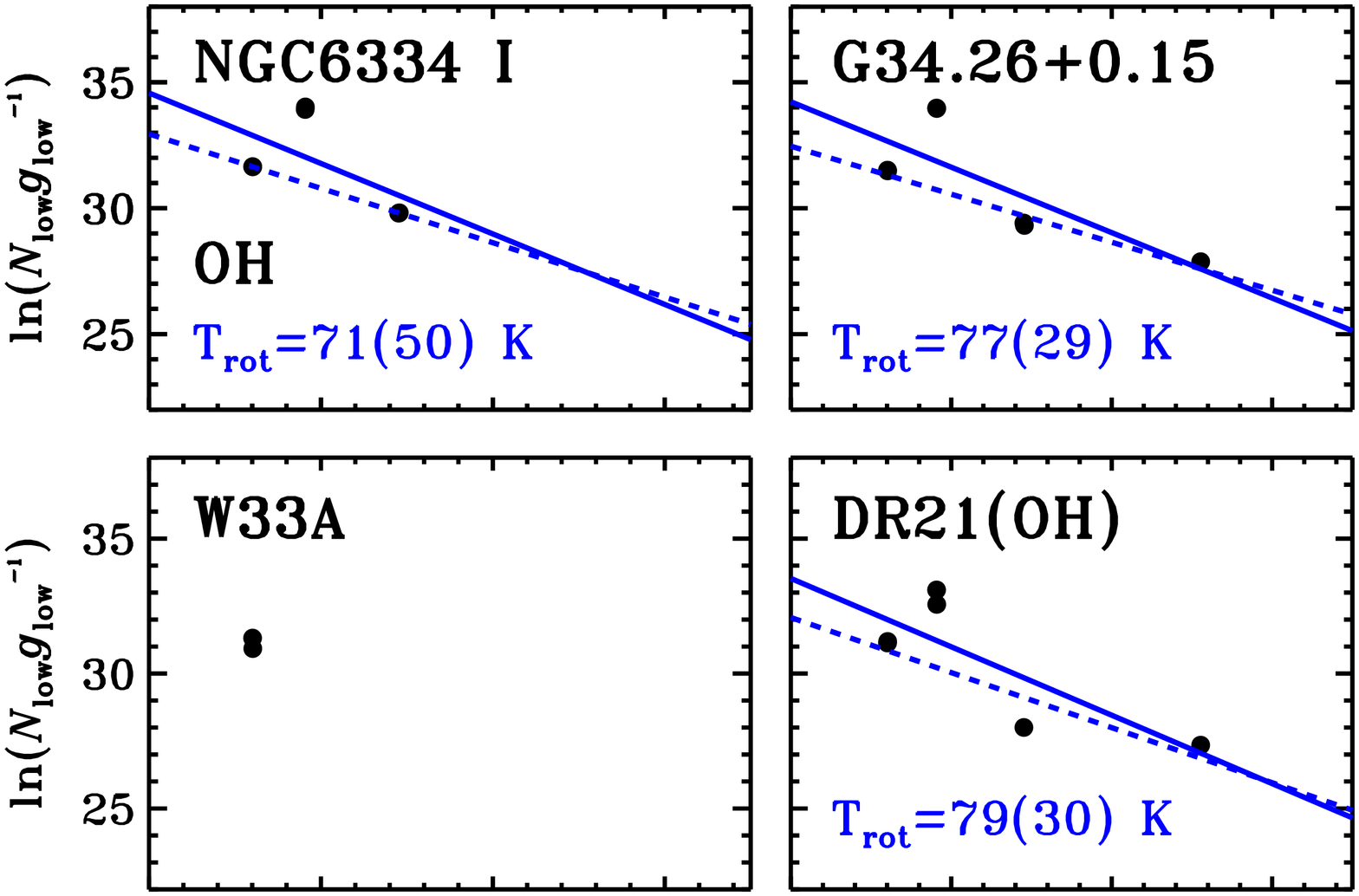}
\vspace{-0.4cm}

\includegraphics[angle=90,height=8cm]{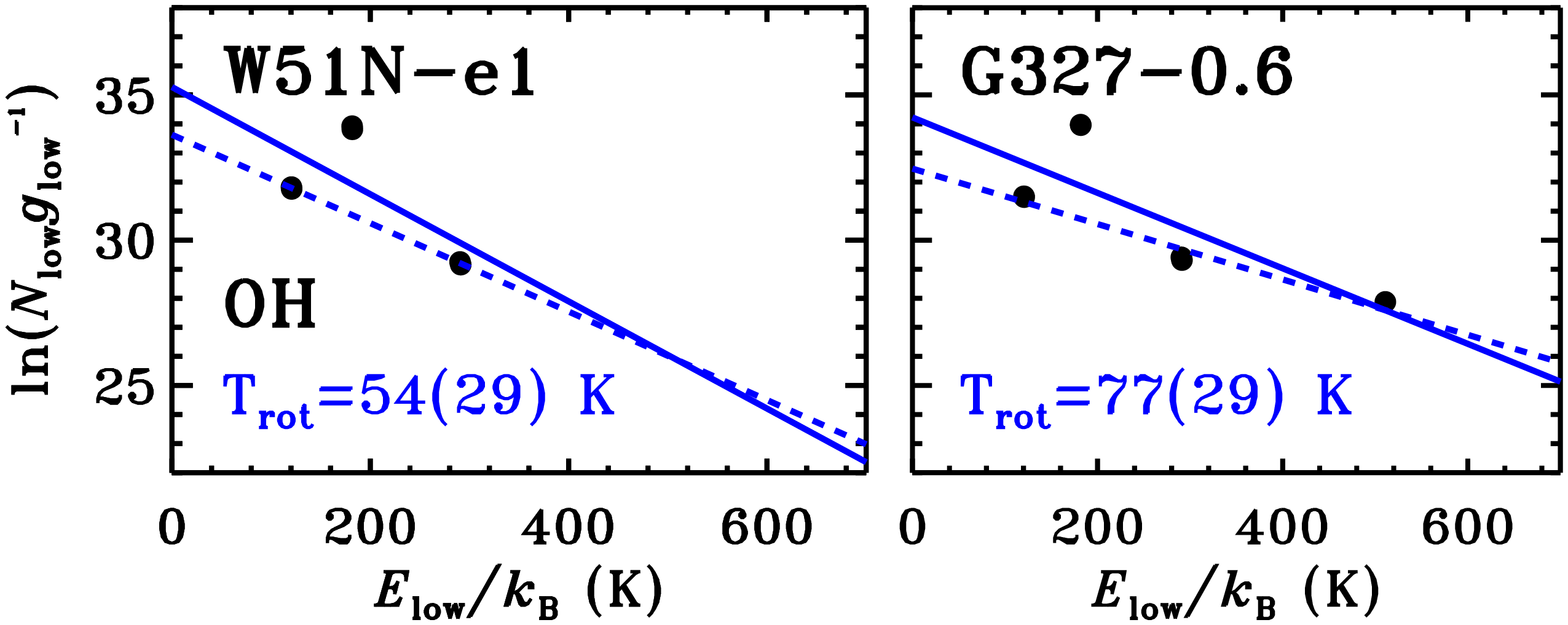}
\vspace{-3cm}

\caption{\label{ohabs} Rotational diagrams of OH calculated using absorption lines, similar 
to Figure \ref{wabs} for H$_2$O. The single component fit to all OH transitions detected in absorption
is shown with a solid line with the corresponding temperature. A separate single component fit 
is done for the OH $^{2}\Pi_{\nicefrac{3}{2}}$ ladder transitions and drawn in a dashed line. 
The respective rotational temperatures are tabulated in Table \ref{tab:exc}.}
\end{center}
\end{figure*}

The largest number of H$_2$O lines is detected in the two most
  evolved sources -- G5.89-0.39 and W3 IRS5. Single component fits to
  12 and 37 water emission lines, respectively, give similar
  rotational temperatures of $\sim250$ K (see Figure \ref{wex}), with no
  systematic differences between $o$-H$_2$O and $p$-H$_2$O
  lines. Rotational temperatures determined for the remaining sources,
  with at least 5 detections of H$_2$O in emission, are higher,
  $T_\mathrm{rot}\sim300-450$ K, but are less accurate.

Rotational temperatures calculated from a single component fit
  to the absorption line diagrams are $\sim200$ K for all
  sources. They are in good agreement with the values obtained from
  the emission diagrams for G5.89-0.39 and W3 IRS5, suggesting that all
  H$_2$O lines originate in the same physical component \citep[see
  e.g.][]{Ce06}. In such a scenario, the column densities should also
  agree, and the comparison of the total number of molecules
  calculated from emission lines and columns determined from the
  absorption lines yields the size of the emitting region, $R$. The
  radius of the H$_2$O emitting area under that condition equals $\sim$3 arcsec
   for G5.89-0.39 and W3 IRS5 (see Table \ref{tab:exc}).

To assess the effects of optical depth and subthermal excitation on the derived temperatures
  from the absorption lines, equivalent widths of lines are calculated
  using the radiative transfer code Radex \citep{RADEX} and translated to
  column densities using Equation (2). The adopted physical conditions 
  of $T_\mathrm{kin}=1000$ K and $n=10^6$ cm$^{-3}$
    are typical of warm, shocked region where water is excited
  \citep{Go12}. The models were calculated using all H$_2$O lines in the PACS range 
  included in the LAMDA database \citep{LAMDA}. The latest available H$_2$O collisional 
  coefficients are used \citep[][and references therein]{Da11}.
  
Figure \ref{wmod} shows the
  `theoretical' rotational diagrams calculated for low and high column densities. 
  A single component fit
   to \textit{all} lines in the high column density model
  gives a temperature of $\sim120$ K, consistent with subthermal excitation,
   and a total column density of 1.2$\cdot10^{15}$ cm$^{-2}$ 
  (ln$N_\mathrm{low}\sim$34.7), an order of magnitude lower than the input column. 
  A separate fit to the lines detected in NGC6334-I (shown in blue) gives a temperature
  of $\sim180$ K and a slightly higher column density (1.9$\cdot10^{15}$ cm$^{-2}$).
  These observed lines are typically highly optically thick with $\tau\sim$few tens, up to 100. 

In the low column density model all lines are optically thin. A fit to all lines gives a temperature
  of $\sim120$ K, similar to the high column density model. The level populations are clearly 
  subthermal ($T_\mathrm{rot}\ll T_\mathrm{kin}$), resulting in
  the scatter in the diagram. These examples illustrate the difficulty in using the inferred 
  rotational temperatures to characterize a complex environment of high-mass star forming regions.

The continuum opacity at PACS wavelengths is typically of the order of a
few in the observed sources, as indicated by the source structures
derived by \cite{vT13}, and becomes higher at shorter
wavelengths. This implies that the absorbing H$_2$O is on the
frontside of the source. Also, any emission at short wavelengths must
originate outside the region where the dust is optically thick.

\subsubsection{OH}
Figure \ref{ohabs} presents rotational diagrams of OH calculated using absorption lines. A single 
component is fitted to all detected lines. A separate fit is done for the lines 
originating in the $^{2}\Pi_{\nicefrac{3}{2}}$ ladder, which are mostly collisionally excited. 
This fit excludes the intra-ladder 79 $\mu$m doublet, connecting with the ground state, that 
readily gets optically thick \citep{Wa13}. A possible line-of-sight contribution by unrelated foreground is expected 
in the ground-state 119 $\mu$m line, which is included in the fit. The resulting rotational temperatures 
for each source are shown in Table \ref{tab:exc}, separately for those two fits.

The average rotational temperature for the $^{2}\Pi_{\nicefrac{3}{2}}$ ladder is very similar 
for all objects and equals 100(12)$\pm$7 K. 
The inclusion of all OH doublets results in lower temperatures, $T_\mathrm{rot,OH}\sim$79(22)$\pm$6 K.
\section{Discussion: from low to high mass}
\subsection{Origin of CO emission}
Several physical components have been proposed as a source of far-IR CO emission in isolated 
low-mass young stellar objects: (i) the inner parts of the quiescent envelope, passively heated by
a central source \citep{CC96,DN97}; (ii) gas in cavity walls heated by UV photons \citep{vK09,vK10,Vi11}; (iii) currently shocked gas 
along the outflow walls produced by the protostellar wind-envelope interaction \citep{vK10,Vi11,Ka13}. 

The quiescent envelope of high-mass protostars is warmer and denser
than for low-mass YSOs and therefore its contribution to the far-IR
CO emission is expected to be larger. In this section, we determine this contribution
for a subsample of our sources using the density and temperature
structure of each envelope obtained by \citet{vT13}. 

In this work, the
continuum emission for all our objects is modeled using a modified 3D
Whitney-Robitaille continuum radiative transfer code
\citep{Ro11,Wh13}. For simplicity, the van der Tak et al. models do
not contain any cavity or disk, and assume a spherically symmetric
power law density structure of the envelope, $n\propto r^{-p}$, where
$p$ is a free parameter. The size and mass of the envelope, and the
power law exponent, $p$, are calculated by best-fit comparison to the
spectral energy distributions and radial emission profiles at 450 and
850 $\mu$m \citep{Sh00}. The models solve for the dust temperature as
function of radius and assume that the gas temperature is equal to the
dust temperature. For more detailed discussion, see \citet{vT13}.

The envelope temperature and density structure from
  \citet{vT13} is used as input to the 1D radiative transfer code
  RATRAN \citep{RATRAN} in order to reproduce simultaneously the
  strenghts of optically thin C$^{18}$O lines from $J=$2-1 to 9-8
  \citep[following the procedure in][]{Yi10,Yi12}. The free parameters
  include C$^{18}$O constant abundance $X_\mathrm{0}$ and the line
  width, FWHM. For three sources, a `jump' abundance profile
  structure is needed, described by the evaporation temperature
  $T_\mathrm{ev}$ and inner abundance X$_\mathrm{in}$.  The parameters
  derived from the fits are summarized in Table \ref{tabenv}; the
  C$^{18}$O observations are taken from \citet{IreneCO}, while the
  modeling details and results for all our objects will be presented
  in I. San~Jos\'{e}-Garc\'{i}a (in prep.).
\begin{table}
\begin{minipage}[t]{\columnwidth}
\caption{Input parameters for the C$^\mathrm{18}$O envelope emission model\tablefootmark{a}}
\label{tabenv}
\centering
\renewcommand{\footnoterule}{}  
\begin{tabular}{lccccccc}
\hline \hline
Object & X$_\mathrm{0}$ & X$_\mathrm{in}$\tablefootmark{b} & $T_\mathrm{ev}$ & FWHM & $^{16}$O/$^{18}$O\tablefootmark{c}\\
~ & ~  & ~ & (K) & (km s$^{-1}$) & ~ \\
\hline
G327-0.6 	& 3.8 10$^{-7}$  &  -- & -- & 5.0 & 387 \\
W51N-e1     & 3.0 10$^{-7}$  & -- & -- & 4.7 & 417 \\
DR21(OH)    & 1.3 10$^{-6}$& -- & -- & 2.4 & 531 \\ 
NGC6334-I   & 0.5 10$^{-7}$  & 2.0 10$^{-7}$ & 35 & 4.2 & 437 \\
G5.89-0.39  & 0.1 10$^{-7}$  & 5.0 10$^{-7}$ & 40 & 4.5 & 460 \\
NGC7538-I1  & 2.1 10$^{-8}$  & 7.5 10$^{-7}$ & 35 & 2.1 & 614 \\
\hline
\end{tabular}
\end{minipage}
\tablefoot{
\tablefoottext{a}{Parameters for the remaining sources will be presented in San~Jos\'{e}-Garc\'{i}a et al. (in prep.).}
\tablefoottext{b}{A jump abundance profile is used to model NGC6334-I, G5.89-0.39 and NGC7538 IRS1.}
\tablefoottext{c}{The ratio depends on the source's distance from the Galaxy center \citep{WR94}.}
}
\end{table}
\begin{table}
\begin{minipage}[t]{\columnwidth}
\caption{Far-IR CO emission: observations and envelope models}
\label{tabco}
\centering
\renewcommand{\footnoterule}{}  
\begin{tabular}{lcccc}
\hline \hline
Object & $L_\mathrm{CO}$(obs) & \multicolumn{2}{c}{$L_\mathrm{CO}$(env)}   \\
~ & (L$_{\odot}$) &  (L$_{\odot}$) & (\%) \\
\hline
\multicolumn{4}{c}{High-mass YSOs\tablefootmark{a}}\\
\hline
G327-0.6 	& 1.9  &  1.6   & 84 \\
W51N-e1     & 25.8 &  14.6 & 56 \\
DR21(OH)    & 1.2  &  0.7  & 58 \\
NGC6334-I   & 3.4  &  2.4  & 71 \\
G5.89-0.39  & 3.9  &  1.8  & 46 \\
NGC7538-I1  & 2.1  &  1.6 & 77 \\
\hline
\multicolumn{4}{c}{Low-mass YSOs\tablefootmark{b}}\\
\hline
NGC1333 I2A & 4.1 10$^{-3}$ & 0.3 10$^{-3}$ & 7 \\
HH46        & 6.9 10$^{-3}$ & 0.5 10$^{-3}$ & 7 \\
DK Cha      & 5.1 10$^{-3}$ & 0.1 10$^{-3}$ & 2 \\
\hline
\end{tabular}
\end{minipage}
\tablefoot{
\tablefoottext{a}{Observed CO luminosities are calculated using detected transitions only,
 from $J=$14-13 to 30-29, depending on the source (see Table \ref{obs1}). The corresponding envelope
 CO luminosities are calculated using the same transitions.}
\tablefoottext{b}{Results from \citet{Vi11}. The observed CO emission is taken to be
the total CO emission originating from all modeled physical components.}
}
\end{table}

The parameters from Table \ref{tabenv} are used as inputs for the 
RATRAN models of $^{12}$CO. The integrated $^{12}$CO line emission 
obtained from RATRAN is convolved with the telescope beam and compared
 with observed line fluxes. 
 
\begin{figure}[!tb]
\begin{center}
\vspace{0.5cm}

\includegraphics[angle=0,height=11cm]{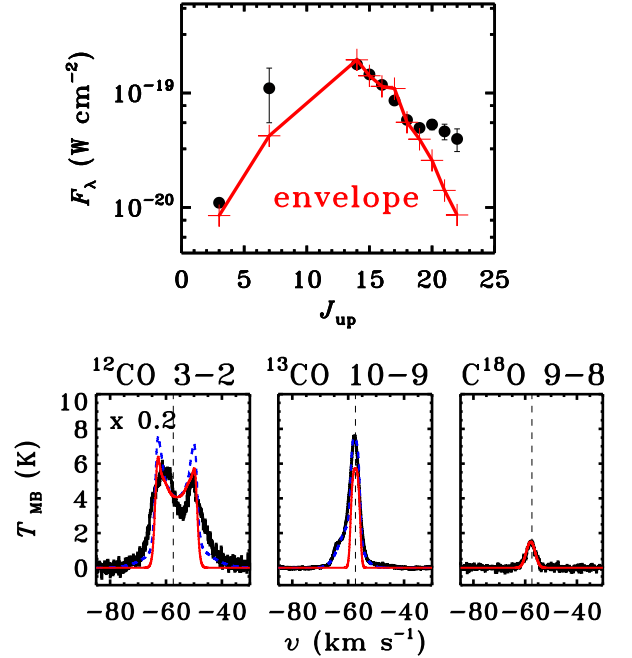}
\vspace{-1.5cm}

\caption{\label{comp2} \textit{Top:} Comparison of integrated line fluxes of $^{12}$CO observed 
by Herschel/PACS and from the ground (black dots 
with errorbars) and the predictions of the quiescent envelope passively 
heated by the luminosity of the source (red crosses) for NGC7538-IRS1. \textit{Bottom:} 
The same model compared with the JCMT $^{12}$CO 3-2 and Herschel/HIFI $^{13}$CO 10-9 and C$^{18}$O 9-8 observed line profiles. 
Additional model including an outflow component is shown in blue dashed line.}
\end{center}
\end{figure}

 Figure \ref{comp2} compares the envelope model for NGC7538 IRS1 with the
 $^{12}$CO $J_\mathrm{up}=14$ to $22$ observations from Herschel/PACS, 
 CO 3-2 \citep[in 14'' beam,][]{IreneCO} and CO 7-6 \citep[in 8'' beam,][]{Bo03b} from the James Clerk Maxwell
  Telescope. By design, the model fits the line profile of C$^{18}$O 9-8 \citep[][from Herschel/HIFI]{IreneCO}
   shown in the bottom of Figure \ref{comp2}. The pure envelope model slightly underproduces the $^{13}$CO 10-9 line, 
  because it does not include any broad entrained outflow component \citep[$T_\mathrm{ex}\sim70$ K,][]{Yi13}. 
  Adding such an outflow to the model \citep[see][]{Mo13} provides an excellent fit to the total line profile. For the case of  
  $^{12}$CO, the pure envelope model reproduces the 3-2 and 7-6 lines within a factor of two, 
  with the discrepancy again being due to the missing outflow in the model.
  This envelope model reproduces the CO integrated intensities for 
  transitions up to $J_\mathrm{up}=18$, but the larger $J$ CO fluxes are underestimated by a large factor.

\begin{figure}[!tb]
\begin{center}
\includegraphics[angle=0,height=11cm]{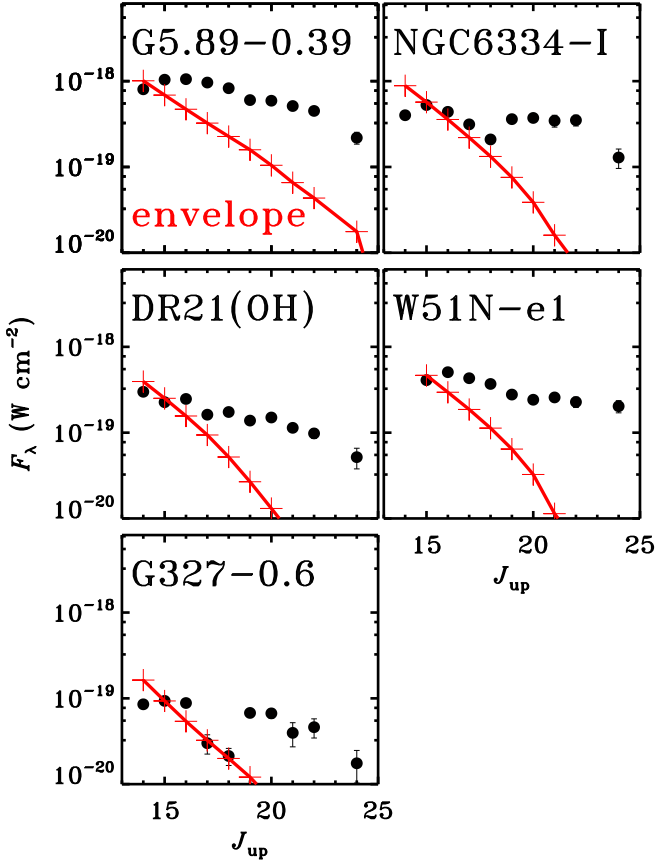}
\vspace{0.5cm}

\caption{\label{comp} Comparison of integrated line fluxes of CO observed by PACS only (black dots 
with errorbars) and the predictions of the quiescent envelope passively 
heated by the luminosity of the source (red crosses).}
\end{center}
\end{figure}

Comparison of the observed and modeled integrated $^{12}$CO line emission for 
the remaining sources from Table \ref{tabenv} is shown in Figure \ref{comp}.
As found for the case of NGC7538 IRS1, the contribution of the quiescent envelope emission
can be as high as 70-100\% of that of the $J$=15-14 line but decreases
sharply for the higher-$J$ transitions. Only 3-22\% of CO $J$=22-21
line emission is reproduced by the envelope models. In total
$\sim50-100$ \% of observed total FIR CO luminosity can be explained
by the envelope emission (see Table \ref{tabco}). 

This contribution is much larger than for low-mass YSOs, where the
quiescent envelope is responsible for only up to 7\% of the total CO emission
\citep{Vi11}. Still, even for the high-mass sources, an additional
physical component is needed to explain the excitation of the
highest-$J$ CO lines. The broad line profiles of high-$J$
  ($J\geq$10) CO lines \citep{IreneCO} argue in favor of a shock
  contribution to the far-IR emission in $^{12}$CO.  There may also be a
  contribution from UV-heating of the outflow cavities by the photons
  from the protostellar accretion shocks or produced by high velocity
  shocks inside the cavities as found for low-mass YSOs \citep{Vi11}
  but this component is best distinguished by high-$J$ $^{13}$CO lines
  \citep{vK09}. Physical models similar to those developed for
  low-mass sources by \citet{Vi11}, which include the different
  physical components, are needed to compare the relative contribution
  of the envelope emission, shocks, and UV-heating in the high-mass
  sources, but this is out of the scope of this paper.
\subsection{Molecular excitation}
\begin{figure}[tb]
\begin{center}
\includegraphics[angle=0,height=12cm]{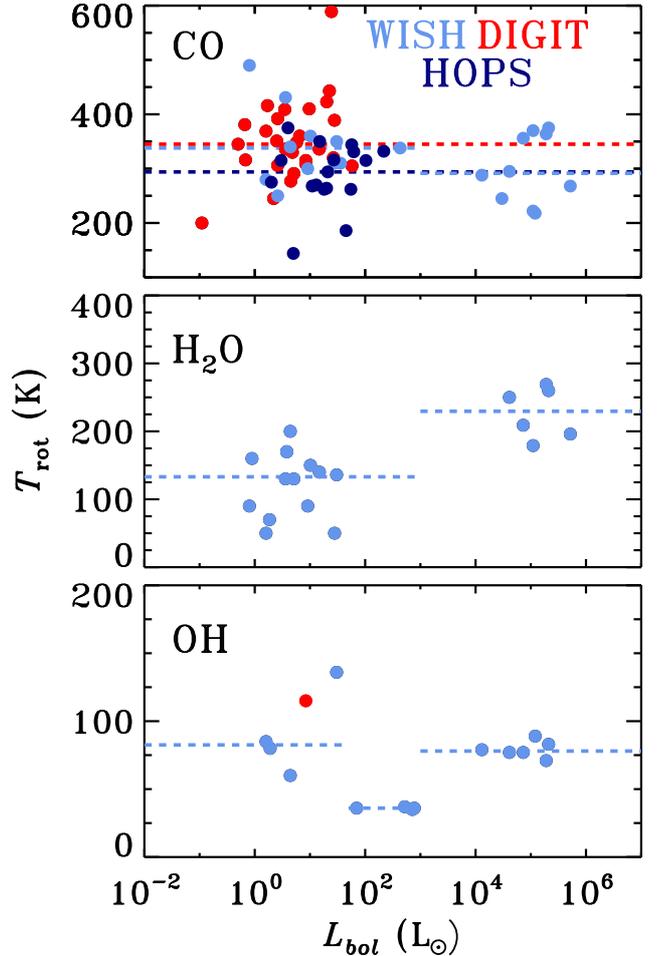}
\vspace{0.5cm}

\caption{\label{alltrot} Rotational temperatures of CO, H$_2$O, and OH for low- to high mass 
star forming regions. WISH, DIGIT and HOPS team's results obtained with PACS are shown in 
blue, red, and navy blue, respectively. Dotted lines show the median values of the rotational 
temperature from each database; for the case of WISH, the median is calculated separately 
for objects with $L_\mathrm{bol}<10^{3}$ L$_\mathrm{\odot}$ and 
$L_\mathrm{bol}>10^{3}$ L$_\mathrm{\odot}$, except for OH where intermediate-mass YSOs 
covering $10^{3}>L_\mathrm{bol}>50$ L$_\mathrm{\odot}$ are also shown separately. The CO and H$_2$O 
excitation of intermediate-mass sources has not yet been surveyed with Herschel.}
\end{center}
\end{figure}
The basic excitation analysis using Boltzmann diagrams in Section 4.2 shows remarkably similar 
rotational temperatures of each molecule for all our high-mass sources, irrespective of their luminosity 
or evolutionary stage. The average values of those temperatures are: 300 K for CO, 220 K for 
H$_2$O, and 80 K for OH. 

Figure \ref{alltrot} presents our results in the context of low- and intermediate-mass 
YSOs studies by \citep{Fi10,He12,Go12,Ma12,Wa13,Ka13,Gr13,Lee13}. 
Rotational temperatures from the Water In Star forming regions with Herschel (WISH), 
the Dust, Ice and Gas in Time (DIGIT), and the Herschel Orion Protostar Survey (HOPS)
 programs are shown separately. OH 
 rotational temperatures of NGC1333 I4B, Serpens SMM1, and L1448 are taken from the literature,
 whereas temperatures for the two additional low-mass YSOs and four intermediate-mass YSOs 
 are calculated in Appendix B based on the line fluxes from \citet{Wa13}. 

 Rotational temperatures of CO are remarkably similar for most sources
 in the luminosity range from 10$^{-1}$ to 10$^{6}$ L$_\mathrm{\odot}$
 and equal to $\sim300-350$ K. For the high-mass sources, this refers
 to the shocked component, not the quiescent envelope component
 discussed in \S 5.1. In order to explain such temperatures in
 low-mass YSOs, two limiting solutions on the physical conditions of
 the gas have been proposed: (i) CO is subthermally excited in hot
 ($T_\mathrm{kin}\geq10^3$ K), low-density ($n$(H$_2$)$\leq10^5$
 cm$^{-3}$) gas (Neufeld 2012, Manoj et al.\ 2013); or (ii) CO is
 close to LTE in warm ($T_\mathrm{kin}$ $\sim$ $T_\mathrm{rot}$) and
 dense ($n$(H$_2$)$> n_\mathrm{crit} \sim 10^{6}$ cm$^{-3}$) gas
 \citep{Ka13}. The low-density scenario (i) in the case of even more
 massive protostars studied in this work is rather unlikely. Even
 though no $^{13}$CO lines are detected in our PACS spectra, three of
 our high-mass protostars were observed in the fundamental $v=1-0$
 vibration-rotation bands of CO and $^{13}$CO \citep{Mi90}. The
 Boltzmann distribution of high-$J$ populations of $^{13}$CO, indicated 
 by a single component on rotational diagrams, 
 implies densities above $10^{6}$ cm$^{-3}$ for W33A and NGC7538-I1 and
 $>10^{7}$ cm$^{-3}$ for W3 IRS5.

 Rotational temperatures of H$_2$O increase for the more massive and
 more luminous YSOs from about 120 K to 220 K (Figure
 \ref{alltrot}). The similarity between the temperatures obtained from
 the absorption and emission lines argues that they arise in the same
 physical component in high-mass YSOs \cite[see also][]{Ce06}. Due to
 the high critical density, the water lines are most likely
 subthermally excited in both low- and high-mass YSOs (see above
 discussion in Section 4.2.2), but in the denser environment of
 high-mass protostars, the gas is closer to LTE and therefore the
 rotational temperatures are higher. High optical depths of H$_2$O lines 
 drive the rotational temperatures to higher values, 
both for the low- and high-mass YSOs. Lines are in emission, when the angular 
size of the emitting region ($\Delta \Omega_L$) multiplied by the blackbody at excitation temperature
 is larger than the continuum flux at the same wavelength,
\begin{equation}
\Delta \Omega_L \times B_\nu(T_{\rm ex}) > F_{\rm cont,\nu}
\end{equation}
and are in absorption in the opposite case. 
   
Rotational temperatures of OH show a broad range of values for
low-mass YSOs (from about 50 to 150 K), whereas they are remarkably
constant for intermediate mass YSOs ($\sim35$ K; see Appendix D) and high-mass YSOs
($\sim80$ K). The low temperatures found towards the intermediate mass
YSOs may be a result of the different lines detected towards those
sources rather than a different excitation mechanism.

\subsection{Correlations}
\begin{figure}[tb]
\begin{center}
\includegraphics[angle=0,height=12cm]{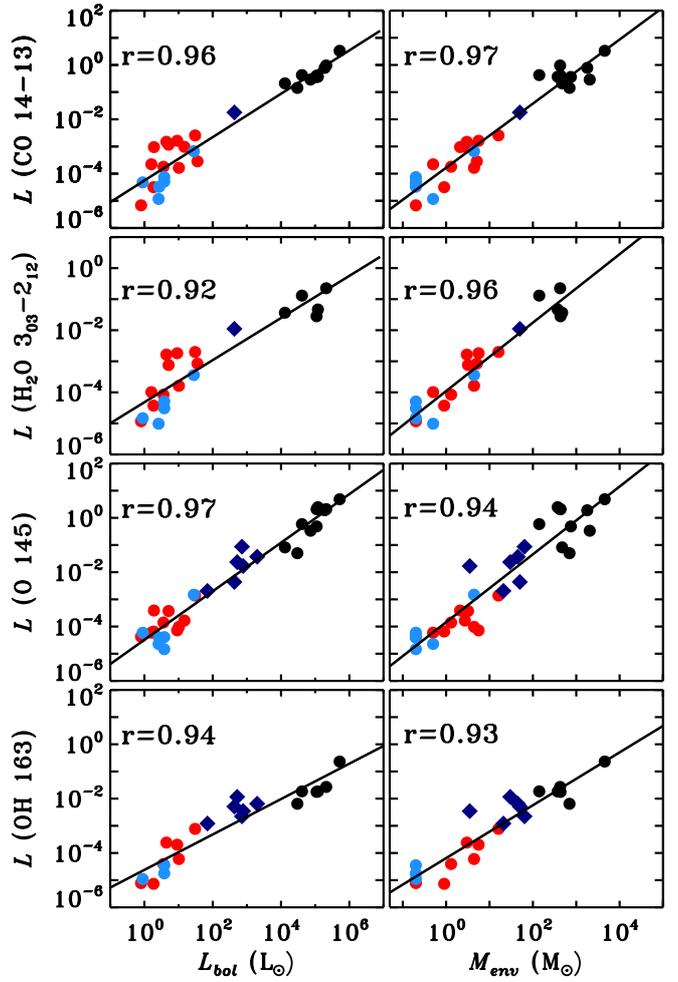}
\vspace{1cm}

\caption{\label{corr} Correlations of line emission with bolometric luminosity
   (left column) and envelope mass (right column) from top to
   bottom: CO 14-13, H$_2$O 3$_{03}$-2$_{12}$, [\ion{O}{i}] at 145
   $\mu$m and OH 163 $\mu$m line luminosities. Low- and intermediate-mass young stellar 
   objects emission is measured over 5$\times$5 PACS maps. Red and blue circles show 
    Class 0 and Class I low-mass YSOs from Karska et al. (2013). Navy diamonds show 
    intermediate mass YSOs from Wampfler et al. (2013; O and OH lines) and from 
    Fich et al. (2010; CO and H$_2$O line).
   High-mass YSOs fluxes are measured in the central position and shown in black. Pearson
   coefficient $r$ is given for each correlation.}
\end{center}
\end{figure}

Figure \ref{corr} shows relations between selected line luminosities of CO, H$_2$O, and [\ion{O}{i}] transitions 
and the physical parameters of the young stellar objects. 
Our sample of objects is extended to the low-mass deeply embedded objects 
studied with PACS in \cite{He12,Go12,Wa13,Ka13} and intermediate-mass objects from \cite{Fi10,Wa13}. 
This allows us to study a broad range of 
luminosities, from $\sim1$ to $10^6$ $L_{\odot}$, and envelope masses, 
from 0.1 to $10^4$ $M_{\odot}$. 

The typical distance to low-mass sources is 200 pc, whereas to
high-mass sources -- 3 kpc. For this comparison, the full PACS array
maps of low-mass regions are taken ($\sim50''$), which corresponds to
spatial scales of $10^4$ AU; only a factor of 3 smaller than the
central spaxel ($\sim10''$) observation of high-mass objects. On the
other hand, the physical sizes of the low-mass sources are smaller
than those of high-mass sources by a factor that is comparable to the
difference in average distance of low- and high-mass sources, so one
could argue that one should compare just the central spaxels for both
cases. Using only the central spaxel for the low-mass YSOs does not
affect the results, however \citep[see Figure 9 in][]{Ka13}.

The choice of CO, H$_2$O, OH, and [\ion{O}{i}] transitions is based on the 
number of detections of those lines in both samples and their emission profiles. 
The strengths of the correlations are quantified using the Pearson coefficient, $r$. For 
the number of sources studied here, the 3 $\sigma$ correlation corresponds to $r\approx0.6$ and 
5 $\sigma$ correlation to $r\approx0.95$.

Figure \ref{corr} shows strong, 5 $\sigma$ correlations between the
selected line luminosities and bolometric luminosities as well as
envelope masses. The more luminous the source, the larger is its 
luminosity in CO, H$_2$O, and [\ion{O}{i}] lines. Similarly, the more
massive is the envelope surrounding the growing protostar, the larger
is the observed line luminosity in those species. The strength of the
correlations over such broad luminosity ranges and envelope masses
suggests that the physical processes responsible for the line emission
are similar.

In the case of low-mass young stellar objects, \citet{Ka13} linked the
CO and H$_2$O emission seen with PACS with the non-dissociative shocks
along the outflow walls, most likely irradiated by the UV photons. The
[\ion{O}{i}] emission, on the other hand, was attributed mainly to the
dissociative shocks at the point of direct impact of the wind on the
dense envelope.  In the high-mass sources the envelope densities and
the strength of radiation are higher, but all in all the origin of the
emission can be similar.

\begin{figure}[tb]
\begin{center}
\includegraphics[angle=0,height=12cm]{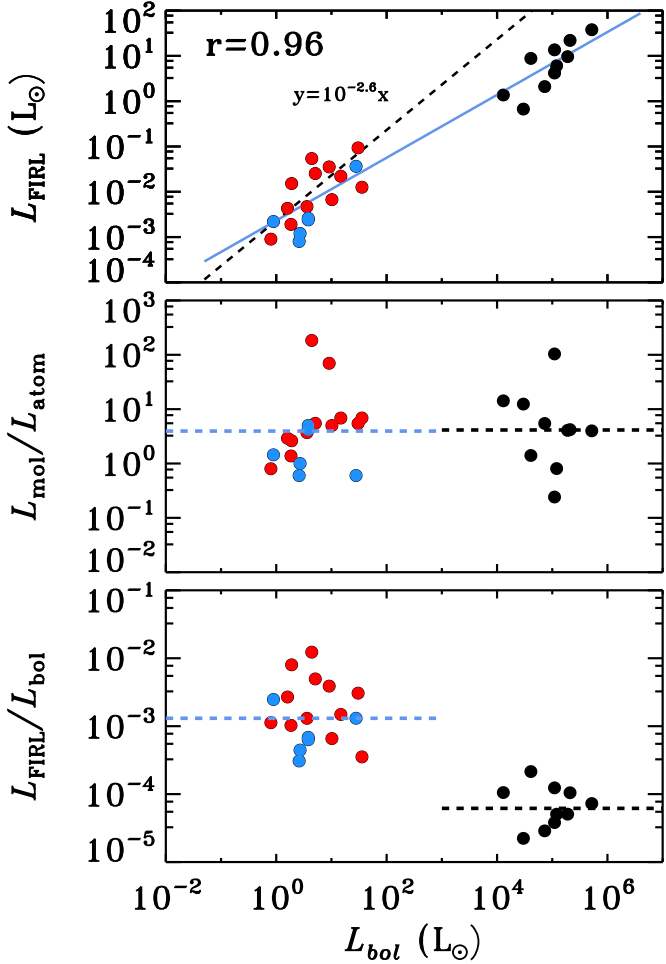}
\vspace{0.5cm}

\caption{\label{firall} From top to bottom: (1) Total far-IR line cooling; 
(2) Ratio of molecular to atomic cooling; (3) Gas to dust cooling ratio, $L_\mathrm{FIRL}/L_\mathrm{bol}$, 
where $L_\mathrm{FIRL}=L_\mathrm{mol}+L_\mathrm{atom}$; 
as a function of bolometric luminosity. Low-mass YSOs are shown in red (Class 0) and 
blue (Class I), whereas high-mass ones are in black.}
\end{center}
\end{figure}
\subsection{Far-IR line cooling}
Figure \ref{firall} compares the total far-IR cooling in lines, its molecular and atomic 
contributions, and the cooling by dust for the YSOs in the luminosity range from  
 $\sim1$ to $10^6$ $L_{\odot}$. 
 
 The far-IR line cooling, $L_\mathrm{FIRL}$, correlates strongly
 (5$\sigma$) with the bolometric luminosity, $L_\mathrm{bol}$, in
 agreement with studies on low-mass YSOs \citep{Ni02,Ka13}. Under the assumption that 
  $L_\mathrm{FIRL}$ is proportional to the shock energy, the strong
 correlation between $L_\mathrm{bol}$ and $L_\mathrm{FIRL}$ has been
 interpreted by \citet{Ni02} as a result of
 the jet power/ velocity being correlated with the escape velocity
 from the protostellar surface or an initial increase of the accretion
 and ejection rate.

 The ratio of molecular and atomic line cooling,
 $L_\mathrm{mol}/L_\mathrm{atom}$, is similar for YSOs of different
 luminosities, although a large scatter is present.  Cooling in
 molecules is about 4 times higher than cooling in oxygen
 atoms. If cooling by [\ion{C}{ii}] was included in the atomic
 cooling, the $L_\mathrm{mol}/L_\mathrm{atom}$ ratio would decrease
 for the high-mass sources. In low-mass YSOs, the [\ion{C}{ii}]
 emission accounts for less than 1\% of the total cooling in lines
 \citep[][for Ser SMM1]{Go12}.  In the high-mass sources, this
 contribution is expected to be higher due to the carbon ionizing and
 CO dissociating FUV radiation. In Section 4.1 we estimate the
 [\ion{C}{ii}] cooling in two high-mass sources to 10-25\% of
 $L_\mathrm{FIRL}$.

 The ratio of cooling by gas (molecular and atomic
 lines) and dust ($L_\mathrm{bol}$) decreases from
 1.3$\cdot10^{-3}$ to 6.2$\cdot10^{-5}$ from low to high-mass YSOs. It
 reflects the fact that H$_2$O, and to a smaller extent OH,
 contributes less to the line luminosity in the high-mass sources,
 because many of its lines are detected in absorption. The detection of H$_2$O
  and OH lines in absorption proves that IR pumping is at least partly
  responsible for the excitation of these molecules and the resulting
  emission lines \citep[e.g.][]{Go06,Wa13}. In the case where
collisions play a marginal role, even the detected emission lines
of those species do not necessarily cool the gas.

 The above numbers do not include cooling from molecules
 outside the PACS wavelength range. This contribution can be
 significant for CO, where for low-mass sources the low-$J$ lines are
 found to increase the total CO line luminosity by about 30\%
 \citep{Ka13}. Thus, the CO contribution to the total gas cooling is
 likely to be even larger than suggested by Table \ref{cool}.

\section{Conclusions}
We have characterized the central position \textit{Herschel}/PACS spectra of 10
high-mass protostars and compared them with the results for low- and intermediate-mass 
protostars analyzed in a similar manner. The conclusions are as follows:
\begin{enumerate}

\item Far-IR gas cooling of high-mass YSOs is dominated by CO (from $\sim15$ to $85$\% of total far-IR 
line cooling, with median contribution of 74\%) and to smaller extent by [\ion{O}{i}] 
(with median value $\sim$20 \%). H$_2$O and OH median contributions to the far-IR cooling 
are less than 1\%. In contrast for low-mass YSOs, the H$_2$O, CO, and [\ion{O}{i}]
contributions are comparable. The effective cooling by H$_2$O is reduced because many 
far-IR lines are in absorption. The [\ion{O}{i}] cooling increases for more evolved sources 
in both mass regimes.

\item Rotational diagrams of CO in the PACS range show a single, \textit{warm component},
corresponding to rotational temperature of \mbox{$\sim$300 K}, consistent with low-mass YSOs. 
Upper limits on 
high$-J$ CO do not exclude the existence of an additional, \textit{hot component} in several 
sources of our sample.

\item Emission from the quiescent envelope accounts for \mbox{$\sim$45-85 \%} of the total CO luminosity 
 observed in the PACS range. The corresponding values for the cooler and less dense envelopes of 
 low-mass YSOs are below 10\%. Additional physical components, most likely shocks, are necessary 
 to explain the highest$-J$ CO lines. 

\item Rotational diagrams of H$_2$O are characterized by
  $T_\mathrm{rot}\sim250$ K for all sources both from emission and
  absorption data.  This temperature is about 100 K higher than for
  low-mass sources, likely due to the higher densities in high-mass sources.
  The diagrams show scatter due to subthermal excitation and optical
  depth effects.
  
\item OH rotational diagrams are described by a single rotational temperatures of $\sim$80 K,
consistent with most low-mass YSOs, but higher by $\sim$45 K than for intermediate-mass objects.
Similar to H$_2$O, lines are sub-thermally excited.
 
\item Fluxes of the H$_2$O $3_{03}-2_{12}$ line and the CO
  $14-13$ line strongly correlate with bolometric luminosities and envelope masses over 
  6 and 7 orders of magnitude, respectively. This correlation suggests 
  a common physical mechanism responsible for the line excitation,
  most likely the non-dissociative shocks based on the studies of
  low-mass protostars.

\item Across the large luminosity range from $\sim1$ to $10^6$
  $L_{\odot}$, the far-IR line cooling strongly correlates with the
  bolometric luminosity, in agreement with studies on low-mass YSOs.
  The ratio of molecular and atomic line cooling is $\sim$4, similar
  for all those YSOs.

\item Because several H$_2$O lines are in absorption, the gas to dust cooling ratio 
decreases from 1.3$\cdot10^{-3}$ to 6.2$\cdot10^{-5}$ from low to high-mass 
YSOs.

\end{enumerate}

\begin{acknowledgements}
Herschel is an ESA space observatory with science instruments provided
by European-led Principal Investigator consortia and with important
participation from NASA. AK acknowledges support from the Christiane 
N\"{u}sslein-Volhard-Foundation, the L$^\prime$Or\'{e}al  
Deutschland and the German Commission for UNESCO via the `Women in Science' prize. 
JRG, LC, and JC thank the Spanish MINECO for funding support from grants AYA2009-07304,
 AYA2012-32032 and CSD2009-00038 and AYA. JRG is supported by a Ramo\'on y Cajal research contract.
Astrochemistry in Leiden is supported by the Netherlands Research
School for Astronomy (NOVA), by a Royal Netherlands Academy of Arts
and Sciences (KNAW) professor prize, by a Spinoza grant and grant
614.001.008 from the Netherlands Organisation for Scientific Research
(NWO), and by the European Community's Seventh Framework Programme
FP7/2007-2013 under grant agreement 238258 (LASSIE).
\end{acknowledgements}   
   
\bibliographystyle{aa}
\bibliography{biblio2009}

\begin{thebibliography}{69}
\expandafter\ifx\csname natexlab\endcsname\relax\def\natexlab#1{#1}\fi

\bibitem[{{Andr\'{e}} {et~al.}(1993){Andr\'{e}}, {Ward-Thompson}, \&
  {Barsony}}]{An93}
{Andr\'{e}}, P., {Ward-Thompson}, D., \& {Barsony}, M. 1993, \apj, 406, 122

\bibitem[{{Andr\'{e}} {et~al.}(2000){Andr\'{e}}, {Ward-Thompson}, \&
  {Barsony}}]{An00}
{Andr\'{e}}, P., {Ward-Thompson}, D., \& {Barsony}, M. 2000, Protostars and
  Planets IV, 59

\bibitem[{{Beuther} {et~al.}(2007){Beuther}, {Churchwell}, {McKee}, \&
  {Tan}}]{Be07}
{Beuther}, H., {Churchwell}, E.~B., {McKee}, C.~F., \& {Tan}, J.~C. 2007,
  Protostars and Planets V, 165

\bibitem[{{Bontemps} {et~al.}(1996){Bontemps}, {Andre}, {Terebey}, \&
  {Cabrit}}]{Bo96}
{Bontemps}, S., {Andre}, P., {Terebey}, S., \& {Cabrit}, S. 1996, \aap, 311,
  858

\bibitem[{{Boonman} {et~al.}(2003){Boonman}, {Doty}, {van Dishoeck}, {Bergin},
  {Melnick}, {Wright}, \& {Stark}}]{Bo03b}
{Boonman}, A.~M.~S., {Doty}, S.~D., {van Dishoeck}, E.~F., {et~al.} 2003, \aap,
  406, 937

\bibitem[{{Boonman} \& {van Dishoeck}(2003)}]{Bo03}
{Boonman}, A.~M.~S. \& {van Dishoeck}, E.~F. 2003, \aap, 403, 1003

\bibitem[{{Ceccarelli} {et~al.}(1996){Ceccarelli}, {Hollenbach}, \&
  {Tielens}}]{CC96}
{Ceccarelli}, C., {Hollenbach}, D.~J., \& {Tielens}, A.~G.~G.~M. 1996, \apj,
  471, 400

\bibitem[{{Cernicharo} {et~al.}(2000){Cernicharo}, {Goicoechea}, \&
  {Caux}}]{Ce00}
{Cernicharo}, J., {Goicoechea}, J.~R., \& {Caux}, E. 2000, \apjl, 534, L199

\bibitem[{{Cernicharo} {et~al.}(2006){Cernicharo}, {Goicoechea}, {Daniel},
  {Lerate}, {Barlow}, {Swinyard}, {van Dishoeck}, {Lim}, {Viti}, \&
  {Yates}}]{Ce06}
{Cernicharo}, J., {Goicoechea}, J.~R., {Daniel}, F., {et~al.} 2006, \apjl, 649,
  L33

\bibitem[{{Cesaroni}(2005)}]{Ce05}
{Cesaroni}, R. 2005, \apss, 295, 5

\bibitem[{{Chavarr{\'{\i}}a} {et~al.}(2010){Chavarr{\'{\i}}a}, {Herpin},
  {Jacq}, {Braine}, {Bontemps}, {Baudry}, {Marseille}, {van der Tak},
  {Pietropaoli}, {Wyrowski}, {Shipman}, {Frieswijk}, {van Dishoeck},
  {Cernicharo}, {Bachiller}, {Benedettini}, {Benz}, {Bergin}, {Bjerkeli},
  {Blake}, {Bruderer}, {Caselli}, {Codella}, {Daniel}, {di Giorgio}, {Dominik},
  {Doty}, {Encrenaz}, {Fich}, {Fuente}, {Giannini}, {Goicoechea}, {de Graauw},
  {Hartogh}, {Helmich}, {Herczeg}, {Hogerheijde}, {Johnstone}, {J{\o}rgensen},
  {Kristensen}, {Larsson}, {Lis}, {Liseau}, {McCoey}, {Melnick}, {Nisini},
  {Olberg}, {Parise}, {Pearson}, {Plume}, {Risacher}, {Santiago-Garc{\'{\i}}a},
  {Saraceno}, {Stutzki}, {Szczerba}, {Tafalla}, {Tielens}, {van Kempen},
  {Visser}, {Wampfler}, {Willem}, \& {Y{\i}ld{\i}z}}]{Ch10}
{Chavarr{\'{\i}}a}, L., {Herpin}, F., {Jacq}, T., {et~al.} 2010, \aap, 521, L37

\bibitem[{{Clegg} {et~al.}(1996){Clegg}, {Ade}, {Armand}, {Baluteau}, {Barlow},
  {Buckley}, {Berges}, {Burgdorf}, {Caux}, {Ceccarelli}, {Cerulli}, {Church},
  {Cotin}, {Cox}, {Cruvellier}, {Culhane}, {Davis}, {di Giorgio}, {Diplock},
  {Drummond}, {Emery}, {Ewart}, {Fischer}, {Furniss}, {Glencross},
  {Greenhouse}, {Griffin}, {Gry}, {Harwood}, {Hazell}, {Joubert}, {King},
  {Lim}, {Liseau}, {Long}, {Lorenzetti}, {Molinari}, {Murray}, {Naylor},
  {Nisini}, {Norman}, {Omont}, {Orfei}, {Patrick}, {Pequignot}, {Pouliquen},
  {Price}, {Nguyen-Q-Rieu}, {Rogers}, {Robinson}, {Saisse}, {Saraceno},
  {Serra}, {Sidher}, {Smith}, {Smith}, {Spinoglio}, {Swinyard}, {Texier},
  {Towlson}, {Trams}, {Unger}, \& {White}}]{LWS}
{Clegg}, P.~E., {Ade}, P.~A.~R., {Armand}, C., {et~al.} 1996, \aap, 315, L38

\bibitem[{{Daniel} {et~al.}(2011){Daniel}, {Dubernet}, \& {Grosjean}}]{Da11}
{Daniel}, F., {Dubernet}, M.-L., \& {Grosjean}, A. 2011, \aap, 536, A76

\bibitem[{{de Graauw} {et~al.}(1996){de Graauw}, {Haser}, {Beintema},
  {Roelfsema}, {van Agthoven}, {Barl}, {Bauer}, {Bekenkamp}, {Boonstra},
  {Boxhoorn}, {Cote}, {de Groene}, {van Dijkhuizen}, {Drapatz}, {Evers},
  {Feuchtgruber}, {Frericks}, {Genzel}, {Haerendel}, {Heras}, {van der Hucht},
  {van der Hulst}, {Huygen}, {Jacobs}, {Jakob}, {Kamperman}, {Katterloher},
  {Kester}, {Kunze}, {Kussendrager}, {Lahuis}, {Lamers}, {Leech}, {van der
  Lei}, {van der Linden}, {Luinge}, {Lutz}, {Melzner}, {Morris}, {van Nguyen},
  {Ploeger}, {Price}, {Salama}, {Schaeidt}, {Sijm}, {Smoorenburg}, {Spakman},
  {Spoon}, {Steinmayer}, {Stoecker}, {Valentijn}, {Vandenbussche}, {Visser},
  {Waelkens}, {Waters}, {Wensink}, {Wesselius}, {Wiezorrek}, {Wieprecht},
  {Wijnbergen}, {Wildeman}, \& {Young}}]{SWS}
{de Graauw}, T., {Haser}, L.~N., {Beintema}, D.~A., {et~al.} 1996, \aap, 315,
  L49

\bibitem[{{de Graauw} {et~al.}(2010){de Graauw}, {Helmich}, {Phillips},
  {Stutzki}, {Caux}, {Whyborn}, {Dieleman}, {Roelfsema}, {Aarts}, {Assendorp},
  {Bachiller}, {Baechtold}, {Barcia}, {Beintema}, {Belitsky}, {Benz}, {Bieber},
  {Boogert}, {Borys}, {Bumble}, {Ca{\"i}s}, {Caris}, {Cerulli-Irelli},
  {Chattopadhyay}, {Cherednichenko}, {Ciechanowicz}, {Coeur-Joly}, {Comito},
  {Cros}, {de Jonge}, {de Lange}, {Delforges}, {Delorme}, {den Boggende},
  {Desbat}, {Diez-Gonz{\'a}lez}, {di Giorgio}, {Dubbeldam}, {Edwards},
  {Eggens}, {Erickson}, {Evers}, {Fich}, {Finn}, {Franke}, {Gaier}, {Gal},
  {Gao}, {Gallego}, {Gauffre}, {Gill}, {Glenz}, {Golstein}, {Goulooze},
  {Gunsing}, {G{\"u}sten}, {Hartogh}, {Hatch}, {Higgins}, {Honingh}, {Huisman},
  {Jackson}, {Jacobs}, {Jacobs}, {Jarchow}, {Javadi}, {Jellema}, {Justen},
  {Karpov}, {Kasemann}, {Kawamura}, {Keizer}, {Kester}, {Klapwijk}, {Klein},
  {Kollberg}, {Kooi}, {Kooiman}, {Kopf}, {Krause}, {Krieg}, {Kramer},
  {Kruizenga}, {Kuhn}, {Laauwen}, {Lai}, {Larsson}, {Leduc}, {Leinz}, {Lin},
  {Liseau}, {Liu}, {Loose}, {L{\'o}pez-Fernandez}, {Lord}, {Luinge}, {Marston},
  {Mart{\'{\i}}n-Pintado}, {Maestrini}, {Maiwald}, {McCoey}, {Mehdi}, {Megej},
  {Melchior}, {Meinsma}, {Merkel}, {Michalska}, {Monstein}, {Moratschke},
  {Morris}, {Muller}, {Murphy}, {Naber}, {Natale}, {Nowosielski}, {Nuzzolo},
  {Olberg}, {Olbrich}, {Orfei}, {Orleanski}, {Ossenkopf}, {Peacock}, {Pearson},
  {Peron}, {Phillip-May}, {Piazzo}, {Planesas}, {Rataj}, {Ravera}, {Risacher},
  {Salez}, {Samoska}, {Saraceno}, {Schieder}, {Schlecht}, {Schl{\"o}der},
  {Schm{\"u}lling}, {Schultz}, {Schuster}, {Siebertz}, {Smit}, {Szczerba},
  {Shipman}, {Steinmetz}, {Stern}, {Stokroos}, {Teipen}, {Teyssier}, {Tils},
  {Trappe}, {van Baaren}, {van Leeuwen}, {van de Stadt}, {Visser}, {Wildeman},
  {Wafelbakker}, {Ward}, {Wesselius}, {Wild}, {Wulff}, {Wunsch}, {Tielens},
  {Zaal}, {Zirath}, {Zmuidzinas}, \& {Zwart}}]{dG10}
{de Graauw}, T., {Helmich}, F.~P., {Phillips}, T.~G., {et~al.} 2010, \aap, 518,
  L6

\bibitem[{{Doty} \& {Neufeld}(1997)}]{DN97}
{Doty}, S.~D. \& {Neufeld}, D.~A. 1997, \apj, 489, 122

\bibitem[{{Fich} {et~al.}(2010){Fich}, {Johnstone}, {van Kempen}, {McCoey},
  {Fuente}, {Caselli}, {Kristensen}, {Plume}, {Cernicharo}, {Herczeg}, {van
  Dishoeck}, {Wampfler}, {Gaufre}, {Gill}, {Javadi}, {Justen}, {Laauwen},
  {Luinge}, {Ossenkopf}, {Pearson}, {Bachiller}, {Baudry}, {Benedettini},
  {Bergin}, {Benz}, {Bjerkeli}, {Blake}, {Bontemps}, {Braine}, {Bruderer},
  {Codella}, {Daniel}, {di Giorgio}, {Dominik}, {Doty}, {Encrenaz}, {Giannini},
  {Goicoechea}, {de Graauw}, {Helmich}, {Herpin}, {Hogerheijde}, {Jacq},
  {J{\o}rgensen}, {Larsson}, {Lis}, {Liseau}, {Marseille}, {Melnick}, {Nisini},
  {Olberg}, {Parise}, {Risacher}, {Santiago}, {Saraceno}, {Shipman}, {Tafalla},
  {van der Tak}, {Visser}, {Wyrowski}, \& {Y{\i}ld{\i}z}}]{Fi10}
{Fich}, M., {Johnstone}, D., {van Kempen}, T.~A., {et~al.} 2010, \aap, 518, L86

\bibitem[{{Fischer} {et~al.}(1999){Fischer}, {Luhman}, {Satyapal},
  {Greenhouse}, {Stacey}, {Bradford}, {Lord}, {Brauher}, {Unger}, {Clegg},
  {Smith}, {Melnick}, {Colbert}, {Malkan}, {Spinoglio}, {Cox}, {Harvey},
  {Suter}, \& {Strelnitski}}]{Fi99}
{Fischer}, J., {Luhman}, M.~L., {Satyapal}, S., {et~al.} 1999, \apss, 266, 91

\bibitem[{{Goicoechea} \& {Cernicharo}(2001)}]{Go01}
{Goicoechea}, J.~R. \& {Cernicharo}, J. 2001, \apjl, 554, L213

\bibitem[{{Goicoechea} {et~al.}(2012){Goicoechea}, {Cernicharo}, {Karska},
  {Herczeg}, {Polehampton}, {Wampfler}, {Kristensen}, {van Dishoeck},
  {Etxaluze}, {Bern{\'e}}, \& {Visser}}]{Go12}
{Goicoechea}, J.~R., {Cernicharo}, J., {Karska}, A., {et~al.} 2012, \aap, 548,
  A77

\bibitem[{{Goicoechea} {et~al.}(2006){Goicoechea}, {Cernicharo}, {Lerate},
  {Daniel}, {Barlow}, {Swinyard}, {Lim}, {Viti}, \& {Yates}}]{Go06}
{Goicoechea}, J.~R., {Cernicharo}, J., {Lerate}, M.~R., {et~al.} 2006, \apjl,
  641, L49

\bibitem[{{Goldsmith} \& {Langer}(1999)}]{GL99}
{Goldsmith}, P.~F. \& {Langer}, W.~D. 1999, \apj, 517, 209

\bibitem[{{Green} {et~al.}(2013){Green}, {Evans}, {J{\o}rgensen}, {Herczeg},
  {Kristensen}, {Lee}, {Dionatos}, {Yildiz}, {Salyk}, {Meeus}, {Bouwman},
  {Visser}, {Bergin}, {van Dishoeck}, {Rascati}, {Karska}, {van Kempen},
  {Dunham}, {Lindberg}, {Fedele}, \& {DIGIT Team}}]{Gr13}
{Green}, J.~D., {Evans}, II, N.~J., {J{\o}rgensen}, J.~K., {et~al.} 2013, \apj,
  770, 123

\bibitem[{{Harwit} {et~al.}(1998){Harwit}, {Neufeld}, {Melnick}, \&
  {Kaufman}}]{Ha98}
{Harwit}, M., {Neufeld}, D.~A., {Melnick}, G.~J., \& {Kaufman}, M.~J. 1998,
  \apjl, 497, L105

\bibitem[{{Helmich} \& {van Dishoeck}(1997)}]{HvD97}
{Helmich}, F.~P. \& {van Dishoeck}, E.~F. 1997, \aaps, 124, 205

\bibitem[{{Herczeg} {et~al.}(2011){Herczeg}, {Brown}, {van Dishoeck}, \&
  {Pontoppidan}}]{He11}
{Herczeg}, G.~J., {Brown}, J.~M., {van Dishoeck}, E.~F., \& {Pontoppidan},
  K.~M. 2011, \aap, 533, A112

\bibitem[{{Herczeg} {et~al.}(2012){Herczeg}, {Karska}, {Bruderer},
  {Kristensen}, {van Dishoeck}, {J{\o}rgensen}, {Visser}, {Wampfler}, {Bergin},
  {Y{\i}ld{\i}z}, {Pontoppidan}, \& {Gracia-Carpio}}]{He12}
{Herczeg}, G.~J., {Karska}, A., {Bruderer}, S., {et~al.} 2012, \aap, 540, A84

\bibitem[{{Hogerheijde} \& {van der Tak}(2000)}]{RATRAN}
{Hogerheijde}, M.~R. \& {van der Tak}, F.~F.~S. 2000, \aap, 362, 697

\bibitem[{{Karska} {et~al.}(2013){Karska}, {Herczeg}, {van Dishoeck},
  {Wampfler}, {Kristensen}, {Goicoechea}, {Visser}, {Nisini}, {San
  Jos{\'e}-Garc{\'{\i}}a}, {Bruderer}, {{\'S}niady}, {Doty}, {Fedele},
  {Y{\i}ld{\i}z}, {Benz}, {Bergin}, {Caselli}, {Herpin}, {Hogerheijde},
  {Johnstone}, {J{\o}rgensen}, {Liseau}, {Tafalla}, {van der Tak}, \&
  {Wyrowski}}]{Ka13}
{Karska}, A., {Herczeg}, G.~J., {van Dishoeck}, E.~F., {et~al.} 2013, \aap,
  552, A141

\bibitem[{{Kessler} {et~al.}(1996){Kessler}, {Steinz}, {Anderegg}, {Clavel},
  {Drechsel}, {Estaria}, {Faelker}, {Riedinger}, {Robson}, {Taylor}, \&
  {Xim{\'e}nez de Ferr{\'a}n}}]{ISO}
{Kessler}, M.~F., {Steinz}, J.~A., {Anderegg}, M.~E., {et~al.} 1996, \aap, 315,
  L27

\bibitem[{{Kristensen} {et~al.}(2012){Kristensen}, {van Dishoeck}, {Bergin},
  {Visser}, {Y{\i}ld{\i}z}, {San Jose-Garcia}, {J{\o}rgensen}, {Herczeg},
  {Johnstone}, {Wampfler}, {Benz}, {Bruderer}, {Cabrit}, {Caselli}, {Doty},
  {Harsono}, {Herpin}, {Hogerheijde}, {Karska}, {van Kempen}, {Liseau},
  {Nisini}, {Tafalla}, {van der Tak}, \& {Wyrowski}}]{Kr12}
{Kristensen}, L.~E., {van Dishoeck}, E.~F., {Bergin}, E.~A., {et~al.} 2012,
  \aap, 542, A8

\bibitem[{{Lee} {et~al.}(2013){Lee}, {Lee}, {Lee}, {Green}, {Evans}, {Choi},
  {Kristensen}, {Dionatos}, {J{\o}rgensen}, \& {the DIGIT team}}]{Lee13}
{Lee}, J., {Lee}, J.-E., {Lee}, S., {et~al.} 2013, ArXiv e-prints

\bibitem[{{Leurini} {et~al.}(2013){Leurini}, {Wyrowski}, {Herpin}, {van der
  Tak}, {G{\"u}sten}, \& {van Dishoeck}}]{Le13}
{Leurini}, S., {Wyrowski}, F., {Herpin}, F., {et~al.} 2013, \aap, 550, A10

\bibitem[{{Manoj} {et~al.}(2013){Manoj}, {Watson}, {Neufeld}, {Megeath},
  {Vavrek}, {Yu}, {Visser}, {Bergin}, {Fischer}, {Tobin}, {Stutz}, {Ali},
  {Wilson}, {Di Francesco}, {Osorio}, {Maret}, \& {Poteet}}]{Ma12}
{Manoj}, P., {Watson}, D.~M., {Neufeld}, D.~A., {et~al.} 2013, \apj, 763, 83

\bibitem[{{Mitchell} {et~al.}(1990){Mitchell}, {Maillard}, {Allen}, {Beer}, \&
  {Belcourt}}]{Mi90}
{Mitchell}, G.~F., {Maillard}, J.-P., {Allen}, M., {Beer}, R., \& {Belcourt},
  K. 1990, \apj, 363, 554

\bibitem[{{Mottram} {et~al.}(2013){Mottram}, {van Dishoeck}, {Schmalzl},
  {Kristensen}, {Visser}, {Hogerheijde}, \& {Bruderer}}]{Mo13}
{Mottram}, J.~C., {van Dishoeck}, E.~F., {Schmalzl}, M., {et~al.} 2013, \aap,
  558, A126

\bibitem[{{Neufeld}(2012)}]{Ne12}
{Neufeld}, D.~A. 2012, \apj, 749, 125

\bibitem[{{Nisini} {et~al.}(2002){Nisini}, {Giannini}, \& {Lorenzetti}}]{Ni02}
{Nisini}, B., {Giannini}, T., \& {Lorenzetti}, D. 2002, \apj, 574, 246

\bibitem[{{Ott}(2010)}]{Ot10}
{Ott}, S. 2010, in Astronomical Society of the Pacific Conference Series, Vol.
  434, Astronomical Data Analysis Software and Systems XIX, ed. Y.~{Mizumoto},
  K.-I. {Morita}, \& M.~{Ohishi}, 139

\bibitem[{{Pilbratt} {et~al.}(2010){Pilbratt}, {Riedinger}, {Passvogel},
  {Crone}, {Doyle}, {Gageur}, {Heras}, {Jewell}, {Metcalfe}, {Ott}, \&
  {Schmidt}}]{Herschel}
{Pilbratt}, G.~L., {Riedinger}, J.~R., {Passvogel}, T., {et~al.} 2010, \aap,
  518, L1

\bibitem[{{Poglitsch} {et~al.}(2010){Poglitsch}, {Waelkens}, {Geis},
  {Feuchtgruber}, {Vandenbussche}, {Rodriguez}, {Krause}, {Renotte}, {van
  Hoof}, {Saraceno}, {Cepa}, {Kerschbaum}, {Agn{\`e}se}, {Ali}, {Altieri},
  {Andreani}, {Augueres}, {Balog}, {Barl}, {Bauer}, {Belbachir}, {Benedettini},
  {Billot}, {Boulade}, {Bischof}, {Blommaert}, {Callut}, {Cara}, {Cerulli},
  {Cesarsky}, {Contursi}, {Creten}, {De Meester}, {Doublier}, {Doumayrou},
  {Duband}, {Exter}, {Genzel}, {Gillis}, {Gr{\"o}zinger}, {Henning},
  {Herreros}, {Huygen}, {Inguscio}, {Jakob}, {Jamar}, {Jean}, {de Jong},
  {Katterloher}, {Kiss}, {Klaas}, {Lemke}, {Lutz}, {Madden}, {Marquet},
  {Martignac}, {Mazy}, {Merken}, {Montfort}, {Morbidelli}, {M{\"u}ller},
  {Nielbock}, {Okumura}, {Orfei}, {Ottensamer}, {Pezzuto}, {Popesso},
  {Putzeys}, {Regibo}, {Reveret}, {Royer}, {Sauvage}, {Schreiber}, {Stegmaier},
  {Schmitt}, {Schubert}, {Sturm}, {Thiel}, {Tofani}, {Vavrek}, {Wetzstein},
  {Wieprecht}, \& {Wiezorrek}}]{Po10}
{Poglitsch}, A., {Waelkens}, C., {Geis}, N., {et~al.} 2010, \aap, 518, L2

\bibitem[{{Robitaille}(2011)}]{Ro11}
{Robitaille}, T.~P. 2011, \aap, 536, A79

\bibitem[{{Rosenthal} {et~al.}(2000){Rosenthal}, {Bertoldi}, \&
  {Drapatz}}]{Ro00}
{Rosenthal}, D., {Bertoldi}, F., \& {Drapatz}, S. 2000, \aap, 356, 705

\bibitem[{{San Jos{\'e}-Garc{\'{\i}}a} {et~al.}(2013){San
  Jos{\'e}-Garc{\'{\i}}a}, {Mottram}, {Kristensen}, {van Dishoeck},
  {Y{\i}ld{\i}z}, {van der Tak}, {Herpin}, {Visser}, {McCoey}, {Wyrowski},
  {Braine}, \& {Johnstone}}]{IreneCO}
{San Jos{\'e}-Garc{\'{\i}}a}, I., {Mottram}, J.~C., {Kristensen}, L.~E.,
  {et~al.} 2013, \aap, 553, A125

\bibitem[{{Sch{\"o}ier} {et~al.}(2005){Sch{\"o}ier}, {van der Tak}, {van
  Dishoeck}, \& {Black}}]{LAMDA}
{Sch{\"o}ier}, F.~L., {van der Tak}, F.~F.~S., {van Dishoeck}, E.~F., \&
  {Black}, J.~H. 2005, \aap, 432, 369

\bibitem[{{Sempere} {et~al.}(2000){Sempere}, {Cernicharo}, {Lefloch},
  {Gonz{\'a}lez-Alfonso}, \& {Leeks}}]{Se00}
{Sempere}, M.~J., {Cernicharo}, J., {Lefloch}, B., {Gonz{\'a}lez-Alfonso}, E.,
  \& {Leeks}, S. 2000, \apjl, 530, L123

\bibitem[{{Shirley} {et~al.}(2000){Shirley}, {Evans}, {Rawlings}, \&
  {Gregersen}}]{Sh00}
{Shirley}, Y.~L., {Evans}, II, N.~J., {Rawlings}, J.~M.~C., \& {Gregersen},
  E.~M. 2000, \apjs, 131, 249

\bibitem[{{Shu} {et~al.}(1987){Shu}, {Adams}, \& {Lizano}}]{Shu87}
{Shu}, F.~H., {Adams}, F.~C., \& {Lizano}, S. 1987, \araa, 25, 23

\bibitem[{{Sturm} {et~al.}(2002){Sturm}, {Lutz}, {Verma}, {Netzer},
  {Sternberg}, {Moorwood}, {Oliva}, \& {Genzel}}]{St02}
{Sturm}, E., {Lutz}, D., {Verma}, A., {et~al.} 2002, \aap, 393, 821

\bibitem[{{van der Tak} {et~al.}(2007){van der Tak}, {Black}, {Sch{\"o}ier},
  {Jansen}, \& {van Dishoeck}}]{RADEX}
{van der Tak}, F.~F.~S., {Black}, J.~H., {Sch{\"o}ier}, F.~L., {Jansen}, D.~J.,
  \& {van Dishoeck}, E.~F. 2007, \aap, 468, 627

\bibitem[{{van der Tak} {et~al.}(2013){van der Tak}, {Chavarr{\'{\i}}a},
  {Herpin}, {Wyrowski}, {Walmsley}, {van Dishoeck}, {Benz}, {Bergin},
  {Caselli}, {Hogerheijde}, {Johnstone}, {Kristensen}, {Liseau}, {Nisini}, \&
  {Tafalla}}]{vT13}
{van der Tak}, F.~F.~S., {Chavarr{\'{\i}}a}, L., {Herpin}, F., {et~al.} 2013,
  \aap, 554, A83

\bibitem[{{van der Tak} {et~al.}(2000){van der Tak}, {van Dishoeck}, {Evans},
  \& {Blake}}]{vT00}
{van der Tak}, F.~F.~S., {van Dishoeck}, E.~F., {Evans}, II, N.~J., \& {Blake},
  G.~A. 2000, \apj, 537, 283

\bibitem[{{van der Wiel} {et~al.}(2013){van der Wiel}, {Pagani}, {van der Tak},
  {Ka{\'z}mierczak}, \& {Ceccarelli}}]{Wi13}
{van der Wiel}, M.~H.~D., {Pagani}, L., {van der Tak}, F.~F.~S.,
  {Ka{\'z}mierczak}, M., \& {Ceccarelli}, C. 2013, \aap, 553, A11

\bibitem[{{van Dishoeck}(2004)}]{EvDISO}
{van Dishoeck}, E.~F. 2004, \araa, 42, 119

\bibitem[{{van Dishoeck} {et~al.}(2011){van Dishoeck}, {Kristensen}, {Benz},
  {Bergin}, {Caselli}, {Cernicharo}, {Herpin}, {Hogerheijde}, {Johnstone},
  {Liseau}, {Nisini}, {Shipman}, {Tafalla}, {van der Tak}, {Wyrowski},
  {Aikawa}, {Bachiller}, {Baudry}, {Benedettini}, {Bjerkeli}, {Blake},
  {Bontemps}, {Braine}, {Brinch}, {Bruderer}, {Chavarr{\'{\i}}a}, {Codella},
  {Daniel}, {de Graauw}, {Deul}, {di Giorgio}, {Dominik}, {Doty}, {Dubernet},
  {Encrenaz}, {Feuchtgruber}, {Fich}, {Frieswijk}, {Fuente}, {Giannini},
  {Goicoechea}, {Helmich}, {Herczeg}, {Jacq}, {J{\o}rgensen}, {Karska},
  {Kaufman}, {Keto}, {Larsson}, {Lefloch}, {Lis}, {Marseille}, {McCoey},
  {Melnick}, {Neufeld}, {Olberg}, {Pagani}, {Pani{\'c}}, {Parise}, {Pearson},
  {Plume}, {Risacher}, {Salter}, {Santiago-Garc{\'{\i}}a}, {Saraceno},
  {St{\"a}uber}, {van Kempen}, {Visser}, {Viti}, {Walmsley}, {Wampfler}, \&
  {Y{\i}ld{\i}z}}]{WISH}
{van Dishoeck}, E.~F., {Kristensen}, L.~E., {Benz}, A.~O., {et~al.} 2011,
  \pasp, 123, 138

\bibitem[{{van Dishoeck} {et~al.}(1998){van Dishoeck}, {Wright}, {Cernicharo},
  {Gonzalez-Alfonso}, {de Graauw}, {Helmich}, \& {Vandenbussche}}]{vD98}
{van Dishoeck}, E.~F., {Wright}, C.~M., {Cernicharo}, J., {et~al.} 1998, \apjl,
  502, L173

\bibitem[{{van Kempen} {et~al.}(2010){van Kempen}, {Kristensen}, {Herczeg},
  {Visser}, {van Dishoeck}, {Wampfler}, {Bruderer}, {Benz}, {Doty}, {Brinch},
  {Hogerheijde}, {J{\o}rgensen}, {Tafalla}, {Neufeld}, {Bachiller}, {Baudry},
  {Benedettini}, {Bergin}, {Bjerkeli}, {Blake}, {Bontemps}, {Braine},
  {Caselli}, {Cernicharo}, {Codella}, {Daniel}, {di Giorgio}, {Dominik},
  {Encrenaz}, {Fich}, {Fuente}, {Giannini}, {Goicoechea}, {de Graauw},
  {Helmich}, {Herpin}, {Jacq}, {Johnstone}, {Kaufman}, {Larsson}, {Lis},
  {Liseau}, {Marseille}, {McCoey}, {Melnick}, {Nisini}, {Olberg}, {Parise},
  {Pearson}, {Plume}, {Risacher}, {Santiago-Garc{\'{\i}}a}, {Saraceno},
  {Shipman}, {van der Tak}, {Wyrowski}, {Y{\i}ld{\i}z}, {Ciechanowicz},
  {Dubbeldam}, {Glenz}, {Huisman}, {Lin}, {Morris}, {Murphy}, \&
  {Trappe}}]{vK10}
{van Kempen}, T.~A., {Kristensen}, L.~E., {Herczeg}, G.~J., {et~al.} 2010,
  \aap, 518, L121

\bibitem[{{van Kempen} {et~al.}(2009){van Kempen}, {van Dishoeck},
  {G{\"u}sten}, {Kristensen}, {Schilke}, {Hogerheijde}, {Boland}, {Nefs},
  {Menten}, {Baryshev}, \& {Wyrowski}}]{vK09}
{van Kempen}, T.~A., {van Dishoeck}, E.~F., {G{\"u}sten}, R., {et~al.} 2009,
  \aap, 501, 633

\bibitem[{{Vastel} {et~al.}(2001){Vastel}, {Spaans}, {Ceccarelli}, {Tielens},
  \& {Caux}}]{Va01}
{Vastel}, C., {Spaans}, M., {Ceccarelli}, C., {Tielens}, A.~G.~G.~M., \&
  {Caux}, E. 2001, \aap, 376, 1064

\bibitem[{{Visser} {et~al.}(2012){Visser}, {Kristensen}, {Bruderer}, {van
  Dishoeck}, {Herczeg}, {Brinch}, {Doty}, {Harsono}, \& {Wolfire}}]{Vi11}
{Visser}, R., {Kristensen}, L.~E., {Bruderer}, S., {et~al.} 2012, \aap, 537,
  A55

\bibitem[{{Wampfler} {et~al.}(2013){Wampfler}, {Bruderer}, {Karska}, {Herczeg},
  {van Dishoeck}, {Kristensen}, {Goicoechea}, {Benz}, {Doty}, {McCoey},
  {Baudry}, {Giannini}, \& {Larsson}}]{Wa13}
{Wampfler}, S.~F., {Bruderer}, S., {Karska}, A., {et~al.} 2013, \aap, 552, A56

\bibitem[{{Whitney} {et~al.}(2013){Whitney}, {Robitaille}, \&
  {Bjorkman}}]{Wh13}
{Whitney}, B., {Robitaille}, T., \& {Bjorkman}, J. 2013, APJS, subm.

\bibitem[{{Wilson} \& {Rood}(1994)}]{WR94}
{Wilson}, T.~L. \& {Rood}, R. 1994, \araa, 32, 191

\bibitem[{{Wright} {et~al.}(2000){Wright}, {van Dishoeck}, {Black},
  {Feuchtgruber}, {Cernicharo}, {Gonz{\'a}lez-Alfonso}, \& {de Graauw}}]{Wr00}
{Wright}, C.~M., {van Dishoeck}, E.~F., {Black}, J.~H., {et~al.} 2000, \aap,
  358, 689

\bibitem[{{Wyrowski} {et~al.}(2006){Wyrowski}, {Menten}, {Schilke},
  {Thorwirth}, {G{\"u}sten}, \& {Bergman}}]{Wy06}
{Wyrowski}, F., {Menten}, K.~M., {Schilke}, P., {et~al.} 2006, \aap, 454, L91

\bibitem[{{Y{\i}ld{\i}z} {et~al.}(2012){Y{\i}ld{\i}z}, {Kristensen}, {van
  Dishoeck}, {Belloche}, {van Kempen}, {Hogerheijde}, {G{\"u}sten}, \& {van der
  Marel}}]{Yi12}
{Y{\i}ld{\i}z}, U.~A., {Kristensen}, L.~E., {van Dishoeck}, E.~F., {et~al.}
  2012, \aap, 542, A86

\bibitem[{{Y{\i}ld{\i}z} {et~al.}(2013){Y{\i}ld{\i}z}, {Kristensen}, {van
  Dishoeck}, {San Jos{\'e}-Garc{\'{\i}}a}, {Karska}, {Harsono}, {Tafalla},
  {Fuente}, {Visser}, {J{\o}rgensen}, \& {Hogerheijde}}]{Yi13}
{Y{\i}ld{\i}z}, U.~A., {Kristensen}, L.~E., {van Dishoeck}, E.~F., {et~al.}
  2013, \aap, 556, A89

\bibitem[{{Y{\i}ld{\i}z} {et~al.}(2010){Y{\i}ld{\i}z}, {van Dishoeck},
  {Kristensen}, {Visser}, {J{\o}rgensen}, {Herczeg}, {van Kempen},
  {Hogerheijde}, {Doty}, {Benz}, {Bruderer}, {Wampfler}, {Deul}, {Bachiller},
  {Baudry}, {Benedettini}, {Bergin}, {Bjerkeli}, {Blake}, {Bontemps}, {Braine},
  {Caselli}, {Cernicharo}, {Codella}, {Daniel}, {di Giorgio}, {Dominik},
  {Encrenaz}, {Fich}, {Fuente}, {Giannini}, {Goicoechea}, {de Graauw},
  {Helmich}, {Herpin}, {Jacq}, {Johnstone}, {Larsson}, {Lis}, {Liseau}, {Liu},
  {Marseille}, {McCoey}, {Melnick}, {Neufeld}, {Nisini}, {Olberg}, {Parise},
  {Pearson}, {Plume}, {Risacher}, {Santiago-Garc{\'{\i}}a}, {Saraceno},
  {Shipman}, {Tafalla}, {Tielens}, {van der Tak}, {Wyrowski}, {Dieleman},
  {Jellema}, {Ossenkopf}, {Schieder}, \& {Stutzki}}]{Yi10}
{Y{\i}ld{\i}z}, U.~A., {van Dishoeck}, E.~F., {Kristensen}, L.~E., {et~al.}
  2010, \aap, 521, L40

\bibitem[{{Zinnecker} \& {Yorke}(2007)}]{ZY07}
{Zinnecker}, H. \& {Yorke}, H.~W. 2007, \araa, 45, 481

\end{thebibliography}

\appendix

\section{Details of PACS observations}
Table \ref{log} shows the observing log of PACS observations used in this paper. 
The observations identifications (OBSID), observation day (OD), date of observation, total integration time, 
primary wavelength ranges, and pointed coordinates (RA, DEC) are listed. All spectra were 
obtained in Pointed / Chop-Nod observing mode. Additional remarks 
are given for several sources. G327-0.6 and W33A observations were mispointed. NGC6334-I, 
W3IRS5, and NGC7538-IRS1 spectra were partly saturated and re-observed (re-obs). Two 
observations of W51N-e1, G34.26, G5.89, and AFGL2591 were done using different pointing.

\begin{table*}
\caption{\label{log} Log of PACS observations}             
\renewcommand{\footnoterule}{}  
\begin{tabular}{lccccccccccccc}     
\hline\hline       
Source & OBSID & OD & Date & Total time & Wavelength ranges & RA & DEC & Remarks\\
~ & ~ & ~ & ~ & (s) & ($\mu$m) & ($^\mathrm{h}$ $^\mathrm{m}$ $^\mathrm{s}$) & ($^{\mathrm{o}}$ $\mathrm{'}$ $\mathrm{''}$) & ~ \\   
\hline
G327-0.6 	& 1342216201 & 659 & 2011-03-04 & 6290 & 102-120 & 15 53 08.8 & -54 37 01.0 & mispointed\\
~		 	& 1342216202 & 659 & 2011-03-04 & 4403 & 55-73 & 15 53 08.8	& -54 37 01.0 & mispointed\\

W51N-e1     & 1342193697 & 327 & 2010-04-06 & 4589 & 55-73 & 19 23 43.7 & +14 30 28.8 & diff. point.\\
~			& 1342193698 & 327 & 2010-04-06 & 4969 & 102-120 & 19 23 43.7 & +14 30 25.3 & diff. point.\\

DR21(OH)    & 1342209400 & 551 & 2010-11-15 & 4401 & 55-73 & 20 39 00.8 & +42 22 48.0 & \\ 
~			& 1342209401 & 551 & 2010-11-16 & 6280 & 102-120 & 20 39 00.8 & +42 22 48.0 & \\ 

W33A	    & 1342239713 & 1018 & 2012-02-25 & 4403 & 55-73 & 18 14 39.1 & -17 52 07.0 & mispointed\\
~			& 1342239714 & 1018 & 2012-02-25 & 3763 & 102-120 & 18 14 39.1 & -17 52 07.0 & mispointed\\
~			& 1342239715 & 1018 & 2012-02-25 & 2548 & 174-210 & 18 14 39.1 & -17 52 07.0 & mispointed\\

G34.26+0.15 & 1342209733 & 542 & 2010-11-07 & 4589 & 55-73 & 18 53 18.8 &	+01 14 58.1 & diff. point.\\
~			& 1342209734 & 542 & 2010-11-07 & 4969 & 102-120 & 18 53 18.7 &	+01 15 01.5 & diff. point.\\

NGC6334-I   & 1342239385 & 1013 & 2012-02-21 & 4403 & 55-73 & 17 20 53.3 & -35 47 00.0 & saturated\\
~			& 1342239386 & 1013 & 2012-02-21 & 3763 & 102-120 & 17 20 53.3 & -35 47 00.0 & saturated\\
~			& 1342239387 & 1013 & 2012-02-21 & 2548 & 174-210 & 17 20 53.3 & -35 47 00.0 & saturated \\
~			& 1342252275 & 1240 & 2012-10-05 & 3771 & 102-120 & 17 20 53.3 & -35 46 57.2 & re-obs\\

NGC7538-I1  & 1342211544 & 589 & 2010-12-24 & 6290 & 102-120 & 23 13 45.3 & +61 28 10.0 & saturated\\
~			& 1342211545 & 589 & 2010-12-24 & 4403 & 55-73 & 23 13 45.3 & +61 28 10.0 & saturated\\
~			& 1342258102 & 1329 & 2013-01-02 & 	3771 & 102-120  & 23 13 45.2 & +61 28 10.4 & re-obs\\

AFGL2591 	& 1342208914 & 549 & 2010-11-14 & 6280 & 102-120 & 20 29 24.7 & +40 11 19.0  & diff. point.\\
~			& 1342208938 & 550 & 2010-11-15 & 4403 & 55-73 & 20 29 24.9	& +40 11 21.0  & diff. point.\\

W3-IRS5	    & 1342191146 & 286 & 2010-02-24 & 6345 & 102-120 & 2 25 40.6 & +62 05 51.0 & saturated \\
~			& 1342191147 & 286 & 2010-02-24 & 4102 & 55-73 & 2 25 40.6 & +62 05 51.0 & saturated  \\
~			& 1342229091 & 860 & 2011-09-21 & 4403 & 55-73 & 2 25 40.6 & +62 05 51.0 & saturated\\
~			& 1342229092 & 860 & 2011-09-21 & 4499 & 102-120 & 2 25 40.6 & +62 05 51.0 & re-obs\\
~			& 1342229093 & 860 & 2011-09-21 & 2249 & 55-73 & 2 25 40.6 & +62 05 51.0 & re-obs\\

G5.89-0.39  & 1342217940 & 691 & 2011-04-05 & 4969 & 102-120 & 18 00 30.5 &	-24 04 00.4 & diff. point.\\
~			& 1342217941 & 691 & 2011-04-05 & 4589 & 55-73 & 18 00 30.5 & -24 04 04.4 & diff. point.\\
\hline
\end{tabular}
\end{table*}

\section{Continuum measurements}
Table \ref{tab:cont} shows the continuum fluxes for all our sources
measured using the full PACS array. The fluxes were used in the
spectral energy distributions presented by \citet{vT13}.

\begin{table*}
\caption{\label{tab:cont} Full-array continuum measurements in 10$^{3}$ Jy}             
\renewcommand{\footnoterule}{}  
{\begin{tabular}{lcccccccccccccccccccccc}     
\hline\hline       
$\lambda$ ($\mu$m) & \multicolumn{10}{c}{Continuum (10$^{3}$ Jy)}\\     
 ~ &  G327-0.6\tablefootmark{a} & W51N-e1  & DR21(OH) & W33A & G34.26 & NGC6334I & NGC7538I1 & AFGL2591 & W3IRS5 & G5.89 \\
\hline
56.8 & 2.2 & 11.0 & 1.7 & 1.9 & 7.9 & 16.1 & 8.6 & 5.5 & 23.0 & 17.7 \\
59.6 & 2.6 & 12.3 & 2.0 & 2.1 & 8.2 & 17.2 & 8.9 & 5.7 & 23.8 & 18.8 \\
62.7 & 3.0 & 13.3 & 2.5 & 2.2 & 9.2 & 18.2 & 9.3 & 5.8 & 24.0 & 19.7 \\
63.2 & 3.0 & 13.6 & 2.6 & 2.3 & 9.4 & 18.3 & 9.3 & 5.8 & 24.2 & 19.9 \\
66.1 & 3.5 & 14.3 & 3.0 & 2.5 & 10.6 & 19.3 & 9.6 & 6.0 & 24.8 & 20.2 \\
69.3 & 3.8 & 15.2 & 3.4 & 2.5 & 10.9 & 19.8 & 9.6 & 6.0 & 24.4 & 20.7 \\
72.8 & 4.6 & 17.8 & 4.1 & 3.0 & 13.2 & 20.5 & 9.7 & 5.9 & 24.8 & 15.7 \\
76.0 & 4.9 & 18.6 & 4.5 & 2.9 & 14.0 & 20.9 & 9.6 & 5.6 & 24.4 & 15.8 \\
79.2 & 5.3 & 18.9 & 4.8 & 3.0 & 14.4 & 20.9 & 9.5 & 5.6 & 23.3 & 15.7 \\
81.8 & 5.6 & 19.2 & 5.1 & 3.2 & 14.8 & 21.1 & 9.6 & 5.7 & 23.1 & 15.7 \\
86.0 & 6.1 & 19.8 & 5.5 & 3.2 & 15.6 & 21.6 & 9.6 & 5.5 & 22.2 & 15.7 \\
90.0 & $>$3.2 & 20.1 & 5.9 & 3.4 & 16.0 & 21.8 & 9.9 & 5.5 & 21.8 & 15.3 \\
93.3 & $>$3.3 & 19.7 & 6.4 & 3.3 & 15.9 & 21.7 & 9.6 & 5.5 & 20.9 & 14.7 \\
108.8 & 8.2 & 18.8 & 7.8 & 3.9 & 16.1 & 23.7 & 10.0 & 5.3 & 19.4 & 13.1 \\
113.5 & 8.1 & 18.5 & 7.8 & 3.9 & 15.9 & 23.0 & 9.8 & 5.2 & 18.2 & 12.5 \\
118.0 & $>$7.1 & 18.1 & 7.9 & 3.8 & 15.6 & 22.2 & 9.5 & 5.0 & 17.3 & 11.9 \\
125.4 & $>$7.2 & 17.5 & 7.9 & 3.7 & 15.2 & 21.2 & 9.0 & 4.7 & 15.8 & 11.0 \\
130.4 & $>$7.2 & 16.7 & 7.8 & 3.6 & 14.6 & 20.2 & 8.5 & 4.4 & 14.6 & 10.2 \\
136.0 & 8.1 & 16.1 & 7.7 & 3.5 & 14.1 & 19.2 & 8.1 & 4.2 & 13.4 & 9.6 \\
145.5 & 8.0 & 14.7 & 7.6 & 3.3 & 13.0 & 17.4 & 7.3 & 3.8 & 11.7 & 8.4 \\
158.5 & 7.7 & 12.6 & 6.9 & 3.0 & 11.1 & 15.1 & 6.3 & 3.3 & 9.5 & 6.8 \\
164.0 & 7.5 & 11.7 & 6.6 & 2.9 & 10.3 & 14.1 & 5.8 & 3.0 & 8.6 & 6.1 \\
169.1 & 7.2 & 11.0 & 6.0 & 2.7 & 9.7 & 13.6 & 5.4 & 2.8 & 7.9 & 5.7 \\
175.8 & 6.9 & 10.0 & 5.9 & 2.7 & 8.9 & 12.3 & 5.0 & 2.6 & 7.0 & 5.0 \\
179.5 & 6.6 & 9.5 & 5.6 & 2.4 & 8.5 & 11.6 & 4.7 & 2.4 & 6.5 & 4.7 \\
186.0 & 5.6 & 8.7 & 4.8 & 2.4 & 7.8 & 10.2 & 4.0 & 2.0 & 5.5 & 4.2 \\
\hline 
\end{tabular}}
\tablefoot{
The calibration uncertainty of 20\% of the flux should be included for comparisons with other 
modes of observations or instruments.
\tablefoottext{a}{One spaxel at N-W corner of the PACS map is saturated around 100 $\mu$m region due to the strong 
continuum; therefore the tabulated values are the lower limits to the total continuum 
flux from the whole map.}
}
\end{table*}

\section{Tables with fluxes and additional figures}
Table \ref{obs1} shows line fluxes and 3 $\sigma$ upper limits of CO lines toward all 
our objects in units of 10$^{-20}$ W cm$^{-2}$. For details, see the table caption.

\begin{figure*}[tb]
\begin{center}
\includegraphics[angle=90,height=13cm]{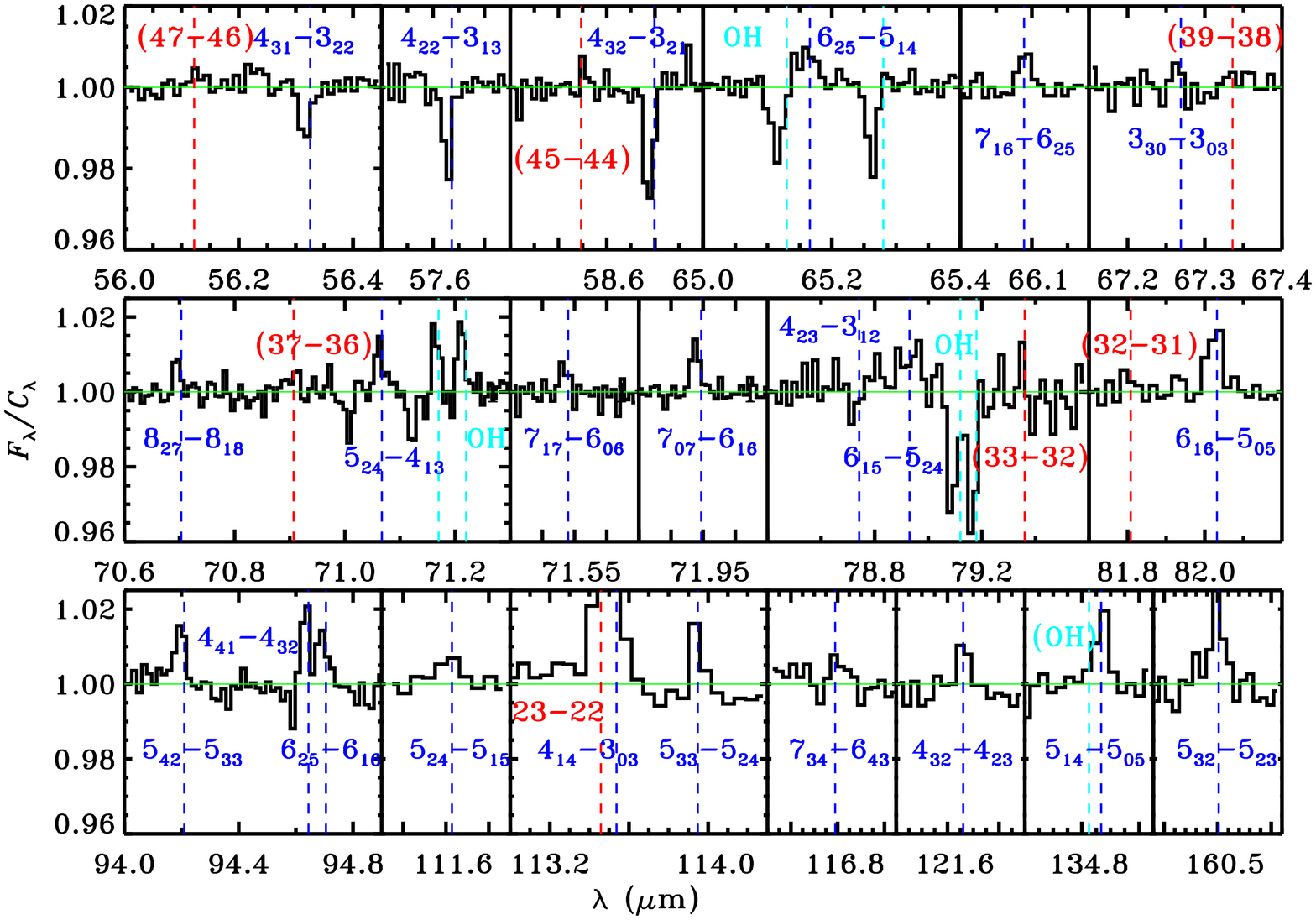}
\caption{\label{wblow} Close-ups of several of the H$_2$O, CO and OH lines in W3IRS5 are shown in Figure 
\ref{spec}. The rest wavelength of each line is indicated by dashed 
lines: blue for H$_2$O, red for CO and light blue for OH. Identifications of the undetected 
lines in the presented spectral regions are shown in brackets.}
\end{center}
\end{figure*}

\begin{figure*}[tb]
\begin{center}
\includegraphics[angle=90,height=13cm]{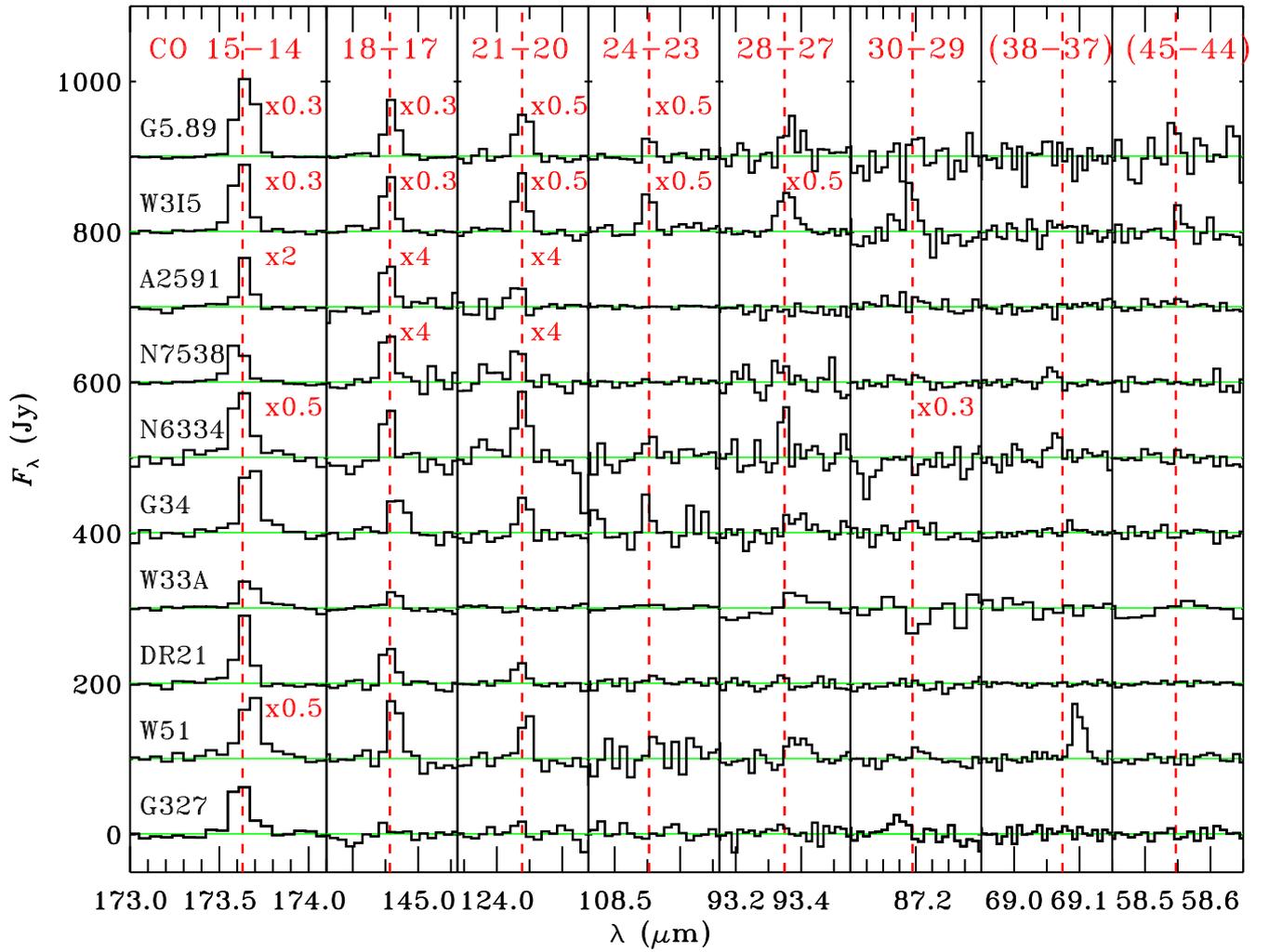}
\vspace{0.5cm}

\caption{\label{cozoom} Close-ups of several transitions of CO lines in the PACS wavelength range towards 
all sources. The spectra are continuum subtracted and shifted vertically for better visualization.}
\end{center}
\end{figure*}

\begin{figure*}[tb]
\begin{center}
\includegraphics[angle=90,height=10cm]{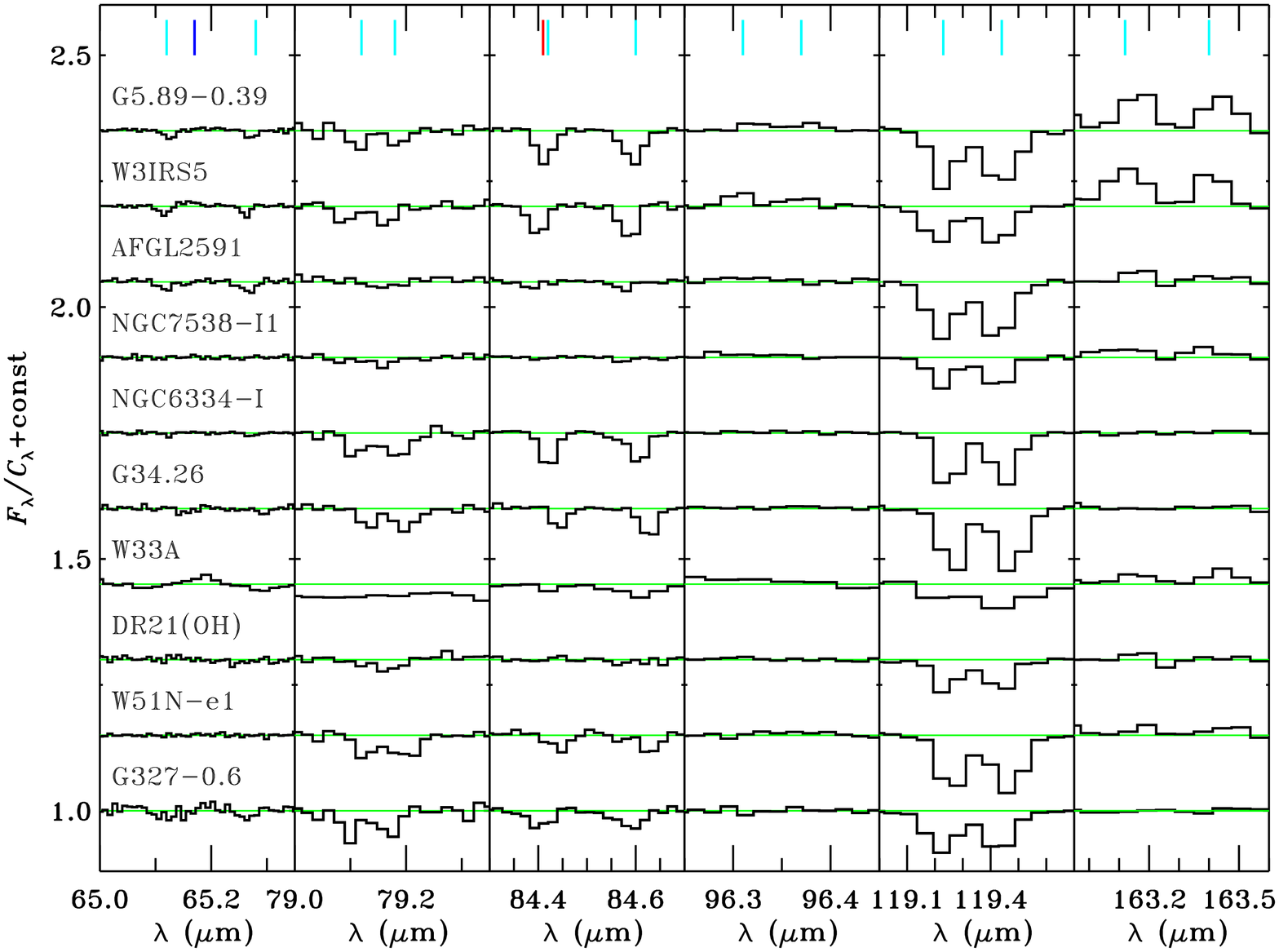}
\vspace{0.5cm}

\caption{\label{ohprof} Normalized spectra of OH doublets for all our sources at central 
position. Doublets at 71 and 98 $\mu$m are excluded because of poor calibration of those 
spectral regions observed with PACS. OH doublet at 84.4 $\mu$m is a blend with the CO 31-30 
line, whereas OH at 65.13 $\mu$m can be affected by H$_2$O 6$_{25}$-5$_{14}$ at 65.17 $\mu$m.}
\end{center}
\end{figure*}

Figure \ref{wblow} show blow-ups of selected spectral regions of W3 IRS5 with high-$J$ CO, 
H$_{2}$O, and OH lines. Figures \ref{cozoom} and \ref{ohprof} show blow-ups of selected CO 
and OH transitions towards all sources. 

\begin{landscape}
\begin{table}
\caption{\label{obs1} CO line fluxes in 10$^{-20}$ W cm$^{-2}$}             
\renewcommand{\footnoterule}{}  
\resizebox{1.3\textheight}{!}
{\begin{tabular}{lrcccccccccccc}     
\hline\hline       
Trans. & $\lambda_\mathrm{lab}$ ($\mu$m) & \multicolumn{10}{c}{Central position flux (10$^{-20}$ W cm$^{-2}$)}\\     
 ~ & ~ & G327-0.6\tablefootmark{a} & W51N-e1  & DR21(OH) & W33A\tablefootmark{a} & G34.26+0.15 & NGC6334-I & NGC7538-I1 & AFGL2591 & W3IRS5 & G5.89-0.39 \\
\hline
CO 14-13 & 185.9990 &  8.52(0.48) & 40.93(1.26) & 29.93(0.67) & 7.96(0.23) & 23.05(1.14) & 40.07(1.60) & 17.79(0.52) & 10.86(0.45) & 76.67(1.38) & 80.36(0.74)\\
CO 15-14 &173.6310 &  9.34(0.36) & 50.65(1.83) & 22.65(1.07) & 4.91(0.32) & 26.30(1.07) & 52.70(2.39) & 14.54(0.33) & 7.72(0.57) & 86.13(1.76) & 103.91(1.41)\\
CO 16-15 &162.8120 &  8.82(0.51) & 43.10(1.43) & 24.71(0.92) & 4.12(0.18) & 22.71(1.06) & 43.88(1.59) & 11.78(0.39) & 8.00(0.27) & 99.73(1.08) & 105.49(1.76)\\
CO 17-16 &153.2670 &  3.00(0.76) & 36.94(1.54) & 16.19(1.21) & 2.74(0.24) & 18.52(1.63) & 31.37(2.46) & 8.60(0.56) & 5.67(0.38) & 86.40(1.31) & 96.31(1.67)\\
CO 18-17 &144.7840 &  2.13(0.48) & 27.78(1.89) & 17.40(1.13) &  4.35(0.39)     & 18.27(1.53) & 20.93(2.02) & 5.82(0.47) & 5.02(0.53) & 83.05(1.69) & 82.73(1.67)\\
CO 19-18 &137.1960 &  6.76(0.64) & 24.18(1.81) & 13.82(1.07) &   3.41(0.37)    & 20.16(1.68) & 36.07(2.02) & 4.96(0.42) & 3.02(0.58) & 79.39(1.54) & 60.19(2.14)\\
CO 20-19 &130.3690 &    6.68(0.81)    & 25.73(1.42) & 15.02(0.93) &  3.54(0.52)    & 21.93(1.93) & 37.25(2.19) & 5.31(0.54) & 4.34(0.64) & 77.08(2.27) & 59.14(1.27)\\
CO 21-20 &124.1930 &    3.96(1.24)    & 22.77(3.15) & 11.38(0.87) &   $<$0.75    & 18.69(2.02) & 34.56(5.43) & 4.62(0.72) & 2.97(0.61) & 60.64(2.13) & 51.31(2.90)\\
CO 22-21 &118.5810 &    4.61(1.16)    & 20.31(3.29) & 9.79(1.23) &   $<$0.77    & 21.66(3.21) & 35.05(5.02) & 3.97(0.89) &   $<$0.90    & 63.44(2.81) & 44.98(3.84)\\
CO 24-23 &108.7630 &    $<$1.78    &   $<7.05$  & 5.18(1.40) &   $<$0.54    & 18.86(5.45) & 12.89(3.28) &   $<$1.68    &   $<$0.86    & 49.51(3.75) & 21.89(3.45)\\
CO 27-26 & 96.7730 &    $<$0.86    & 4.71(0.79) & 3.68(0.56) &   $<$1.50    & 4.34(1.22) & 9.46(1.76) & $<$2.49 &   $<$1.01    & 23.29(2.65) & 6.16(0.81)\\
CO 28-27 & 93.3490 &   $<$1.75    & 6.28(1.21) &   $<$0.94    &   $<$0.91    & 5.06(1.35) & 11.75(1.99) & $<$4.97 &   $<$1.04    & 24.72(1.75) & 9.71(1.63)\\
CO 29-28 & 90.1630 &    $<$1.24    & 5.23(1.54) &   $<$1.23    &   $<$1.25    & 3.67(1.30) & 10.80(5.36) & $<$6.35 &   $<$1.70    & 20.23(1.94) & 5.38(3.49)\\
CO 30-29 & 87.1900 &    $<$1.10    & 3.59(0.72) &   $<$1.00    &   $<$1.72    & 4.48(1.10) &   $<$7.00    & $<$2.10 &   $<$1.57    & 14.38(2.03) & 6.32(2.59)\\
CO 32-31 & 81.8060 &    $<$1.19    & $<$5.84 &   $<$1.19    &   $<$1.43    &   $<$2.31    &   $<$4.63    &   $<$1.84    &   $<$2.03    &   $<$9.49    &   $<$3.70   \\
CO 33-32 & 79.3600 &    $<$3.99    &   $<$11.04 &   $<$1.70    &   $<$2.28    &   $<$3.76    &   $<$18.46    &   $<$2.09    &   $<$1.99    &   $<$3.40    &   $<$13.59   \\
CO 34-33 & 77.0590 &    $<$1.88    &   $<$4.57  &   $<$1.83    &   $<$1.83    &   $<$3.17    &   $<$2.71    &   $<$1.40    &   $<$2.16    &   $<$5.27    &   $<$6.93   \\
CO 35-34 & 74.8900 &    $<$2.06    &   $<$2.90  &   $<$0.99    &   $<$1.27    &   $<$2.03    &   $<$3.09    &   $<$2.50    &   $<$1.20    &   $<$1.83    &   $<$18.61   \\
CO 36-35 & 72.8430 &    $<$0.55    &   $<$0.63    &   $<$0.29    &   $<$0.70    &   $<$1.17    &   $<$1.74    &   $<$0.69    &   $<$0.37    & $<$7.03 &   $<$3.00 \\  
CO 38-37 & 69.0740 &    $<$1.36    &   $<$1.06    &   $<$1.93    &   $<$0.67    &   $<$1.77    &   $<$1.21    & $<$3.23 &   $<$2.09    &   $<$2.16    &   $<$2.74   \\
CO 39-38 & 67.3360 &    $<$1.20    &   $<$0.84    &   $<$0.40    &   $<$1.86    &   $<$0.80    &   $<$2.50    & $<$2.40 &   $<$0.89    &   $<$1.99    &   $<$3.54   \\
CO 40-39 & 65.6860 &    $<$0.57    &   $<$2.40    &   $<$0.43    &   $<$1.05    &   $<$2.40    &   $<$1.94    &   $<$2.17    &   $<$0.94    &   $<$2.63    &   $<$2.91   \\
CO 41-40 & 64.1170 &    $<$1.06    &   $<$1.11    &   $<$0.57    &   $<$1.80    &   $<$1.54    &   $<$2.91    &   $<$1.59    &   $<$1.80    & $<$5.56 &   $<$4.60 \\  
CO 42-41 & 62.6240 &    $<$1.03   &   $<$2.74    &   $<$0.44    &   $<$1.08    &   $<$2.50    &   $<$2.19   &   $<$1.91    &   $<$2.34    &   $<$4.01    &   $<$5.01   \\
CO 43-42 & 61.2010 &    $<$0.90    &   $<$1.39    &   $<$0.58   &   $<$1.05    &   $<$1.20    &   $<$4.59    &   $<$2.10    &   $<$2.41    &   $<$4.94    &   $<$6.21   \\
CO 44-43 & 59.8430 &    $<$0.67    &   $<$1.61    &   $<$0.39    &   $<$0.68    &   $<$1.41    &   $<$3.38    & $<$2.30 &   $<$1.54    &   $<$2.85    &   $<$3.49 \\  
CO 45-44 & 58.5470 &    $<$0.96    &   $<$1.24    &   $<$0.55    &   $<$1.41    &   $<$1.42    &   $<$1.34    &   $<$1.69    &   $<$1.08    & $<$6.77 &   $<$4.89   \\
CO 46-45 & 57.3080 &    $<$2.33   &   $<$10.54   &   $<$1.03    &   $<$1.81    &   $<$4.01    &   $<$6.49    &   $<$3.75    &   $<$2.65    &   $<$6.46    &   $<$8.54   \\
CO 47-46 & 56.1220 &    $<$0.76    &   $<$1.08    &   $<$0.50    &   $<$2.29    &   $<$ 1.85    &   $<$2.51    &   $<$1.39    &   $<$1.71    & $<$5.78 &   $<$4.79 \\  
\hline
\end{tabular}}    
\tablefoot{
The uncertainties are 1$\sigma$ measured in the continuum on both sides of each line; 
calibration uncertainty of 20\% of the flux should be included for comparisons with other 
modes of observations or instruments. 3$\sigma$ upper limits calculated using 
wavelength dependent values of full-width high maximum for a point source observed with PACS
are listed for non-detections. CO 23-22 and CO 31-30 fluxes are not listed due to severe blending with
the H$_2$O 4$_{14}$-3$_{03}$ line at 113.537 $\mu$m and the OH $\frac{7}{2}$,$\frac{3}{2}$-$\frac{5}{2}$,$\frac{3}{2}$ 
line at 84.4 $\mu$m, which are often in absorption. CO 25-24, CO 26-25 and CO 37-36 transitions 
are located in the regions of overlapping orders, where the flux calibration is unreliable. 
\tablefoottext{a}{A mispointed observation. The fluxes are calculated from a sum of two spaxel closest to the 
true source position, see Section 2.}
}
\end{table}
\end{landscape}

\section{OH in low and intermediate mass sources}
\begin{figure*}[!tb]
\begin{center}
\includegraphics[angle=0,height=10cm]{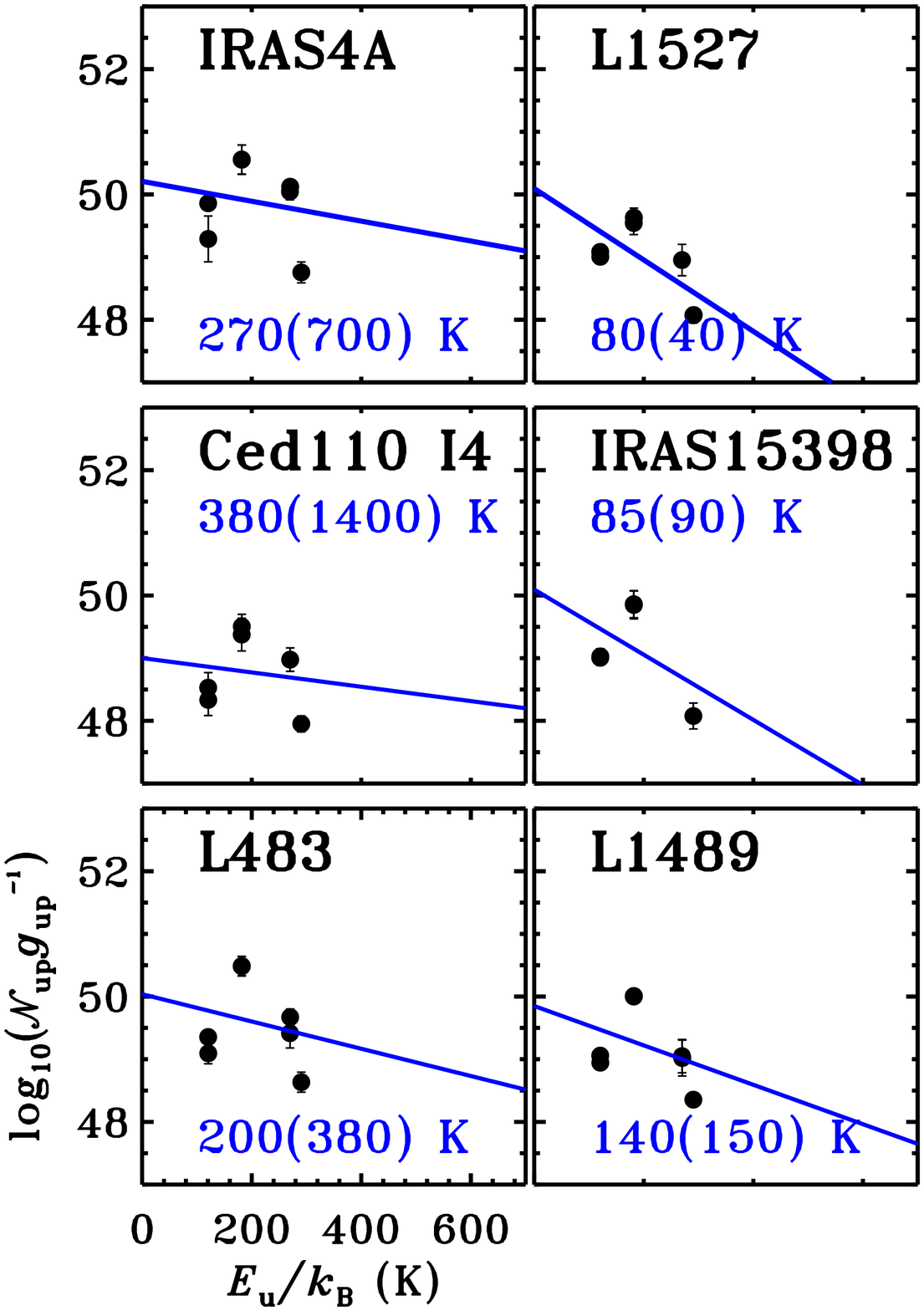}
\includegraphics[angle=0,height=10cm]{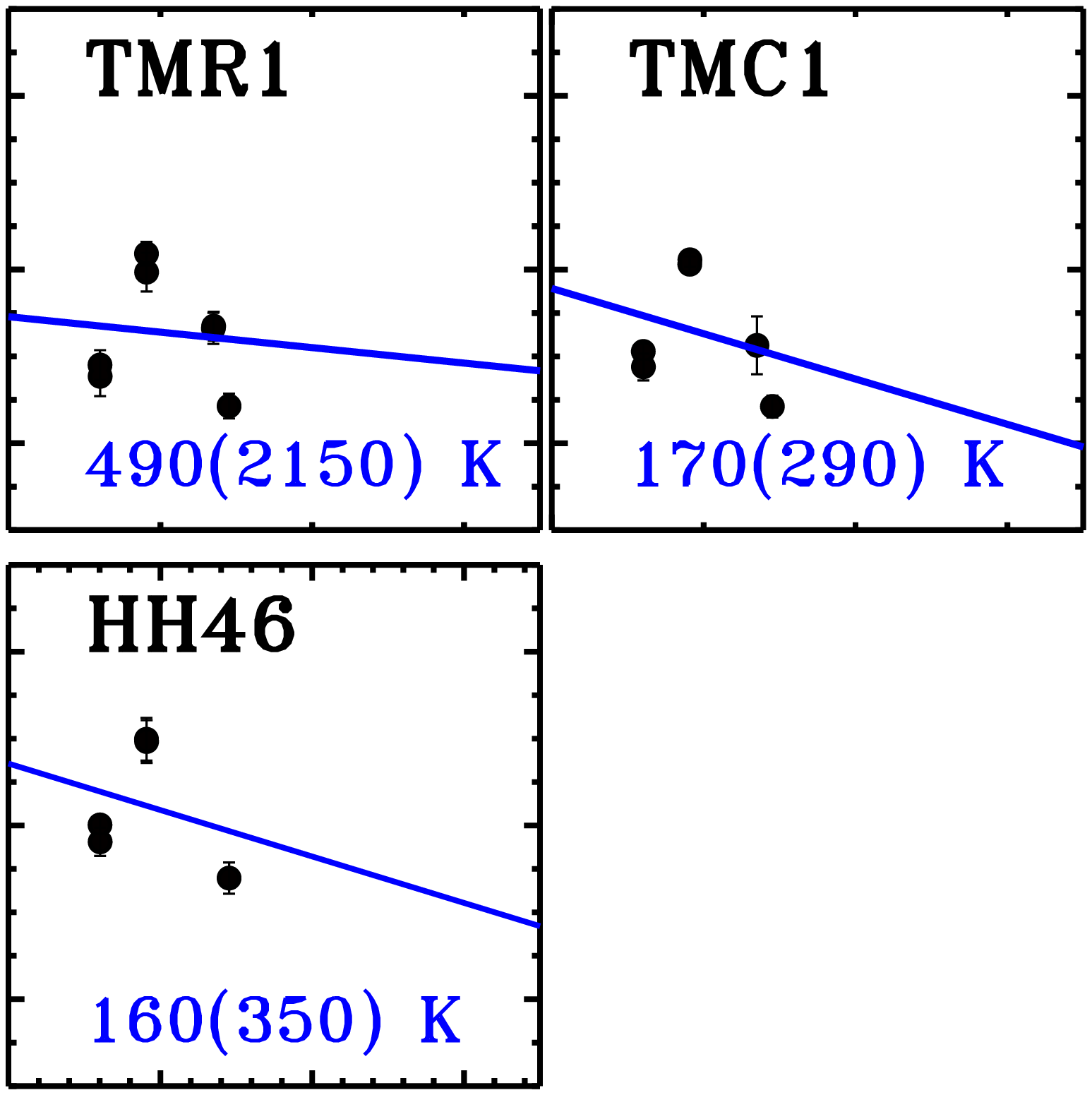}
\vspace{1cm}

\caption{\label{ohem} OH rotational diagrams (from emission lines) for low-mass young 
stellar objects (fluxes from Wampfler et al. 2013).}
\end{center}
\end{figure*}

\begin{figure}[!tb]
\begin{center}
\includegraphics[angle=0,height=10cm]{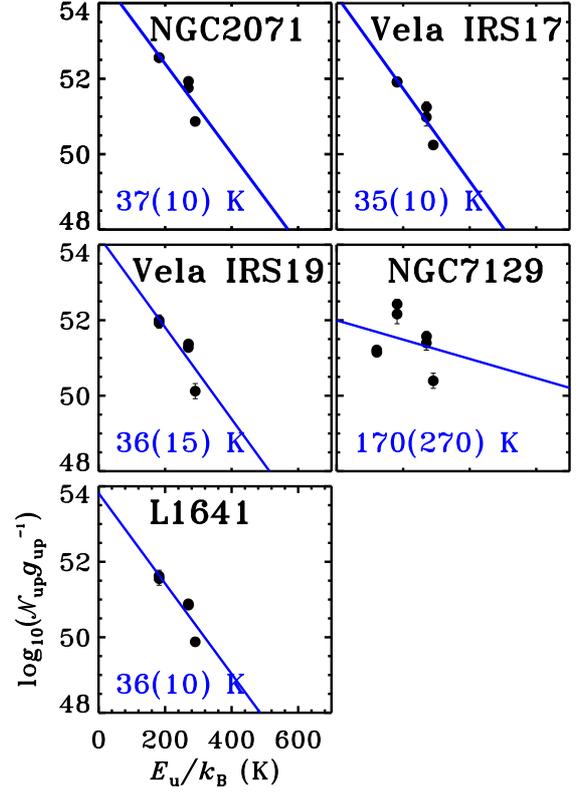}
\vspace{1cm}

\caption{\label{ohem2} OH rotational diagrams (from emission lines) for intermediate-mass young 
stellar objects (fluxes from Wampfler et al. 2013).}
\end{center}
\end{figure}

Figures \ref{ohem} and \ref{ohem2} show rotational diagrams of OH for low- and intermediate-mass 
young stellar objects based on the fluxes presented in \citet{Wa13}. Only the sources with 
at least 3 detected doublets in emission (out of 4 targeted in total) are shown in diagrams. 
Rotational temperatures and total numbers of emitting molecules are summarized in Table \ref{tab:exc3}.
In case of low-mass YSOs, a single component fit is usually not a good approximation 
(with the exception of L1527 and IRAS15398). In the intermediate-mass YSOs, on the other hand,
such approximation holds and results in a very similar rotational temperatures of 
OH $T_\mathrm{rot}\sim35$ K for all sources except NGC7129 FIRS2.
\begin{table}
\caption{\label{tab:exc3}OH rotational excitation and number of emitting molecules 
$\mathcal{N}_\mathrm{u}$ based on emission lines for low- and intermediate-mass sources}
\centering
\renewcommand{\footnoterule}{}  
\begin{tabular}{lcccc}
\hline \hline
Source & $T_\mathrm{rot}$(K) & $\mathrm{log}_\mathrm{10}\mathcal{N}$ \\
\hline
\multicolumn{3}{c}{Low-mass YSOs} \\
\hline 
NGC1333 I4A & 270(700)  & 52.4(0.9) \\
L1527       & \textbf{80(40)} & 51.5(0.6) \\
Ced110 I4   & 380(1400) & 51.4(0.9) \\
IRAS15398   & \textbf{85(90)} & 51.5(1.0) \\
L483 	    & 200(380) & 52.0(0.9) \\
L1489 		& 140(150) & 51.6(0.7) \\
TMR1		& 490(2150) & 52.0(0.8) \\
TMC1		& 170(290) & 51.7(0.9) \\
HH46		& 160(350) & 52.6(1.1) \\
\hline
\multicolumn{3}{c}{Intermediate-mass YSOs} \\
\hline 
NGC2071		& 37(10) & 55.7(0.8) \\
Vela IRS17  & 35(10) & 55.1(0.7) \\
Vela IRS19  & 36(15) & 55.1(1.1) \\
NGC7129 FIRS2 & 170(270) & 53.9(0.9) \\
L1641 S3MMS1 & 36(10) & 54.7(0.8) \\
\hline
\end{tabular}
\tablefoot{Rotational diagrams are shown in Figures \ref{ohem} and \ref{ohem2}. 
Objects with at least 3 detected doublets in Wampfler et al. (2013) are presented.
Rotational temperatures of OH calculated with error less than 100 K are shown in boldface.}
\end{table}


\end{document}